\documentclass[a4paper,11pt]{article}
\pdfoutput=1 

\usepackage{jheppub} 

\usepackage[T1]{fontenc} 
\usepackage{graphicx}
\usepackage{amsmath}
\usepackage{amssymb}
\usepackage{multirow}

\newcommand{\mrm}[1]{\mathrm{#1}}

\renewcommand{\c}{{\mathrm c}}
\renewcommand{\d}{{\mathrm d}}
\newcommand{\e}{{\mathrm e}}

\newcommand{\g}{{\mathrm g}}

\newcommand{\p}{{\mathrm p}}
\newcommand{\q}{{\mathrm q}}
\newcommand{\s}{{\mathrm s}}
\renewcommand{\t}{{\mathrm t}}
\renewcommand{\u}{{\mathrm u}}

\renewcommand{\H}{{\mathrm H}}

\newcommand{\W}{{\mathrm W}}
\newcommand{\Z}{{\mathrm Z}}

\newcommand{\cbar}{\overline{\mathrm c}}
\newcommand{\dbar}{\overline{\mathrm d}}

\newcommand{\pbar}{\overline{\mathrm p}}
\newcommand{\qbar}{\overline{\mathrm q}}

\newcommand{\sbar}{\overline{\mathrm s}}
\newcommand{\tbar}{\overline{\mathrm t}}
\newcommand{\ubar}{\overline{\mathrm u}}

\newcommand{\Pom}{\mathbb{P}}

\newcommand{\as}{\alpha_{\mathrm{s}}}

\newcommand{\shat}{\hat{s}}

\newcommand{\sigmahat}{\hat{\sigma}}

\newcommand{\kT}{k_{\perp}}

\newcommand{\pT}{p_{\perp}}
\newcommand{\pTs}{p^2_{\perp}}

\newcommand{\pTmin}{p_{\perp\mathrm{min}}}

\newcommand{\pTo}{p_{\perp 0}}

\title{Hard Diffraction\\ with Dynamic Gap Survival}

\author{Christine O. Rasmussen}
\author{and Torbj\"orn Sj\"ostrand}
\affiliation{Theoretical Particle Physics\\ Department of
Astronomy and Theoretical Physics\\Lund Unicersity,\\S\"olvegatan
14A\\SE-223 62 Lund, Sweden}

\emailAdd{christine.rasmussen@thep.lu.se}
\emailAdd{torbjorn@thep.lu.se}

\abstract{
We present a new framework for the modelling of hard diffraction in 
$\p\p$ and $\p\pbar$ collisions. It starts from the the approach 
pioneered by Ingelman and Schlein, wherein the single diffractive 
cross section is factorized into a Pomeron flux and a Pomeron PDF. 
To this it adds a dynamically calculated rapidity gap survival
factor, derived from the modelling of multiparton interactions.
This factor is not relevant for diffraction in $\e\p$ collisions,
giving non-universality between HERA and Tevatron diffractive event
rates. The model has been implemented in \textsc{Pythia}~8
and provides a complete description of the hadronic state associated
with any hard single diffractive process. Comparisons with $\p\pbar$ 
and $\p\p$ data reveal improvement in the description of single 
diffractive events.}

\begin{document} 
\maketitle
\flushbottom

\section{Introduction}

The nature of diffractive excitation in hadron-hadron collisions
remains a bit of a mystery. We may motivate \textit{why} it happens, 
e.g.\ based on the optical analogy that lies behind its name, or in
the related Good-Walker formalism \cite{Good:1960ba}.
But to explain \textit{how} diffractive events are produced, and 
with what properties, is a longer story. In a first step the single 
diffractive cross section should be describable as a function of the 
diffractive mass $M$ and the squared momentum transfer $t$. 
In a second step the generic properties of a diffractive system 
of a given mass should be explained: multiplicity distributions, 
rapidity and transverse momentum spectra and other event 
characteristics. 
In a third step the existence and character of exclusive diffractive 
processes and the underlying events associated therewith should be 
understood.

Over the years much data has accumulated, and many models have been
presented, but so far without any model that explains all aspects of 
the data, and without any consensus which models are the most relevant
ones. It is beyond the scope of the current article to review all
the data and models; for a selection of relevant textbooks and reviews 
see \cite{Perl1974,Forshaw:1997dc,Jung:1998ed,Barone:2002cv,Donnachie:2002en,%
Jung:2002mx,Albrow:2010yb}. 

For the path we will follow in this article, Regge theory provides the
basic mathematical framework. In it, poles in the plane of complex 
spin $\alpha$ may be viewed as the manifestations of hadronic 
resonances in the crossed channels. A linear trajectory of poles 
$\alpha(t) = \alpha(0) + \alpha' t$ corresponds to a 
$\sigma_{\mathrm{tot}} \sim s^{\alpha(0)-1}$. Several trajectories 
appear to exist, but for high-energy applications the most important 
is the Pomeron ($\Pom$) one, which with its $\alpha(0) > 1$ is 
deemed responsible for the observed rise of the total cross section, 
and in modern terminology would correspond to a set of glueball 
states. With single-Pomeron exchange as the starting point, higher 
orders involve multiple Pomeron exchanges, and also interactions 
between the Pomerons being exchanged, driven by a triple-Pomeron 
vertex. Out of this framework the cross section for various 
diffractive topologies can be derived, differentially in mass and 
$t$, given a set of numbers that must be extracted from data.

Such models do not address the structure of the diffractive system.
The fireball models of older times implied isotropic decay in the rest 
frame of the diffractive system, or possibly elongated along the 
collision axis, but without internal structure. 
The Ingelman-Schlein (IS) model \cite{Ingelman:1984ns} made the bold 
assumption that the exchanged Pomeron could be viewed as a hadronic 
state, and that therefore a diffractive system could be described as a 
hadron-hadron collision at a reduced energy. This implied the existence
of Parton Distribution Functions (PDFs) for the Pomeron. Thereby also 
hard processes became available, confirmed by the observation of 
jet production in diffractive systems \cite{Bonino:1988ae}. The PomPyt 
program \cite{Bruni:1993is} combined Pomeron fluxes and PDFs, largely 
determined from HERA data, with the \textsc{Pythia} event generator 
of the time \cite{Sjostrand:1993yb} to produce complete hadronic final 
states, and PomWig \cite{Cox:2000jt} did similarly for Herwig 
\cite{Corcella:2000bw}.

One limitation of these models is that they are restricted to the 
exchange of one Pomeron per hadron-hadron collision, not the multiple
ones expected in Regge theory. Translated into a QCD-based,  
more modern view of such collisions, Multiple Partonic Interactions
(MPIs) occur between the incoming hadrons \cite{Sjostrand:1987su}.
That is, since hadrons are composite objects, there is the possibility 
for several partons from a hadron to collide, predominantly by 
semisoft $2 \to 2$ QCD interactions. These together create colour flows
(strings \cite{Andersson:1983ia}) criss-crossing the event, typically
filling up the whole rapidity range between the two beam particles with
hadron production. Thereby a ``basic'' process containing a rapidity 
gap can lose that by MPIs. (MPIs and soft colour exchanges could also be 
sources of gaps \cite{Buchmuller:1995qa,Edin:1995gi}, a possibility 
we will not study further in this article, so as to keep the discussion 
focussed.)

A spectacular example is Higgs production by gauge-boson fusion, 
$\W^+\W^- \to \H^0$ and  $\Z^0\Z^0 \to \H^0$, where the naive process 
should result in a large central gap only populated by the Higgs decay 
products, since no colour exchange is involved. Including MPIs, this 
gap largely fills up \cite{Dokshitzer:1991he}, although a fraction of 
the events should contain no further MPIs \cite{Bjorken:1992er}, a 
fraction denoted as the Rapidity Gap Survival Probability (RGSP). 
Such a picture has been given credence by the observation of 
``factorization breaking'' between HERA and the Tevatron: the 
Pomeron flux and PDFs determined at HERA predicts about an order of 
magnitude more QCD jet production than observed at the Tevatron, e.g.\
\cite{Affolder:2000vb}.

In this article the intention is to provide a dynamical description 
of such factorization breaking, as a function of the hard process 
studied and its kinematics, and to predict the resulting event 
structure for hard diffraction in hadronic collisions. This is done 
in three steps. Firstly, given a hard process selected based on the 
inclusive PDFs, the fraction of a PDF that should be associated with 
diffraction is calculated, as a convolution of the Pomeron flux and 
its PDFs. Secondly, the full MPI framework of \textsc{Pythia}, 
including also the effects of initial- and final-state radiation, 
is applied to find the fraction of events without any further MPIs. 
Those events that survive these two steps define the diffractive event
fraction, while the rest remain as regular nondiffractive events.
Thirdly, diffractive events may still have MPIs within the $\Pom\p$
subsystem, and therefore the full hadron-hadron underlying-event
generation machinery is repeated for this subsystem. The nondiffractive
events are kept as they are in this step.

One should not expect perfect agreement with data in this approach;
there are too many uncertainties that enter in the description. 
Neveretheless a qualitative description can be helpful, not only to 
understand the trend of existing data, but also to pave the way for
future studies. The new framework we present here has been implemented
as an integrated part of the \textsc{Pythia}~8.2 event generator
\cite{Sjostrand:2014zea}, and can be switched on for any standard
hard process. It thereby complements the already existing modelling
of soft diffraction, i.e.\ of diffractive events with no discernible
hard process. The dividing line between these two descriptions is not
sharp, and in the future we will explore tensions between the two.

As should be clear from this introduction, our model is ``just'' a
combination of the existing IS and RGSP ideas, and thus not anything
fundamentally new. The devil lies in the details, however, and to
the best of our knowledge nobody has previously worked out a complete 
model of this character. 

The article is structured as follows. In section 2 we introduce the 
new model framework, which then is validated in section 3. Some tentative 
comparisons with data are presented in section 4. The article concludes
with a summary and outlook in section 5.  

\section{The model}

In this article we study hard diffraction, so this means we assume 
the presence of some hard process in the events of interest. 
Standard examples would be jet, $\Z^0$ and $\W^{\pm}$ production. 
By factorization a cross section involving partons $i,j$ from incoming 
beams $A,B$ can be written as 
\begin{equation}
\sigma = \sum_{i,j} \iint \d x_1 \, \d x_2 \, 
f_{i/A}(x_1, Q^2) \, f_{j/B}(x_2, Q^2) \, 
\sigmahat_{ij}( \shat = x_1 x_2 s, Q^2) ~,
\end{equation}
where $\sigmahat$ is the parton-level cross section, integrated over 
relevant further degrees of freedom, like a $\pT$ range for jets.   

Assuming Pomerons to have some kind of existence inside the proton,
in the Ingelman-Schlein spirit, we introduce a Pomeron flux
$f_{\Pom/\p}(x_{\Pom},t)$, where $x_{\Pom}$ is the $\Pom$ momentum 
fraction and $t$ its (spacelike) virtuality. The $\Pom$ has a partonic 
substructure, just like a hadron, and thus we can define PDFs 
$f_{i/\Pom}(x, Q^2)$. The PDF could also depend on the $t$ scale, 
just like the photon has a PDF strongly dependent on its virtuality. 
For lack of a model for such a dependence we assume the $\Pom$ PDF
is a suitable average over the $t$ range probed. As a consequence  
we will not need $t$ for most of the studies, and so it can be 
integrated out of the flux, 
$f_{\Pom/\p}(x_{\Pom}) = \int f_{\Pom/\p}(x_{\Pom},t) \, \d t$.

Given the ansatz with Pomeron flux and PDF, the PDF of a proton can 
be split into one regular nondiffractive (ND) and one $\Pom$-induced 
diffractive (D) part,
\begin{equation}
f_{i/\p}(x, Q^2) = f_{i/\p}^{\mathrm{ND}}(x, Q^2) 
+ f_{i/\p}^{\mathrm{D}}(x, Q^2) ~,
\end{equation}
where
\begin{align}
\label{eq:fipD}
f_{i/\p}^{\mathrm{D}}(x, Q^2) &=  \int_0^1 \d x_{\Pom} \, 
  f_{\Pom/\p}(x_{\Pom})\,  \int_0^1 \d x' \, 
  f_{i/\Pom}(x', Q^2) \, \delta(x - x_{\Pom} x') \nonumber \\
&=  \int_x^1  \frac{\d x_{\Pom}}{x_{\Pom}} \,  
  f_{\Pom/\p}(x_{\Pom}) \, f_{i/\Pom} \left( 
  \frac{x}{x_{\Pom}}, Q^2 \right) ~.
\end{align}
The assumption that the diffractive part 
$f_{i/\p}^{\mathrm{D}}(x, Q^2; x_{\Pom}, t)$ of the full PDF can be 
decomposed in this way is in approximate agreement with the HERA data
\cite{Aaron:2012hua}. 

For two incoming protons (or antiprotons, or other hadrons)
$A$ and $B$, an initial probability for diffraction 
$\mathcal{P}^{\mathrm{D}} \approx \mathcal{P}_A^{\mathrm{D}}
+ \mathcal{P}_B^{\mathrm{D}}$ is obtained from the ratio 
of diffractive to inclusive PDFs,
\begin{align}
\label{eq:Probabilities}
\mathcal{P}_A^{\mathrm{D}} &= \frac{f_{i/B}^{\mathrm{D}}(x_B, Q^2)}
{f_{i/B}(x_B, Q^2)}~~~\mathrm{for}~~~AB \to XB~, \nonumber \\
\mathcal{P}_B^{\mathrm{D}} &= \frac{f_{i/A}^{\mathrm{D}}(x_A, Q^2)}
{f_{i/A}(x_A, Q^2)}~~~\mathrm{for}~~~AB \to AX~,
\end{align}
where $\mathcal{P}_{A/B}^{\mathrm{D}}$ is the probability for side $A/B$ 
to be the diffractive system, thus being dependent on the variables of 
the opposite side.

This probability is used to determine, on an event-by-event basis, 
the nature of the selected hard scattering, whether diffractive or not.
Currently we concentrate on single diffraction. 
A natural extension would be to associate the product 
$\mathcal{P}_A^{\mathrm{D}}\mathcal{P}_B^{\mathrm{D}}$ with central 
diffraction (CD), where two Pomerons collide and one parton is 
extracted from each $\Pom$. It would also be possible to 
extend the formalism such that part of the SD rate is reassigned
as double diffraction (DD), where the hard collision happens 
inside one of the two diffractive systems. Neither CD nor DD are 
considered in this first study, however. Instead, for the fraction
$\mathcal{P}_A^{\mathrm{D}}\mathcal{P}_B^{\mathrm{D}}$ of events,
which normally is small anyway, a random choice is made between 
$AB \to AX$ and $AB \to XB$.  
 
The key aspect of the model is now that it contains a dynamical gap 
survival. This means that we do not allow any further MPIs to occur 
between the two incoming hadrons, so as to ensure that the gap survives. 
In practise the tentative classification as diffractive, based on 
eq.~(\ref{eq:Probabilities}), initially has no consequences: 
all events are handled as if they were nondiffractive hadron-hadron 
collisions. 

Only if no additional MPIs occur does a diffractive 
classification survive and only then is the $\mathbb{P}\mathrm{p}$ 
subsystem set up. Specifically the $x_{\Pom}$ value is selected 
according to the distribution implied by eq.~(\ref{eq:Probabilities}), 
and also a $t$ value is selected for the outgoing proton. 
Technically, it is only at this stage that ``pure'' samples of 
diffractive events can be selected, should one
wish to single out such events.  

Once the $\Pom\p$ system has been set up, it is  allowed to develop 
a partonic structure just like any hadron-hadron collision. Both 
initial-state radiation (ISR) and final-state radiation (FSR) 
thereby dress the original hard process by the emission of further
softer partons. Also further MPIs inside this system are allowed,
based on the $f_{i/\Pom}(x, Q^2)$ PDFs, successively modified to 
take into account the momentum and flavours already carried away by 
the MPI, ISR and FSR activity at $\pT$ scales above the currently 
considerd one, just like for nondiffractive systems. 

The ISR/FSR/MPI description is based on the perturbative parton picture.
Nonperturbative aspects have to be added to this. Beam remnants carry 
the momentum not kicked out of the incoming $\Pom$ and $\p$. For the 
former a fictitious ``valence quark'' content of either $\d\dbar$ 
or $\u\ubar$ is chosen at random for each new event. It is essentially
equivalent to having a gluon as remnant, but is slightly more convenient.
All outgoing partons are colour-connected by colour flux lines 
- strings - that fragment to produce the primary hadrons of the
final state. The colour flow in an event is not unambiguously 
determined, however, and data suggest that colours tend to be 
more correlated than naively comes out of the perturbative picture,
a phenomenon known as Colour Reconnection (CR).  
 
We can by combining these two simple ideas give an explanation of 
the discrepancies between Tevatron and HERA. The dynamical 
gap survival introduces an additional suppression factor, reducing 
the number of diffractive events without any additional parameters. 

\subsection{Pomeron fluxes and PDFs} 

For numerical studies it is necessary to specify Pomeron flux and
PDF parametrizations. There are currently seven parametrizations/models 
for the former and five for the latter available in \textsc{Pythia}. 

The parametrizations for the Pomeron flux 
$f_{\Pom/\p}(x_{\Pom},t)$ are 
\begin{itemize}
\item Schuler-Sj\"ostrand model (SaS) \cite{Schuler:1993wr}, 
\item the Bruni-Ingelman model \cite{Bruni:1993ym}, 
\item the Streng-Berger model \cite{Jung:1993gf},
\item the Donnachie-Landshoff model \cite{Donnachie:1984xq}, 
\item the Minimum Bias Rockefeller model (MBR)
\cite{Ciesielski:2012mc} with an option to renormalize the flux, and
\item the H1 models Fit A and B \cite{Aktas:2006hy, Aktas:2006hx}. 
\end{itemize}
All have to obey an approximate form 
$f_{\Pom/\p}(x_{\Pom}) \sim 1/ x_{\Pom}$ in order to obtain an 
approximate diffractive mass spectrum $\sim \d M_X^2 / M_X^2$,
as required by Regge theory and by data. Just like the rise of the
total cross section requires a supercritical Pomeron 
$\alpha(0) = 1 + \epsilon > 1$, with $\epsilon \approx 0.08$,
several of the fluxes have adapted this steeper slope
$f_{\Pom/\p}(x_{\Pom}) \sim 1/ x_{\Pom}^{1 + 2\epsilon}$ (where the factor 
of 2 in front of $\epsilon$ comes from the optical theorem). There are 
also some attempts to account for an excess in the low-mass resonance 
region. The $t$ dependence is typically parametrized as a single 
exponential $f_{\Pom/\p}(x_{\Pom}, t) \sim \exp(B_{\mrm{SD}}t)$, but also 
as a sum of two exponentials, or as a (power-like) dipole form
factor. The MBR model differs from the others, since the model
renormalizes the flux to unity. This renormalization suppresses
the flux, thus making the dynamical gap survival obsolete.
In order to make direct comparisons to the other available
flux-models, we have implemented the renormalization as an
option with the default being the non-renormalized flux. 

The parametrizations for the Pomeron PDFs $f_{i/\Pom}(x, Q^2)$ are 
\begin{itemize}
\item PomFix, a simple (toy) $Q^2$-independent parametrization,  
\item the H1 Fit A and B NLO PDFs \cite{Aktas:2006hy}, 
\item the H1 Jets NLO PDF \cite{Aktas:2007bv}, and 
\item the H1 Fit B LO PDF \cite{Aktas:2006hy}, 
\item the ACTW B PDF with $\epsilon=0.14$ \cite{Alvero:1998ta},  
\item the ACTW D PDF with $\epsilon=0.14$ \cite{Alvero:1998ta},  
\item the ACTW SG PDF with $\epsilon=0.14$ \cite{Alvero:1998ta},  
\item the ACTW D PDF with $\epsilon=0.19$ \cite{Alvero:1998ta}.  
\end{itemize}
The first of these has a momentum sum of unity, whereas the latter
are not normalized to any specific value, the argument used being 
that the Pomeron is not a real particle and so does not obey that kind
of constraints \cite{Donnachie:1987gu, Martin:2006td}. 
(Technically H1 chose to normalize 
the $\Pom$ flux to unity at $x_{\Pom} = 0.003$, and then let the PDF 
normalization float. No normalisation constraints are included in
the ACTW PDFs, as this is primarily set by the normalisation of
the DL flux. Thus the momentum sum of these PDFs range
from 0.5 to 2, depending on fit.) 
Pragmatically it could be argued that what is 
measured is the convolution of the $\Pom$ flux and the $\Pom$ PDF, 
so that is is feasible to shuffle any constant number between the two. 
Unfortunately this makes it less trivial to mix freely, and makes it  
almost a necessity to combine H1 PDFs with H1 fluxes. In
\textsc{Pythia}~8, it is only allowed to combine the ACTW PDFs with 
the DL flux, as these have been fitted together, and each of the
fits uses different $\epsilon$ values. 

No attempts have been made to exclude or validate different flux--PDF 
combinations in the light of more recent HERA data than available 
at the time of the fits; this woud be a separate project. We do note, 
however, that a more recent ZEUS article \cite{Chekanov:2009aa} 
compares a new ZEUS DPDF SJ fit with the H1 Fit B, showing disagreements
on the 10--20\% level. For our purposes this is an acceptable uncertainty,
and we will often use Fit B as a reference, but keep an open mind to 
variations. 

This is not the end of the story from an event-generator point of view,
however. In most of the available PDF parametrizations the momentum sums to 
approximately 0.5, but this does not mean that half of the $\Pom$ momentum 
in the 
$\Pom\p$ collision can just be thought away. At the very least this 
other half has to be considered as an inert component that sails 
through without interacting, but is present in the beam remnant. 
A further complication arises when MPIs are introduced. Normally
these are generated in a sequence of decreasing $\pT$, with the 
PDFs for an MPI adjusted to take into account the momentum 
and flavours carried away by the preceding MPIs. So if 0.4 of the 
$\Pom$ momentum has already been taken, does it mean that 0.1 or 0.6 
of it remains? This is an issue that did not exist at HERA, where 
MPIs are negligible outside of the photoproduction region.  
The choice made in \textsc{Pythia} is to assume that the full $\Pom$ 
momentum is available for MPIs. Furthermore we allow the option to
rescale the PDFs by a constant factor so as to change the momentum,
notably by a factor of two to restore (approximately) the momentum
sum rule. This should then be compensated by a corresponding rescaling
of the $\Pom$ flux in the opposite direction. That way the $\Pom$ can be 
brought closer to an ordinary hadron, and more $\Pom$ flux-PDF
combinations can be used.

Another problem is that most PDF fits are NLO ones. Given the sparsity 
of data, it should be clear that ``NLO accuracy'' does not mean the same 
thing as it does for the inclusive proton PDF. Since \textsc{Pythia} 
only contains LO matrix elements (MEs) for QCD processes there is no 
extra bonus for using NLO PDFs. Worse, it is well known that the 
gluon PDF (of the proton) is much smaller in NLO than in LO for small 
$x$ and $Q^2$; in principle it can even become negative. This behaviour 
compensates for the NLO MEs being larger than the LO ones in this region, 
but the compensation is nontrivial. Therefore an all-LO description, 
for all its weaknesses, is more robust in the small-$\pT$ region, 
which is where the MPI machinery largely operates. 
The default choice thus is  H1 Fit B LO. 
  
Finally also the inclusive proton PDF $f_{i/\p}(x, Q^2)$ should be chosen.
Here several options come with \textsc{Pythia}, and many more can be
obtained via the interfaces to LHAPDF5 and LHAPDF6 
\cite{Whalley:2005nh,Buckley:2014ana}.
The current default set is the NNPDF 2.3 QCD+QED LO one with 
$\as(M_{\Z}) = 0.130$ \cite{Ball:2012cx}. The argument for using LO has already 
been outlined above. Since the proton PDF is much better constrained  
than that of the $\Pom$, there is less of a point in varying it between 
different options consistent with current $\p$ data.
Note that, for diffractive events, the dependence
on the original choice of proton PDF is largely removed on the 
$\Pom$ side by applying eq.~(\ref{eq:Probabilities}). It does remain
on the proton side, and in the dynamical calculation of rapidity gap
survival, however.

\subsection{MPI phenomenology}

The QCD $2 \to 2$ processes are dominated by $t$-channel gluon exchange,
which gives a perturbative cross section 
$\d\sigmahat /\d\pTs \sim \as^2(\pTs) / \pT^4$ that diverges in the
$\pT \to 0$ limit. Two modifications are needed to make sense out of
this divergence. 

Firstly a divergent integrated QCD cross section should not necessarily 
be construed as a divergent total $\p\p$ cross section. Rather a 
\mbox{$\mu = \sigma^{\mrm{tot}}_{\mrm{2 \to 2}}/\sigma^{\mrm{tot}}_{\p\p} > 1$} 
for $\pT > \pTmin$ should be interpreted as implying  an average of 
$\mu$ such partonic interactions per $\p\p$ collision. Overall 
energy-momentum conservation will reduce the naively calculated rate,
but would still kick out essentially all beam momentum if we allow
$\pTmin \to 0$, in contradiction with the presence of well-defined
beam jets wherein a single particle can carry an appreciable fraction
of the incoming beam momentum. 

Secondly, therefore, it is important to note the presence of a 
screening mechanism: whereas standard perturbation theory is based 
on asymptotically free incoming states, reality is that partons are 
confined inside colour singlet states. This introduces a nonperturbative
scale of the size of a hadron, or rather of the average distance $d$ 
between two opposite-colour charges. In this spirit we introduce a 
free parameter $\pTo \sim 1/d$ that is used to dampen the cross section
\begin{equation}
\label{eq:dampMPI}
\frac{\d\sigma}{\d\pTs} \propto \frac{\as^2(\pTs)}{\pT^4}
\longrightarrow \frac{\as^2(\pTo^2 + \pTs)}{(\pTo^2 + \pTs)^2} ~.
\end{equation}
Technically the dampening is implemented as an extra factor multiplying
the standard QCD $2 \to 2$ cross sections, but could equally well have
been associated with a dampening of the PDFs; it is only the product 
of these that enters in measurable quantities.

Empirically, a $\pTo$ scale of 2 -- 3 GeV is required to describe data.
This scale is larger than expected from the proton size alone, and is
also in a regime where normally one would expect perturbation theory 
to be valid. The $\pTo$ scale appears to increase slowly with energy,
which is consistent with the growth of the number of gluons at smaller
$x$ values, leading to a closer-packing of partons and thereby a 
reduced screening distance $d$. A similar parametrization is chosen 
as for the rise of total cross section
\begin{equation}
\label{eq:pToMPI}
\pTo(E_{\mathrm{CM}}) = p_{\perp 0}^{\mathrm{ref}} \times 
  \left( \frac{E_{\mathrm{CM}}}{E_{\mathrm{CM}}^{\mathrm{ref}}} 
  \right)^{E_{\mathrm{CM}}^{\mathrm{pow}}} ,
\end{equation}
with $E_{\mathrm{CM}}^{\mathrm{pow}}$ and $p_{\perp 0}^{\mathrm{ref}}$ being
tunable parameters and $E_{\mathrm{CM}}^{\mathrm{ref}}$ a reference energy 
scale.

With the protons being extended objects, the amount of overlap between 
two incoming ones strongly depends on the impact parameter $b$. 
A small $b$ will allow for many parton-parton collisions, 
i.e.\ a high level of MPI activity, and a close-to-unity probability 
for the incoming protons to interact. A large $b$, on the other hand,
gives less average activity and a higher likelihood that the protons 
pass by each other unaffected. Diffractive events predominantly occur 
in peripheral collisions, a concept well-known already from the optical
point of view. In our approach it comes out naturally since we only 
allow one interaction to occur, namely the hard process of interest; 
if there is a second one this will fill the rapidity gap and kill the 
diffractive nature.    

The shape of the proton and the resulting overlap -- the convolution
of the two incoming proton distributions -- is not known in any detail.
The proton electric charge distribution may give some hints, but
measures quarks only and not gluons, and is in the static limit.
Instead a few different simple parametrizations can be chosen:  
\begin{itemize}
\item a simple Gaussian, offering no free parameters,
\item a double Gaussian, i.e. a sum of two Gaussians with different radii
and proton momentum fractions, and 
\item an overlap of the form $\exp(-b^p)$ (which does not correspond 
to a simple shape for the individual proton), with $p$ a free parameter.
\end{itemize}
(A further option is a Gaussian with an $x$-dependent width, but this 
has not been implemented in a diffractive context.)
All are normalized to unit momentum sum for the incoming partons, and 
an overall radius normalization factor is fixed by the total cross section.

The more uneven the matter distribution, the broader will the charged
multiplicity distribution be. Notably the higher the overlap for central 
collisions, the higher the tail to very large multiplicities.  
Also other measures, like forward-backward correlations, probe the 
distribution. Unfortunately it is always indirectly, and closely 
correlated with other model details. As an example we can mention that  
the earliest tunes worked with a much lower $\pTo$ than today and with
double Gaussians rather far away from the single-Gaussian behaviour.
This changed when more modern PDFs started to assume a steeper rise of 
the gluon PDF at small $x$, and when the \textsc{Pythia} parton showers
were extended to apply to all MPIs rather than only the hardest one, 
and for some other improvements over the years. Currently best fits are 
not very far away from a simple Gaussian, e.g. with an overlap like 
$\exp(-b^{1.85})$, but still on the side of more peaked than a Gaussian.

An event that contains a high-$\pT$ interaction is likely to be more 
central than one that does not, since the former has more MPIs and
therefore more chances that the hardest of these reaches a high $\pT$.
This bias effect is included in the choice of a $b$ for an event where
the hardest interaction has been given, and is used in the subsequent
generation of MPIs. For the current study of hard diffraction this means
that the hard process is initially picked biased towards smaller $b$
values, but afterwards the central $b$ region is strongly suppressed 
because the likelihood of several MPIs is so big there.   

Starting from a hard interaction scale, and a selected $b$, the
probability for an MPI at a lower scale has the characteristic form
\begin{equation}
\label{eq:probMPI}
\frac{\d\mathcal{P}}{\d\pTs} = O(b) \, \frac{1}{\sigma_{\mrm{ref}}}
\, \frac{\d\sigma_{\mrm{QCD}}}{\d\pTs} ~.
\end{equation}
Here $O(b)$ is the overlap enhancement/depletion factor,
$\d\sigma_{\mrm{QCD}}$ the differential cross section for
all $2 \to 2$ QCD processes, and $\sigma_{\mrm{ref}}$ the total cross
section for the event classes affected by the QCD processes. 
Historically $\sigma_{\mrm{ref}}$ has been equated with the nondiffractive
cross section in \textsc{Pythia}, on the assumption that diffraction
only corresponds to a negligible fraction of $\d\sigma_{\mrm{QCD}}$.
Within the current framework a reformulation to use the full inelastic
cross section would make sense, but would require further work and 
retuning, and is therefore left aside for now.

Given eq.~(\ref{eq:probMPI}) as a starting point, MPIs can be generated 
in a falling $\pT$ sequence, using a Sudakov-style formalism akin to 
what is used in parton showers. Actually, in the complete generation
the MPI, ISR and FSR activity is interleaved into one common 
$\pT$-ordered chain of interactions and branchings, with one common
``Sudakov form factor'', down to the respective cutoff scales.

In the current case, the MPI formalism is used twice. Firstly, to
determine whether an event is diffractive, and if not to generate 
the complete nondiffractive event. Secondly, for diffractive events, 
to determine the amount of MPI activity within the $\Pom\p$ system.
Here eq.~(\ref{eq:probMPI}) can be reused, but with new meaning for
the components of the equation. 
\begin{itemize}
\item The $\d\sigma_{\mrm{QCD}}/\d\pT$ is now evaluated using the 
$\Pom$ PDF on one side, but with the same damping as in 
eq.~(\ref{eq:dampMPI}), where $E_{\mrm{CM}}$ in eq.~(\ref{eq:pToMPI}) 
is now the $\Pom\p$ invariant mass. If the $\Pom$ 
is supposed to have a smaller size than the proton then this could be 
an argument for a higher $\pTo$ in this situation, but we have not
here pursued this.
\item The $\sigma_{\mrm{ref}}$ now represents the $\Pom\p$ total cross
section, an unknown quantity that relates to the normalizations of 
the $\Pom$ flux and $\Pom$ PDF. By default is is chosen to have a
fixed value of 10~mb, higher than is normally quoted in literature. 
This way, with other quantities at their default settings, the charged 
multiplicity of a $\Pom\p$ collision agrees reasonably well with
that of a nondiffractive $\p\p$ one at the same invariant mass. 
This may not be the best of arguments, but is a reasonable first 
choice that is experimentally testable, at least in principle.
\item The $O(b)$ factor may be changed, see next.
\end{itemize}

The impact parameter $b_{\Pom\p}$ of the $\Pom\p$ subcollision does 
not have to agree with the $b_{\p\p}$ of the whole $\p\p$ collision.
It introduces the transverse matter profile of the Pomeron, even 
less known than that of the proton. Generally a Pomeron is supposed 
to be a smaller object in a localized part of the proton, but one should 
keep an open mind. For lack of better, three possibilities have been
implemented, which can be compared to gauge the impact of this uncertainty. 
\begin{itemize}
\item $b_{\Pom\p} = b_{\p\p}$. This implicitly assumes that a Pomeron 
is as big as a proton and centered in the same place. Since small 
$b_{\p\p}$ values already have been suppressed, by the MPI selection 
step, it implies that few events will have high enhancement factors. 
\item $b_{\Pom\p} = \sqrt{b_{\p\p}}$ (where normalization is such that 
$\langle b \rangle = 1$ for minimum-bias events). This can crudely be 
motivated as follows. In the limit that the $\Pom$ is very tiny, such 
that the proton matter profile varies slowly over the width of the $\Pom$,
then what matters is where the Pomeron strikes the other proton. 
Thus the variation of $O(b)$ with $b$ is that of one proton, not two, 
and so the square root of the normal variation, loosely speaking. 
Technically this is messy to implement, but the current simple 
recipe provides the main effect of reducing the variation, bringing 
all $b$ values closer to the average. 
\item Pick a completely new $b_{\Pom\p}$, as was done with $b_{\p\p}$
in the first place. This allows a broad spread from central to peripheral 
values, and thereby also a larger and more varying MPI activity inside 
the diffractive system than the other two options, and thereby offers
a useful contrast. 
\end{itemize}

\section{Validation}

In this section we summarise some of the tests and sanity checks we 
have performed on the model implementation. This provide insight 
into how the model operates and with what general results, but also 
highlights the uncertain nature of many of the components of the 
model. 

In the model we have two options for when an event is classified as 
diffractive: either right after the event has passed the PDF selection 
criterion, eq.~(\ref{eq:Probabilities}), or after passing the further 
MPI criterion. Results using only the former will from now on be denoted 
``PDF selected'', and with the latter in addition ``MPI selected''. 
Our full model for hard diffraction corresponds to the latter, but the 
intermediate level is helpful in separating the effects of these two
rather different physics components.

Notably, many distributions tend to be mainly determined by one of the 
two criteria. Those mainly sensitive to the PDF selection include the 
$x_{\mathbb{P}}$ and thereby the mass of the diffractive system, and
the squared momentum transfer $t$ of the process and thereby the 
scattering angle $\theta$ of the outgoing proton. In particular we 
will explore the dependence on Pomeron fluxes and PDFs. Aspects 
that depend on the details of the MPI model include several particle
distributions, such as multiplicities, and that will also be highlighted.

The key number where both components are comparably important is the 
overall diffractive rate, where each of them gives an order-of-magnitude
suppression, resulting in a $\sim$1\% fraction of hard events being 
of a diffractive nature. This number thereby receives a considerable 
overall uncertainty.

\subsection{The Pomeron flux and PDF}

\begin{figure}[tbp]
\begin{minipage}[t]{0.5\textwidth}
\centering
\includegraphics[scale=0.4]{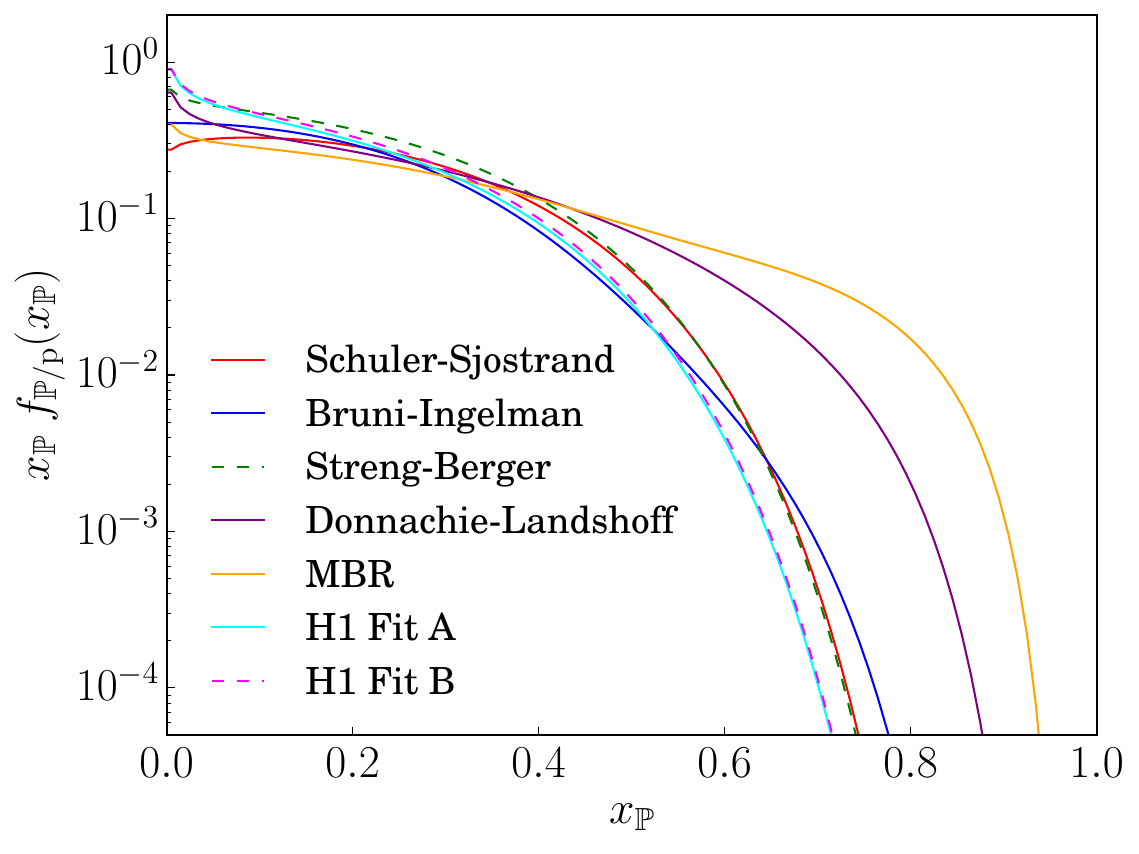}\\
(a)
\end{minipage}
\hfill
\begin{minipage}[t]{0.5\textwidth}
\centering
\includegraphics[scale=0.4]{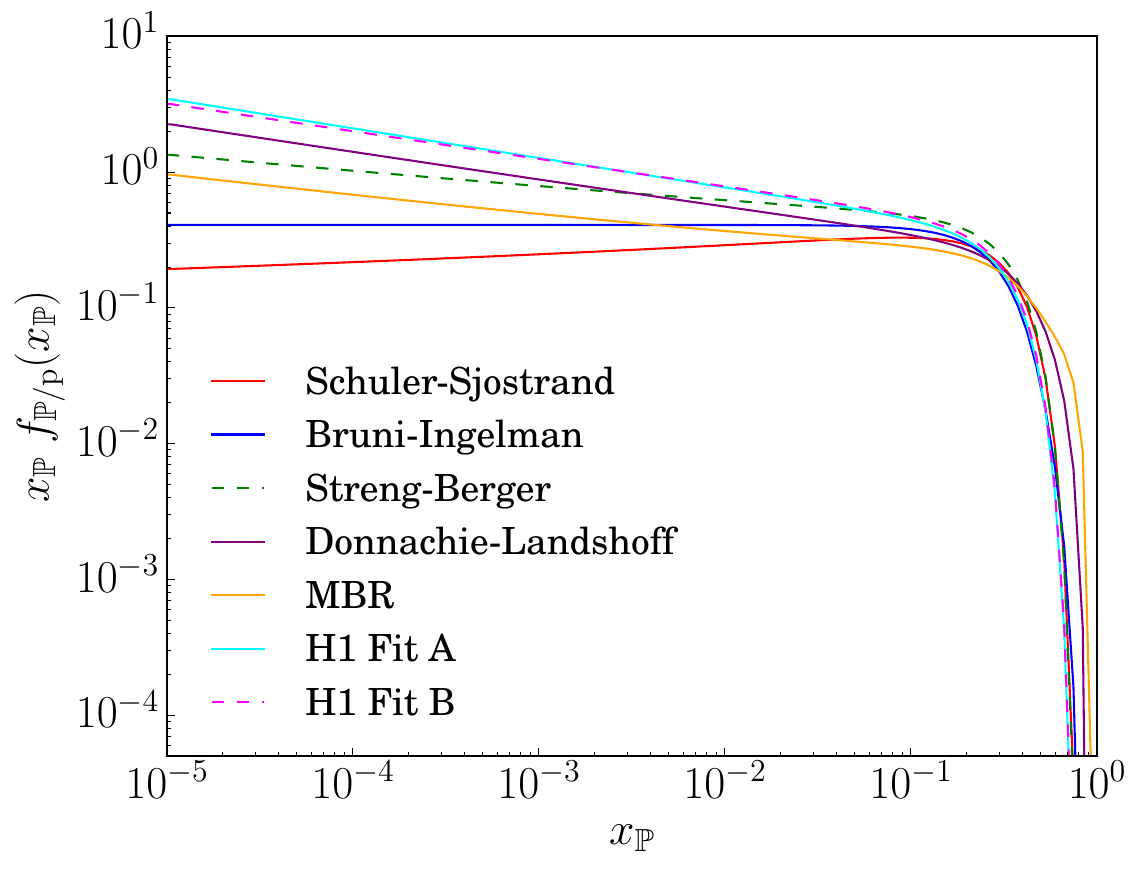}\\
(b)
\end{minipage}
\caption{\label{Fig:PomeronFluxes}
The seven different Pomeron fluxes included in \textsc{Pythia} 
on linear (a) and logarithmic scale (b). Note that the MBR flux
has not been renormalized (see \cite{Ciesielski:2012mc}).}
\end{figure}

We begin by studying the effects of variations of the $\Pom$ 
parametrizations. In figures~\ref{Fig:PomeronFluxes}a and 
\ref{Fig:PomeronFluxes}b the seven different 
Pomeron fluxes are compared. As can be seen there is a considerable 
spread. Even in the region of medium $x_{\Pom}$ values, 
$x_{\Pom} \sim 0.1$, this corresponds to more than a factor of two 
between the extremes. The dramatic differences at large $x_{\Pom}$ 
are not readily visible, since a 
large-$x_{\Pom}$ event usually corresponds to a small rapidity gap 
and therefore is difficult to discern from non-diffractive events. 
The limit of small $x_{\Pom}$ generally is more interesting, tying in
with the intercept of the Pomeron trajectory, but plays a lesser role 
for the current study of hard diffraction.

\begin{figure}[tbp]
\begin{minipage}[t]{0.5\textwidth}
\centering
\includegraphics[scale=0.4]{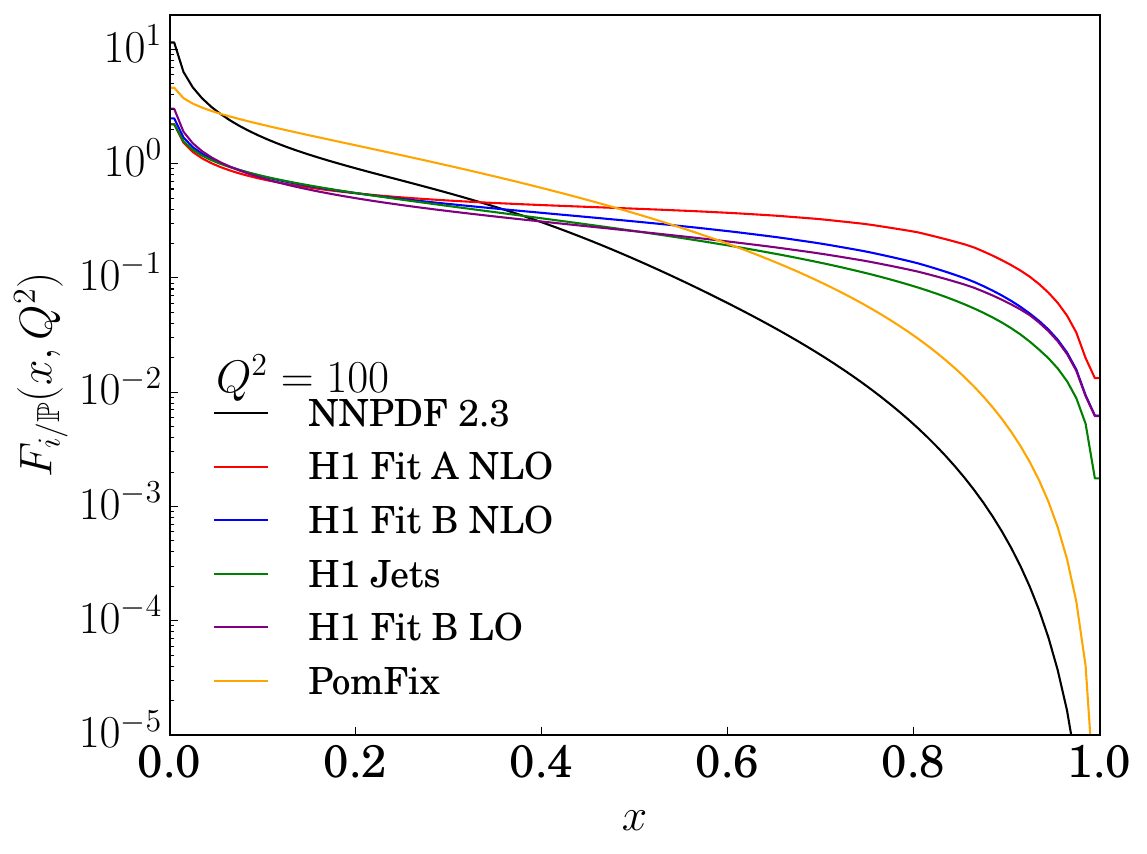}\\
(a)
\end{minipage}
\hfill
\begin{minipage}[t]{0.5\textwidth}
\centering
\includegraphics[scale=0.4]{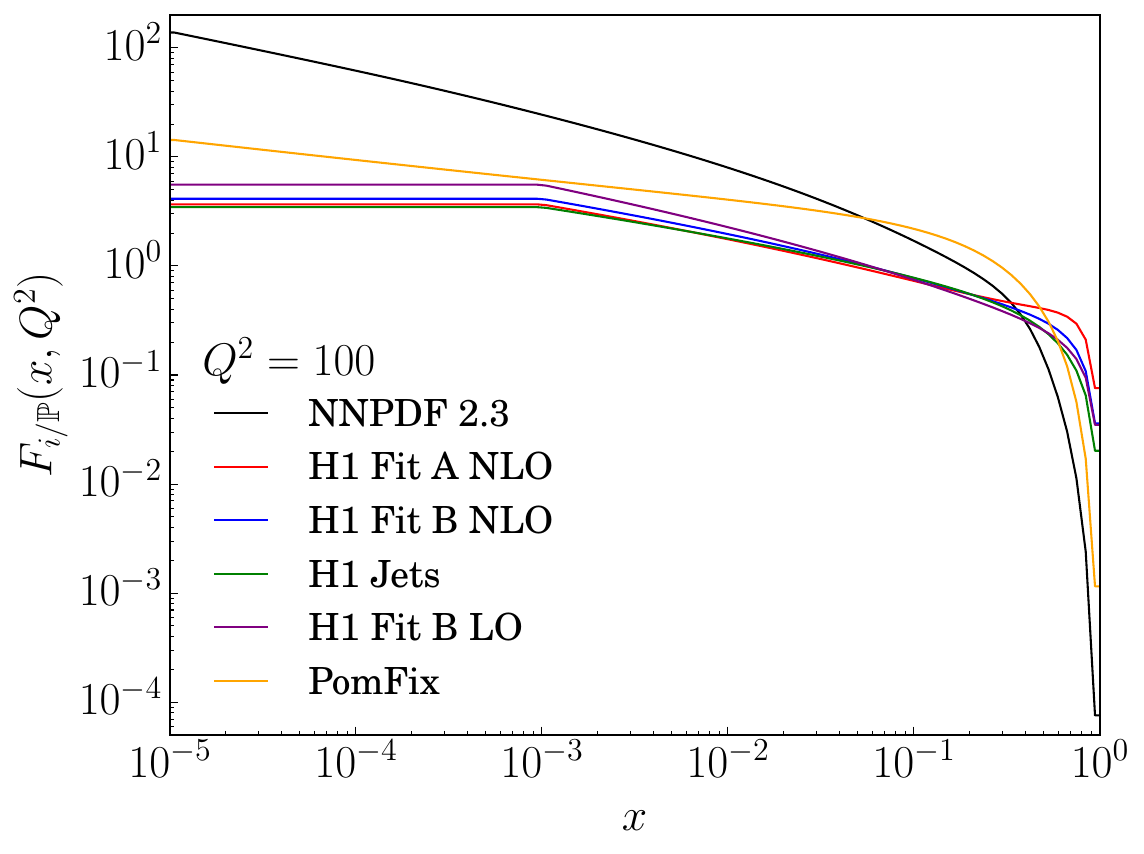}\\
(b)
\end{minipage}
\begin{minipage}[t]{0.5\textwidth}
\centering
\includegraphics[scale=0.4]{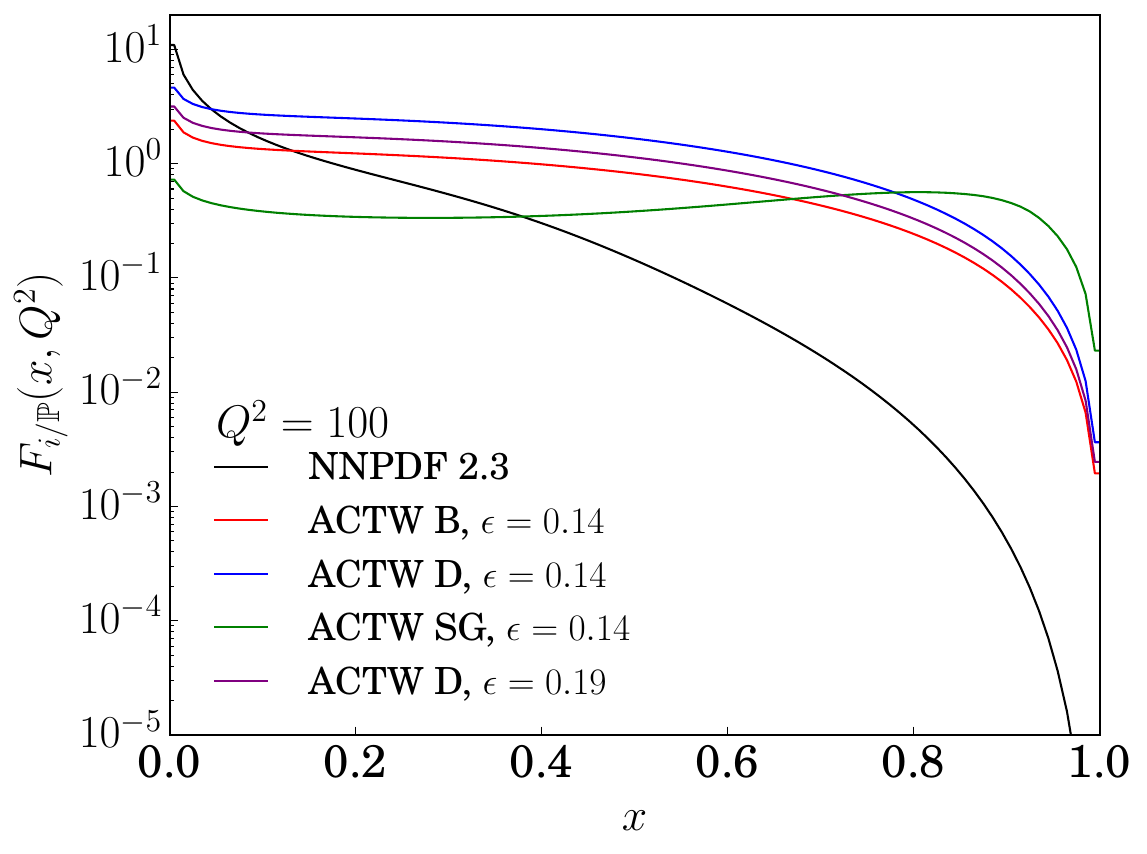}\\
(c)
\end{minipage}
\hfill
\begin{minipage}[t]{0.5\textwidth}
\centering
\includegraphics[scale=0.4]{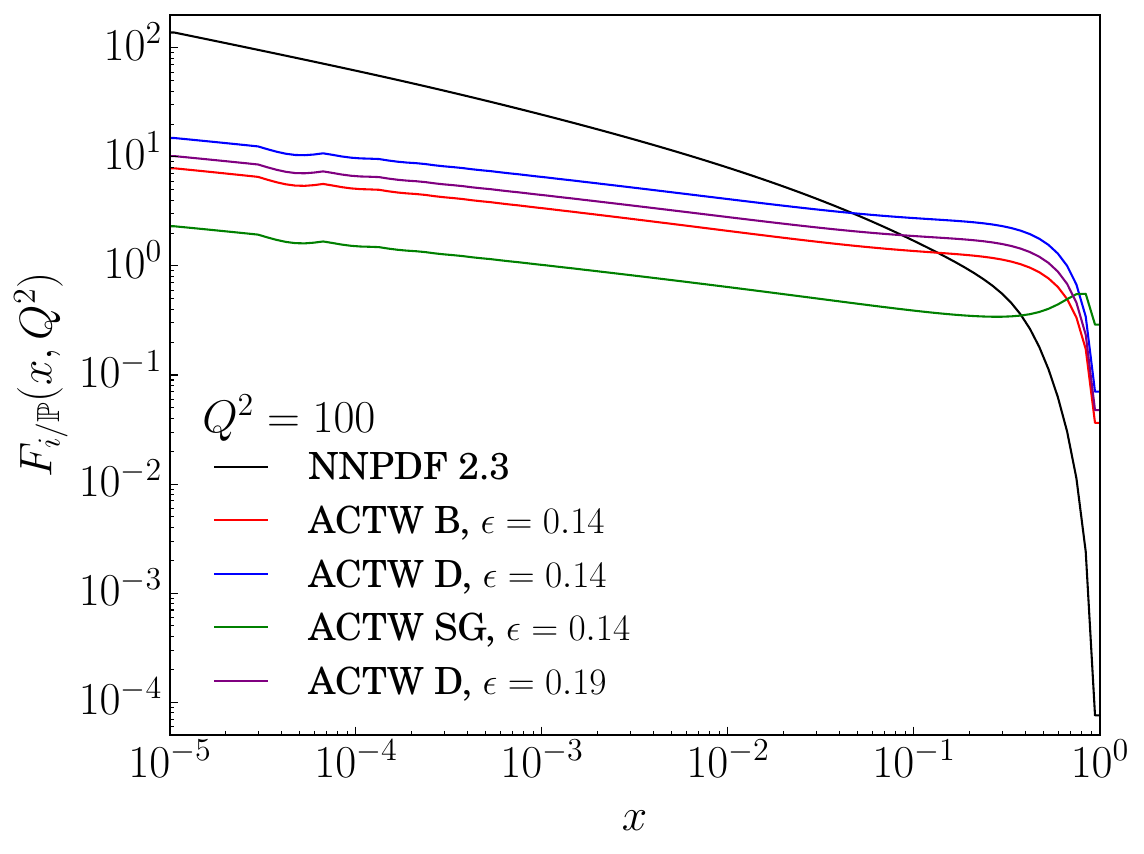}\\
(d)
\end{minipage}
\caption{\label{Fig:PomeronPDFs}
The QCD charge-weighted sum, eq.~(\ref{eq:QCDWeightedPDF}), 
of the H1 PDFs and the toy PDF PomFix compared to the NNPDF 2.3 
proton PDF on linear (a) and logarithmic scale (b).
The QCD charge-weighted sum, eq.~(\ref{eq:QCDWeightedPDF}), 
of the ACTW PDFs compared to the NNPDF 2.3 
proton PDF on linear (c) and logarithmic scale (d).}
\end{figure}

\begin{figure}[tbp]
\begin{minipage}[t]{0.5\textwidth}
\centering
\includegraphics[scale=0.4]{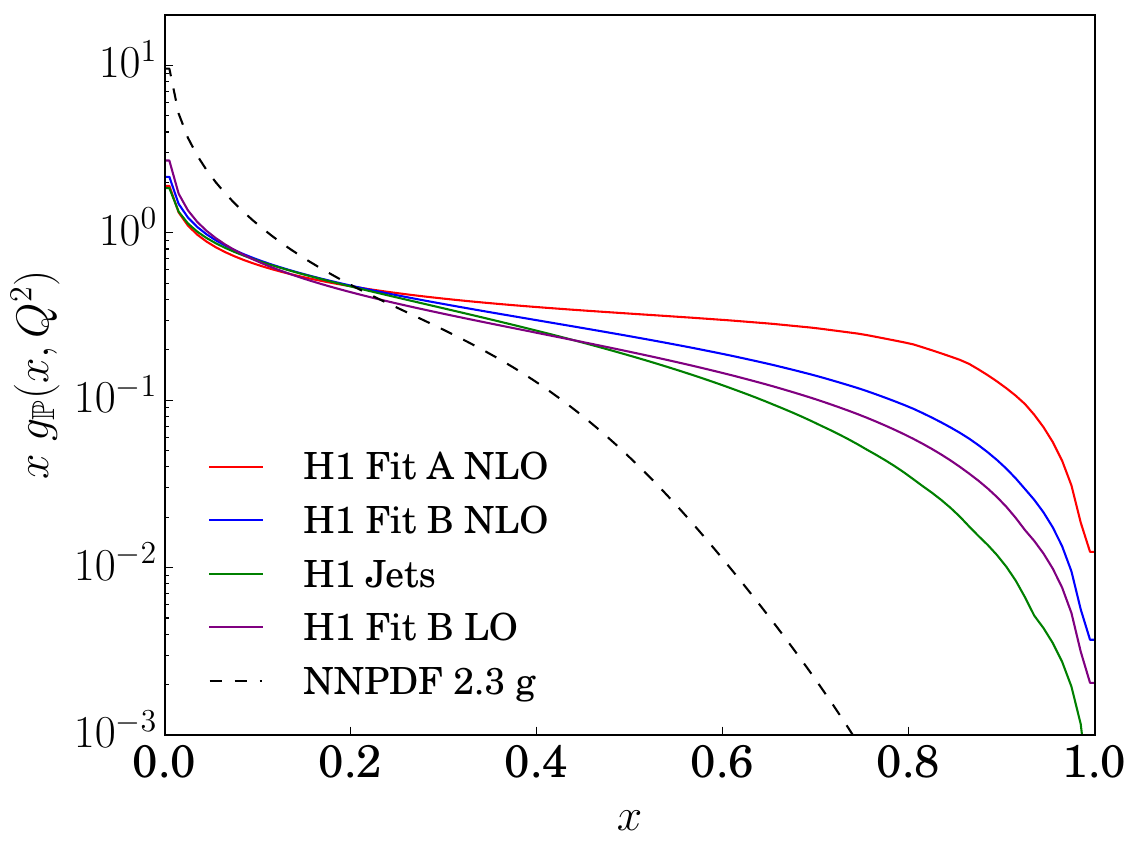}\\
(a)
\end{minipage}
\hfill
\begin{minipage}[t]{0.5\textwidth}
\centering
\includegraphics[scale=0.4]{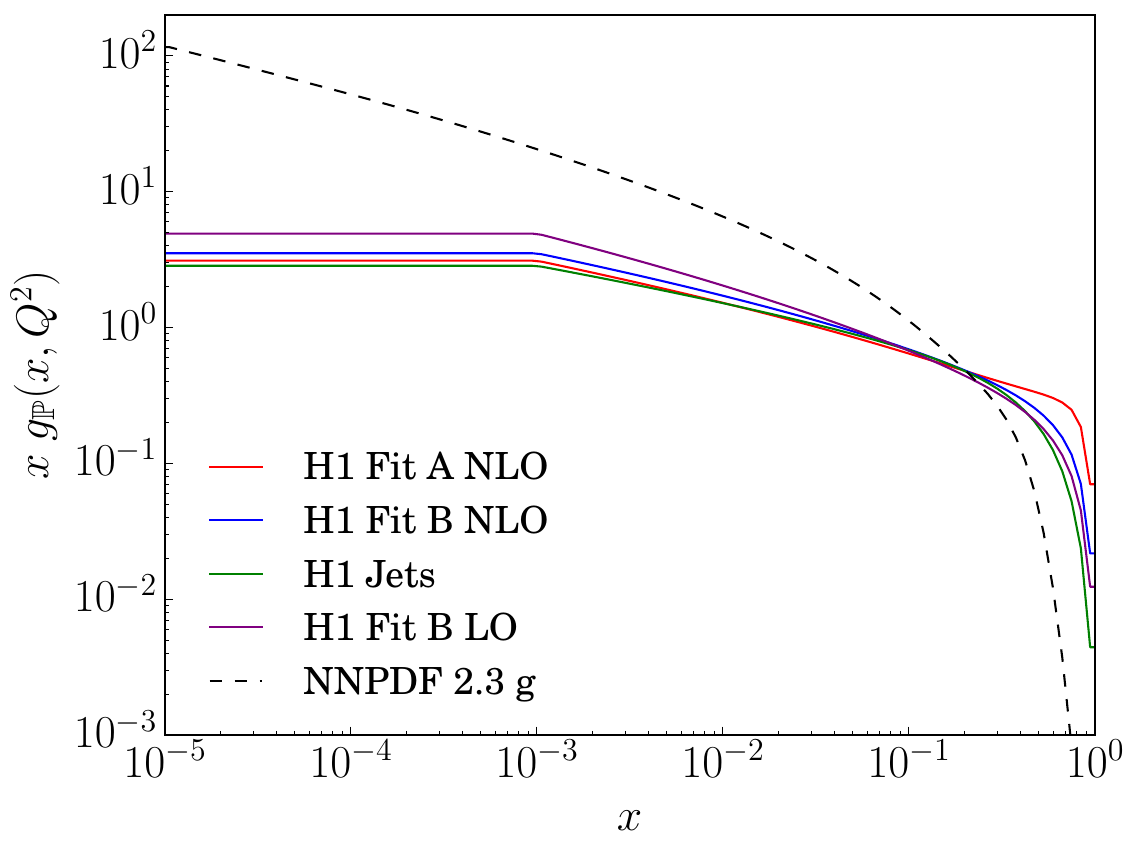}\\
(b)
\end{minipage}
\begin{minipage}[t]{0.5\textwidth}
\centering
\includegraphics[scale=0.4]{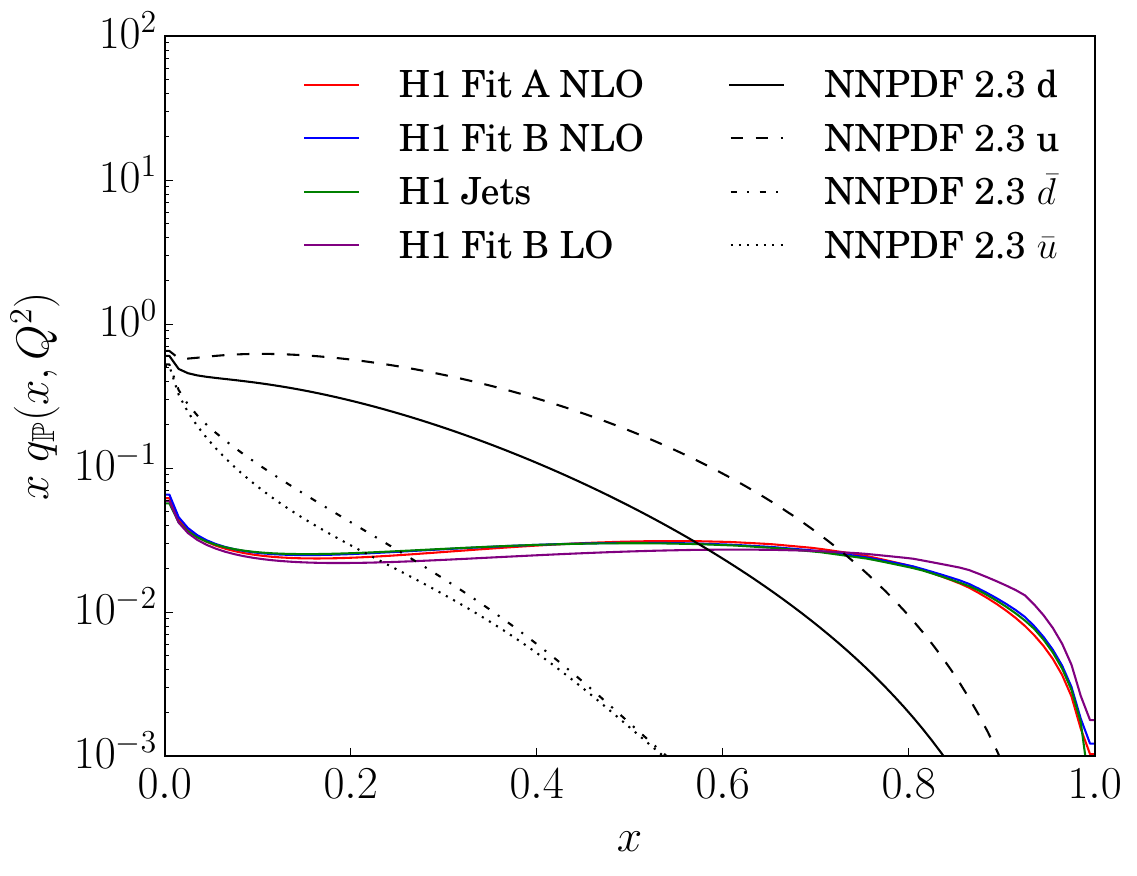}\\
(c)
\end{minipage}
\hfill
\begin{minipage}[t]{0.5\textwidth}
\centering
\includegraphics[scale=0.4]{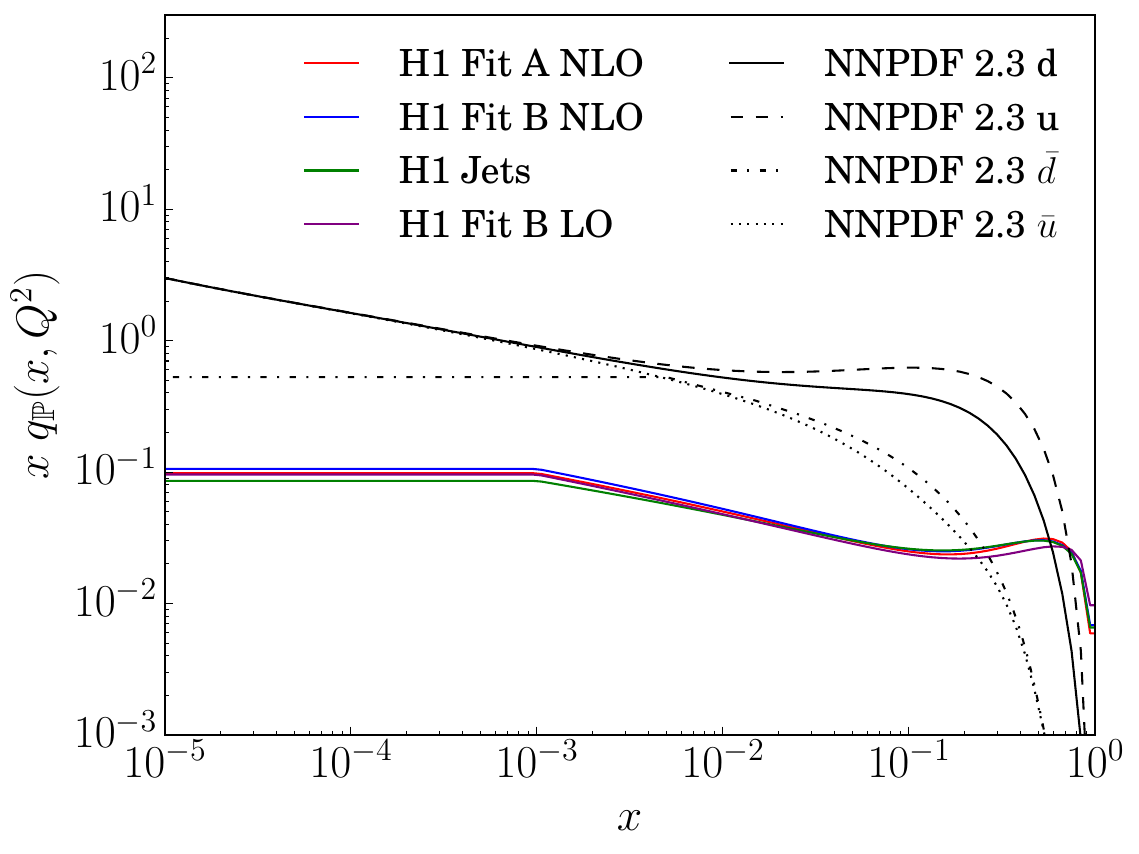}\\
(d)
\end{minipage}
\caption{\label{Fig:PartonDistributionsInPomeronHERA}
The H1 $\Pom$ gluon distribution on linear (a) and logarithmic (b) scale. 
The H1 $\Pom$ quark and antiquark distributions on linear (c) and
logarithmic (d) scale. Both compared to the NNPDF 2.3 proton PDF distributions. 
Note that for the $\Pom$ we have $\d=\u=\s=\dbar=\ubar=\sbar(=\c=\cbar)$, 
where the $\c,~\cbar$ are only included in H1Jets.}
\end{figure}

\begin{figure}[tbp]
\begin{minipage}[t]{0.5\textwidth}
\centering
\includegraphics[scale=0.4]{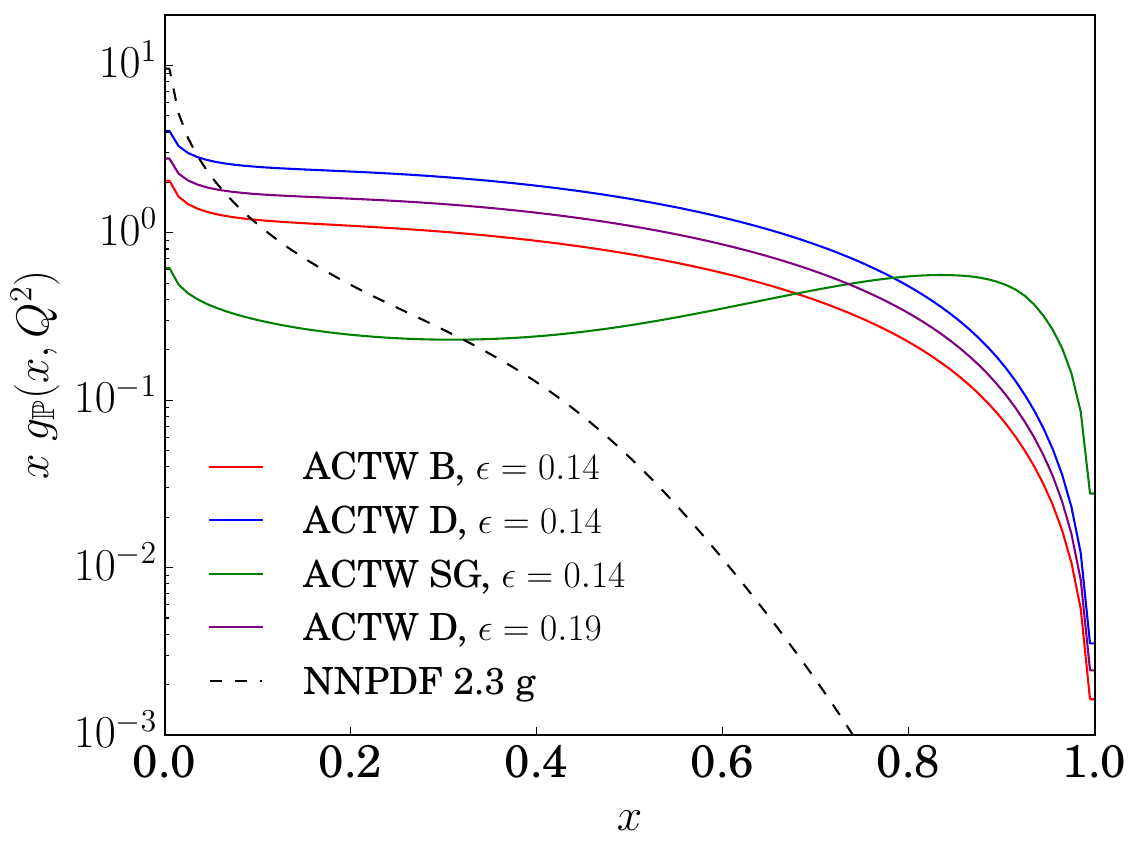}\\
(a)
\end{minipage}
\hfill
\begin{minipage}[t]{0.5\textwidth}
\centering
\includegraphics[scale=0.4]{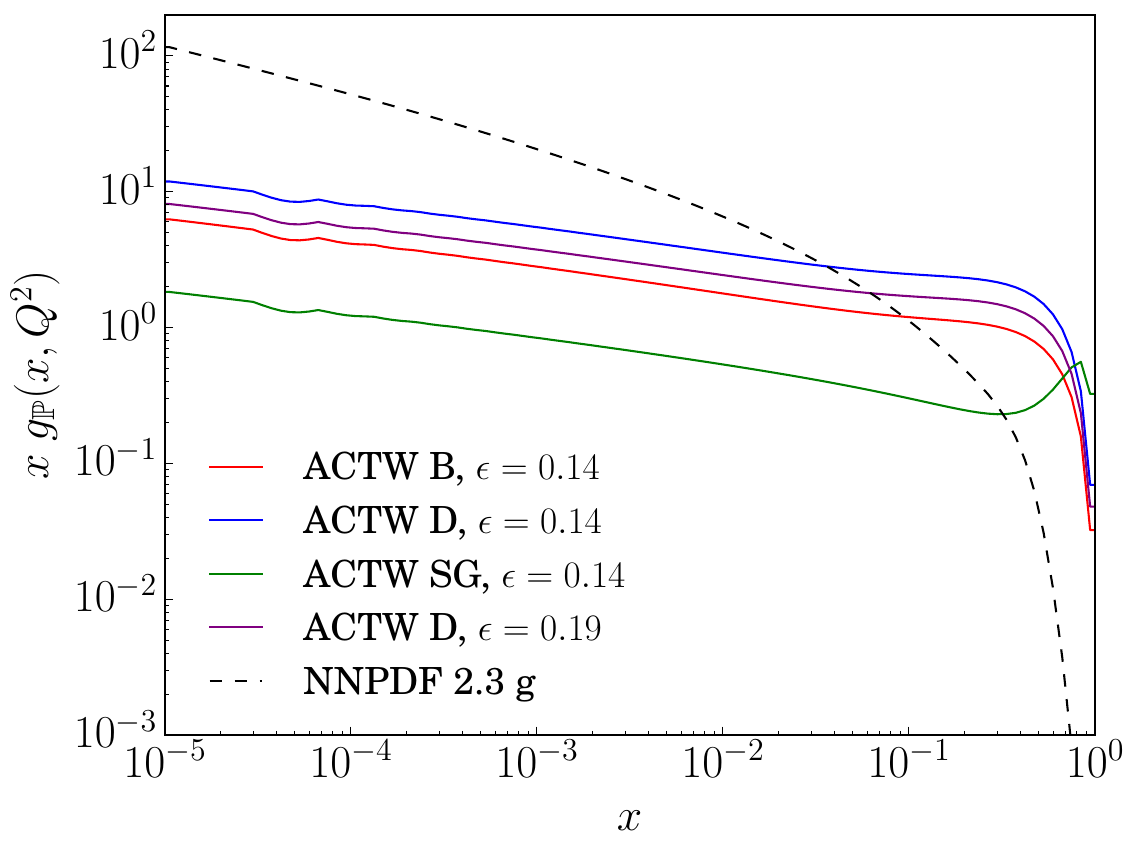}\\
(b)
\end{minipage}
\begin{minipage}[t]{0.5\textwidth}
\centering
\includegraphics[scale=0.4]{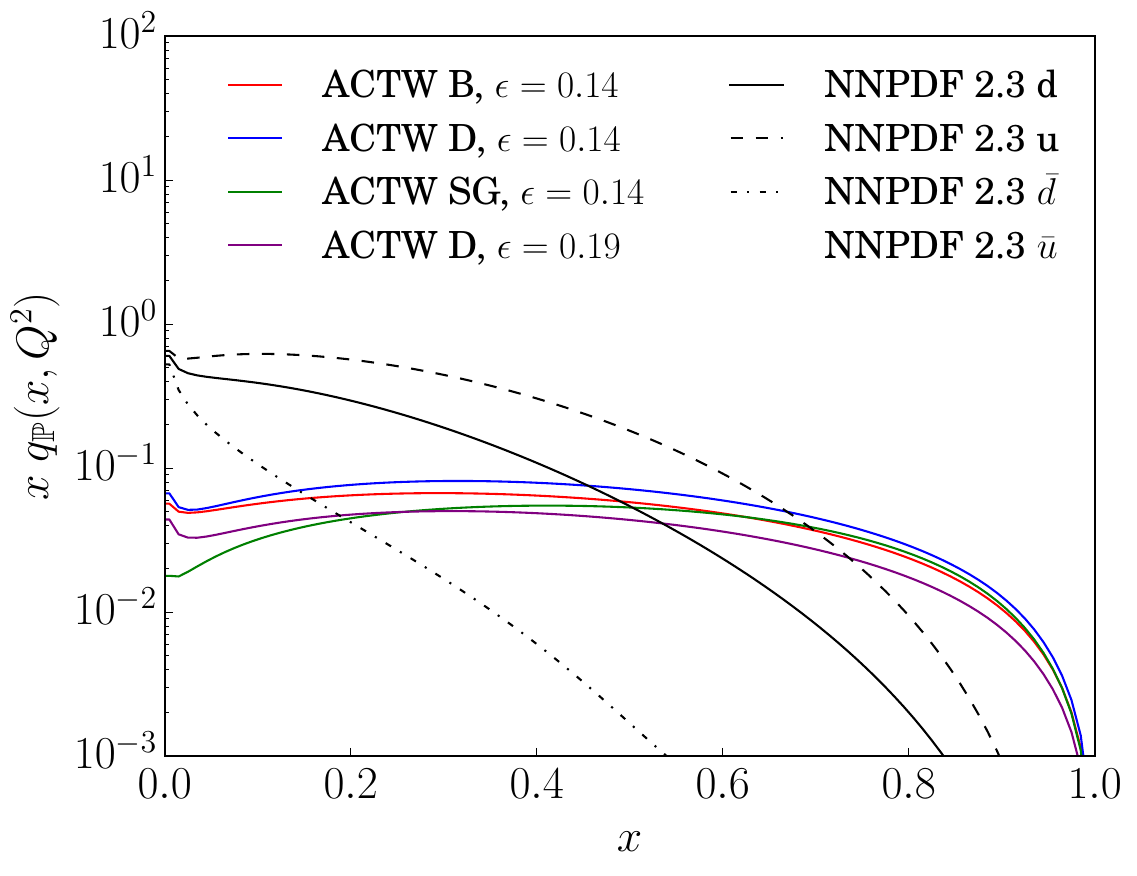}\\
(c)
\end{minipage}
\hfill
\begin{minipage}[t]{0.5\textwidth}
\centering
\includegraphics[scale=0.4]{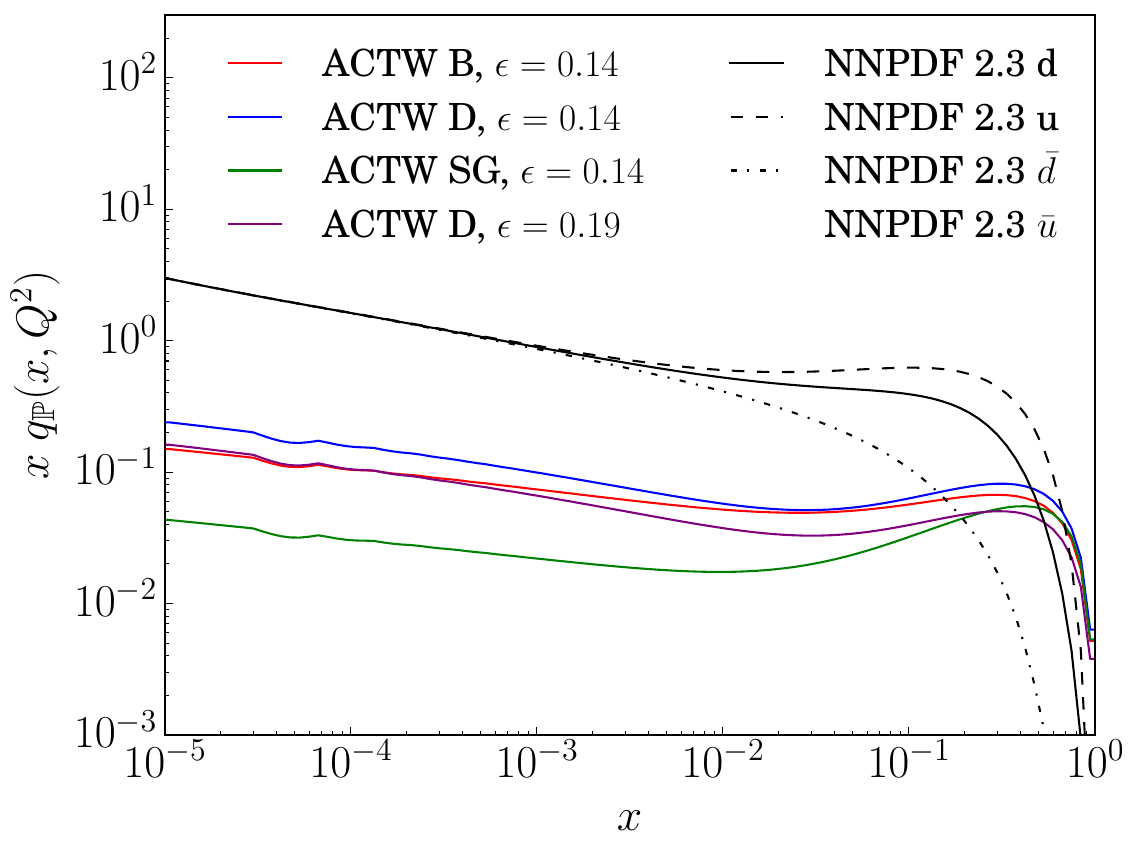}\\
(d)
\end{minipage}
\caption{\label{Fig:PartonDistributionsInPomeronACTW}
The ACTW $\Pom$ gluon distribution on linear (a) and logarithmic (b) scale. 
The ACTW $\Pom$ quark and antiquark distributions on linear (c) and
logarithmic (d) scale. Both compared to the NNPDF 2.3 proton PDF distributions. 
Note that for the $\Pom$ we have $\d=\u=\s=\dbar=\ubar=\sbar(=\c=\cbar)$, 
where the $\c,~\cbar$ are only included in H1Jets.}
\end{figure}

Turning to the Pomeron PDFs, a detailed comparison would entail 
the separate quark and gluon distributions at varying $Q^2$ scales.
To simplify we show the QCD-charge-weighted sum 
\begin{equation}\label{eq:QCDWeightedPDF}
F_{\Pom}(x, Q^2) = \frac{4}{9} \sum_{i = \q, \qbar} 
  x f_{i/\Pom}(x, Q^2) + x g_{\Pom}(x, Q^2) 
\end{equation}
at a single value $Q^2 = 100$ GeV$^2$,
figures~\ref{Fig:PomeronPDFs}a to \ref{Fig:PomeronPDFs}d. 
We notice that they all tend to be significantly harder than the 
corresponding proton PDF, here represented by the NNPDF 2.3 QCD+QED LO 
one. (The PomFix option is just a toy one, shown for completeness, but
not used in the following.) For the gluon on its own, the $\Pom$ is
significantly harder than the $\p$, consistent with the idealized
picture of a $\Pom$ as a glueball state with two ``valence gluons'',
figures~\ref{Fig:PartonDistributionsInPomeronHERA}a, b and
\ref{Fig:PartonDistributionsInPomeronACTW}a, b.
Surprisingly, also the quark PDFs of the $\Pom$ 
(figures~\ref{Fig:PartonDistributionsInPomeronHERA}c, d and 
\ref{Fig:PartonDistributionsInPomeronACTW}c, d) are harder than proton 
ones, suggesting the presence of ``valence quarks'' in the $\Pom$,
although an order of magnitude below the gluons. Another observation 
is that the $\Pom$ PDF sets we compare are all primarily based on H1 analyses, 
with largely the same data and with correlated assumptions for the 
definition of diffractive events. This is especially notable in the 
quark distributions, which are close to identical. Also the close 
affinity of gluons at lower $x$ values should not be overstressed.
The slightly larger variations in the ACTW PDFs are due to both
the different values of the flux-parameter $\epsilon$, as well as
different parametrisations of the PDFs.
Finally, note that the H1 parametrizations only apply down to 
$x = 10^{-3}$, and are frozen below that. This is likely to underestimate 
the low-$x$ rise of PDFs, which as well could have been of the same shape 
as in the proton. A small kink in the ACTW PDFs around $x=10^{-4}$ is 
due to regions of phase space where the parametrization of the initial
quark distribution would become negative and has been reset to vanish.

\begin{figure}[tbp]
\begin{minipage}[t]{0.5\textwidth}
\centering
\includegraphics[scale=0.4]{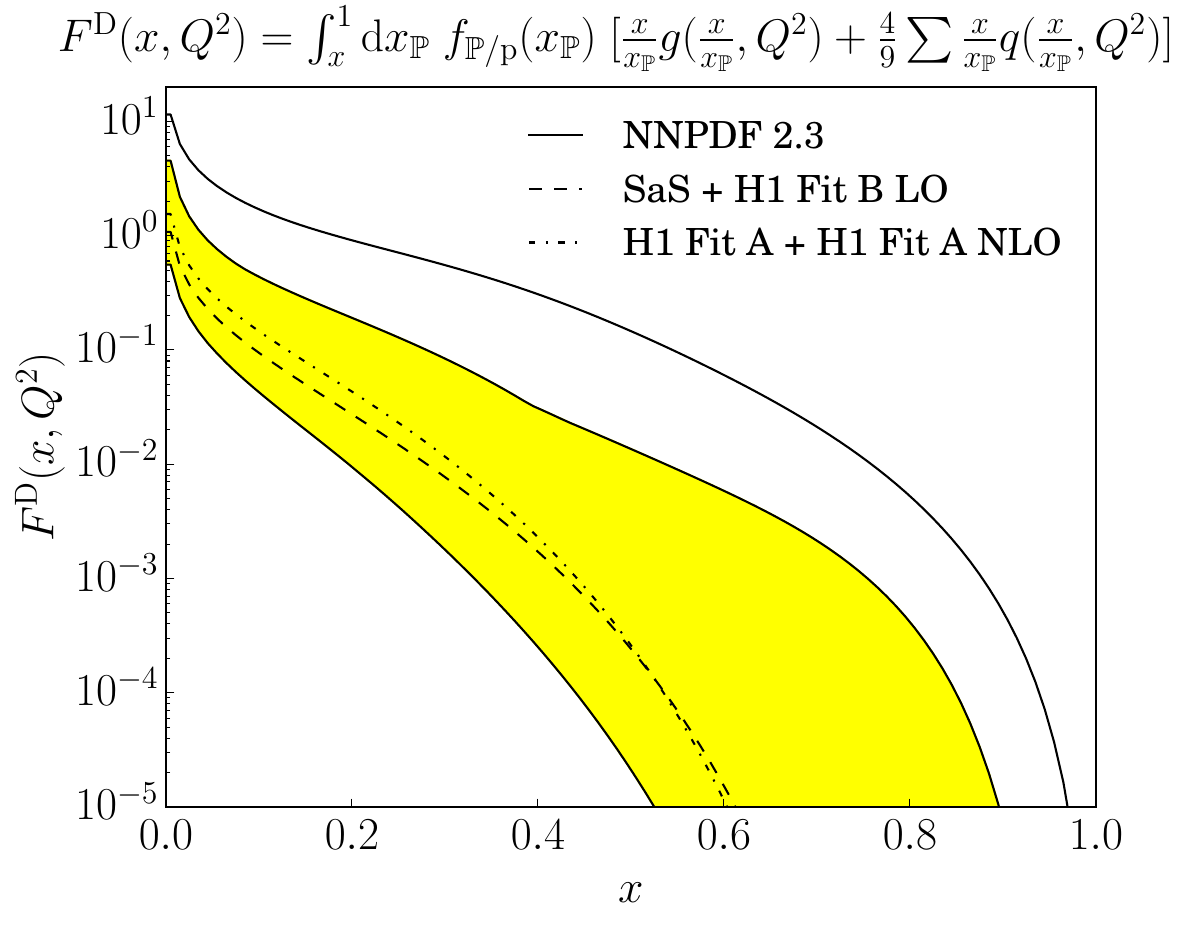}\\
(a)
\end{minipage}
\hfill
\begin{minipage}[t]{0.5\textwidth}
\centering
\includegraphics[scale=0.4]{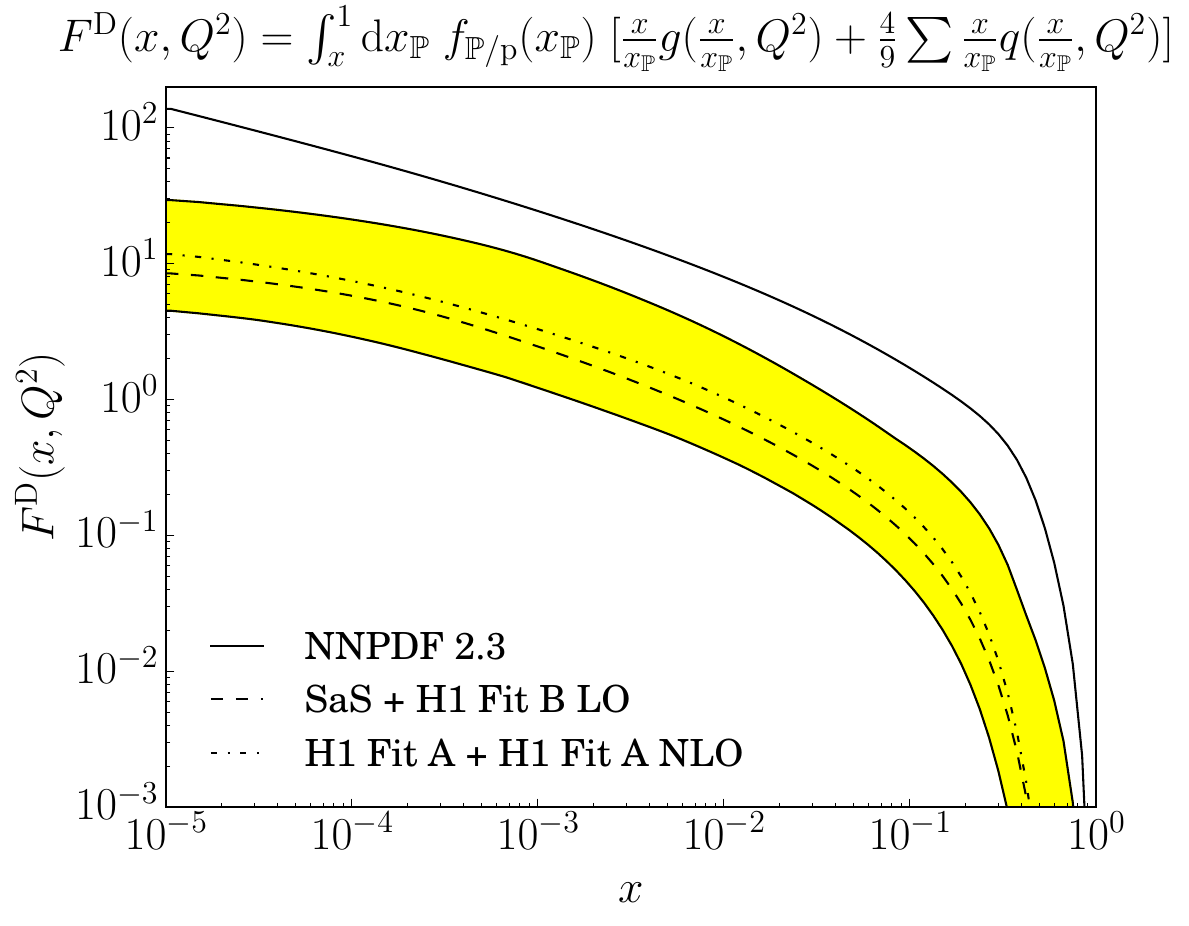}\\
(b)
\end{minipage}
\begin{minipage}[t]{0.5\textwidth}
\centering
\includegraphics[scale=0.4]{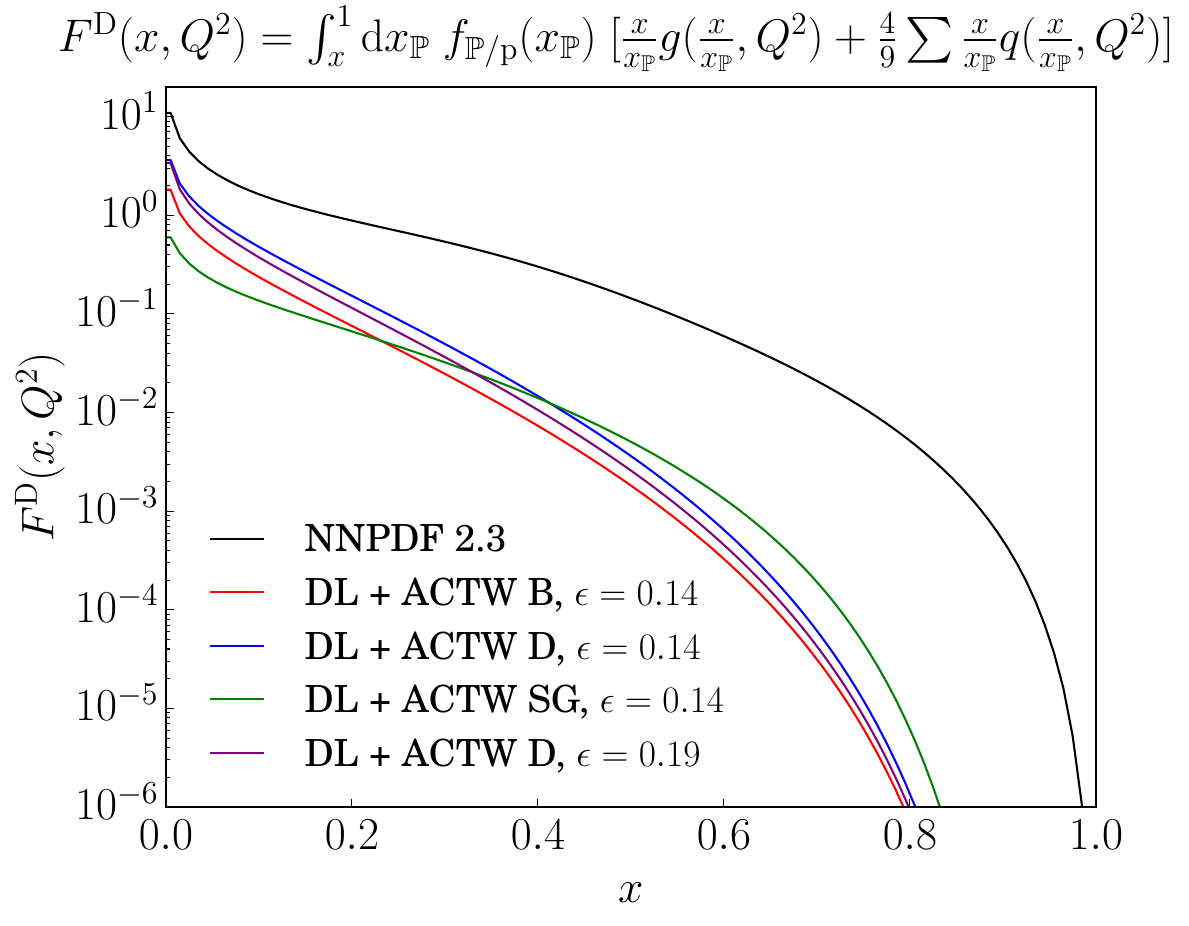}\\
(c)
\end{minipage}
\hfill
\begin{minipage}[t]{0.5\textwidth}
\centering
\includegraphics[scale=0.4]{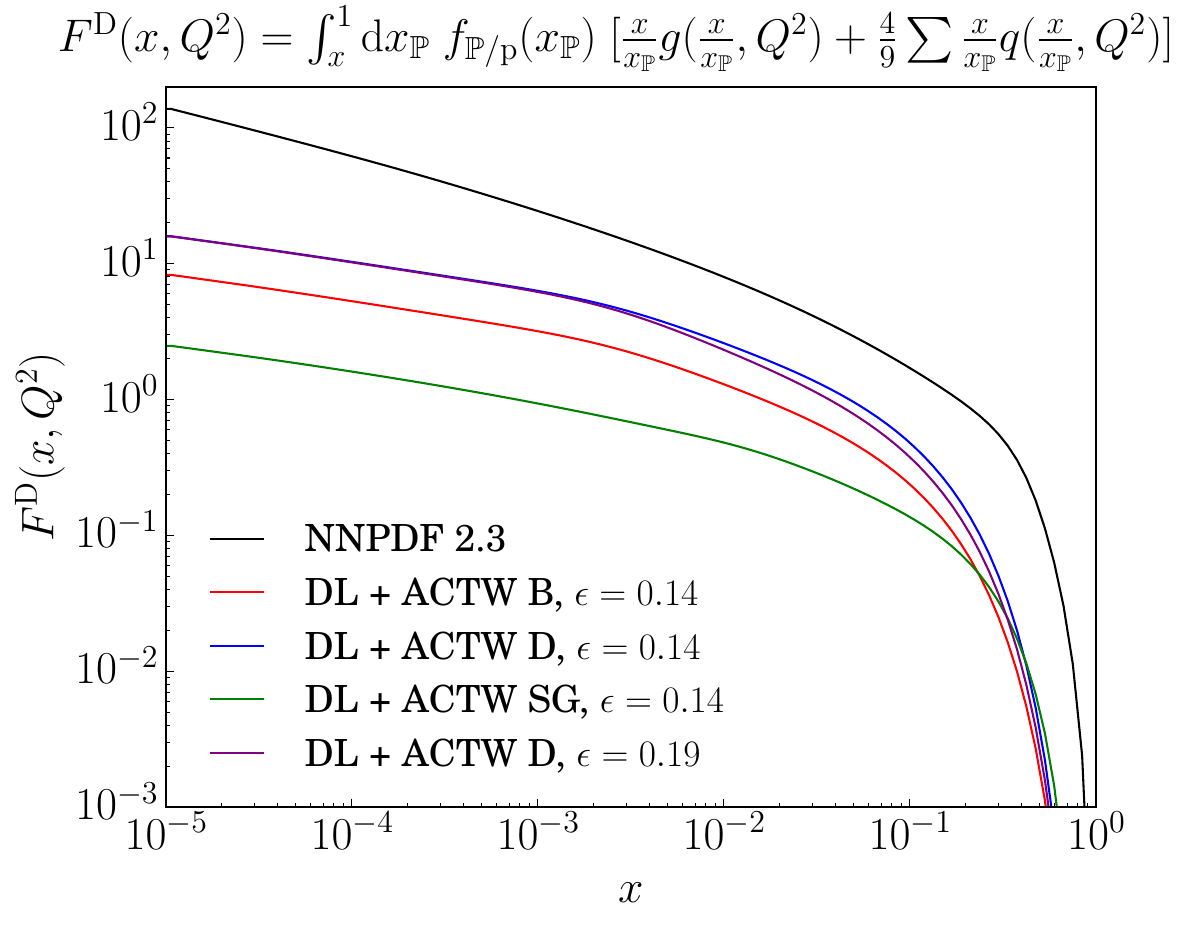}\\
(d)
\end{minipage}
\caption{\label{Fig:PomeronConvolution}
The convolution of Pomeron fluxes and H1 PDFs for a few cases, with the 
range between the extremes marked in yellow; (a) linear and (b) 
logarithmic $x$ scale. The convolution of DL flux and ACTW PDFs
on (c) linear
and (d) logarithmic $x$ scale.}
\end{figure}

In the end, what matters is the convolution of the $\Pom$ flux with 
its PDFs, and that is shown in figure~\ref{Fig:PomeronConvolution}.
There would be too many combinations possible to show individually,
so we only indicate the range of possibilities and a few specific 
combinations. This may be on the extreme side, since some fluxes and 
PDFs come as fixed pairs, not really intended to be mixed freely.
The key feature to note is that in this convolution the Pomeron 
part is now falling steeper at large $x$ than the proton as a whole.
This has the immediate consequence that diffractive hard subcollisions 
are not necessarily going to be produced more in the forwards direction
than the bulk of corresponding nondiffractive ones, but on the contrary
may be more central. The difference is not all that dramatic, however.
It is also  partly compensated by a somewhat slower increase of the 
$\Pom$ towards lower $x$ values, a feature that for the H1 $\Pom$ PDF
derives from the artificial freezing of  below $x = 10^{-3}$. 
Note that the four different ACTW PDFs differ by up to an
order of magnitude. The two D fits are similar in shape and size
as expected, but especially the SG fit stands out being up to a
factor 10 smaller than the D fits. Most of this discrepancy is
also seen in figure~\ref{Fig:PomeronPDFs}c,d but also arise from 
the difference in normalisation, the D and SG fits 
having momentum sums of $\sim1.8$ and $\sim0.5$, respectively.
The lack of 
major shape differences between the $\Pom$ part and the rest will be 
visible in the more detailed studies later on. 
Because of the close similarity of most of the different (but related) $\Pom$ 
PDFs at low-to-medium $x$, the bulk of the differences come from the 
$\Pom$ fluxes. We have chosen to exemplify this for $2 \rightarrow 2$ 
QCD processes with $\pT >20$ GeV in $\sqrt{s}=8$ TeV $\p\p$ 
collisions, with the diffractive fractions for a few combinations 
shown in table~\ref{Tab:DiffractiveFractions}. 

\begin{table}[tbp]
\centering
\begin{tabular}{|c|c|c|}
\hline
\multicolumn{3}{|c|}{Diffractive fractions}\\
\multicolumn{3}{|c|}{$\p\p$ collisions at $\sqrt{s}=8$ TeV}\\
\hline
\begin{tabular}{c}$\Pom$ PDF \\ $\Pom$ flux \end{tabular}
& PDF selection & MPI selection\\
\hline
H1 Fit B LO & & \\
SaS & (14.33 $\pm$ 0.11) \% & (0.98 $\pm$ 0.03) \%\\ 
\hline
H1 Fit B LO & & \\
MBR                 & (14.79 $\pm$ 0.11) \% & (0.96 $\pm$ 0.03) \%\\ 
\hline
H1 Jets & & \\
SaS & (13.70 $\pm$ 0.11) \% & (0.92 $\pm$ 0.03) \%\\ 
\hline
H1 Fit A NLO & & \\
H1 Fit A            & (20.55 $\pm$ 0.13) \% & (1.35 $\pm$ 0.04) \%\\ 
\hline
H1 Fit B LO & & \\
H1 Fit A            & (18.49 $\pm$ 0.12) \% & (1.32 $\pm$ 0.04) \%\\ 
\hline
ACTW D14 & & \\
DL                  & (46.54 $\pm$ 0.16) \% & (3.18 $\pm$ 0.06) \%\\ 
\hline
ACTW SG14 & & \\
DL                  & (11.82 $\pm$ 0.10) \% & (0.81 $\pm$ 0.03) \%\\ 
\hline
ACTW D19 & & \\
DL                  & (42.09 $\pm$ 0.16) \% & (2.89 $\pm$ 0.05) \%\\ 
\hline
\end{tabular}
\caption{\label{Tab:DiffractiveFractions}
Diffractive fractions for the $2\rightarrow2$ QCD processes 
with $\pT > 20$ GeV obtained with \textsc{Pythia}~8. 
The samples have been produced without any phase-space cuts.}
\end{table}

Note that changing the Pomeron parametrizations changes the 
fraction of events passing the PDF selection, but that the 
suppression factor introduced by the dynamical gap survival is 
about $\sim0.07$ for all combinations in table~\ref{Tab:DiffractiveFractions}.
This reflects the fact that neither the MPI model nor the proton PDF are 
influenced by the Pomeron parametrization, hence the probability for 
obtaining no additional MPIs in the $\p\p$ system should not change. 
(This does not have to hold in general, but here we compare very similar 
distributions of $x$ and $\pT$ values of the hard interaction, and then 
also the MPI effects are closely the same.) Note also that some of the
ACTW PDFs gives substantially larger fractions than the HERA
PDFs. This is related to the fact that the intercept of the
$\Pom$ trajectory is larger in ACTW fits than in the H1 ones,
$\epsilon = 0.14 - 0.19$ vs. 0.085. This gives a larger flux at 
high-energy hadron colliders. A similar flux increase can
of course be obtained for the H1 PDFs, with the
caveat that the flux might not be able to describe the total 
cross section and other associated quantities. 
Additionally the gluon is only probed indirectly in DIS, and so is 
poorly constrained, while it dominates for QCD jet rates. 

\begin{figure}[tbp]
\begin{minipage}[t]{0.5\textwidth}
\centering
\includegraphics[scale=0.4]{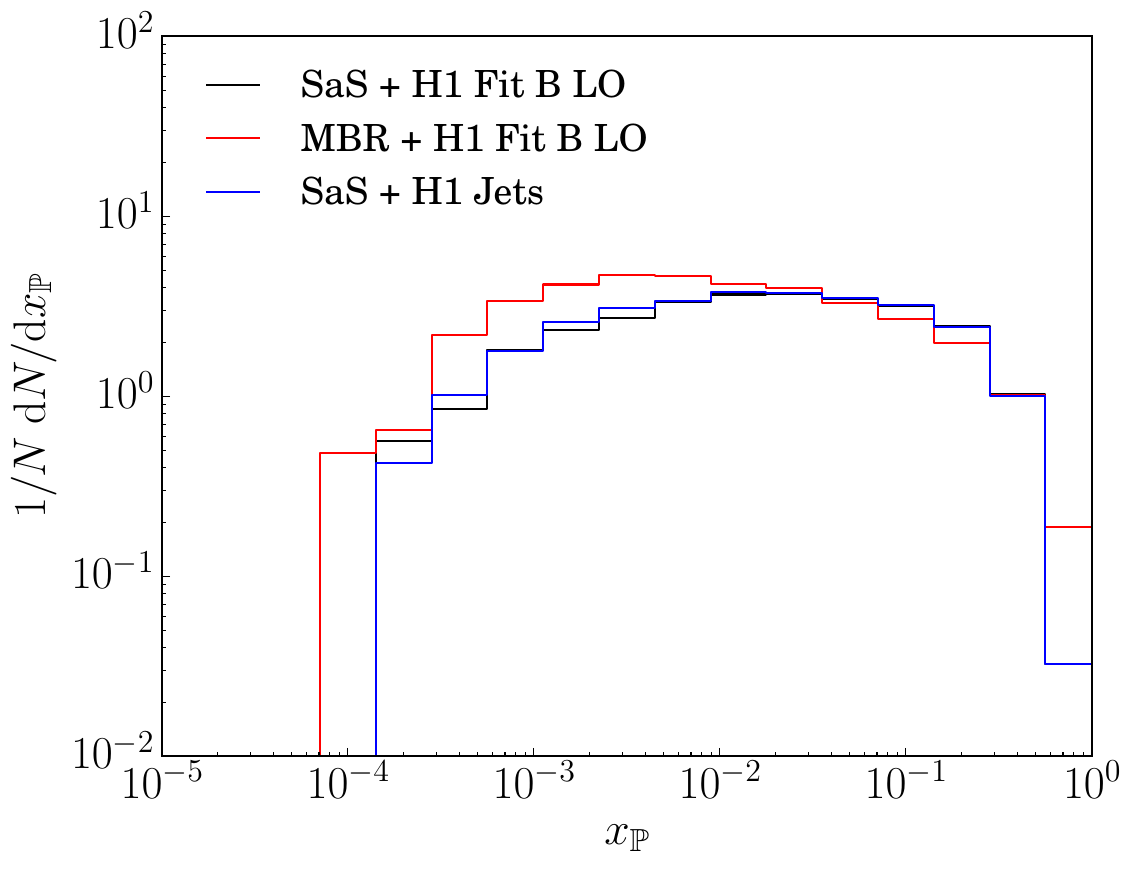}\\
(a)
\end{minipage}
\hfill
\begin{minipage}[t]{0.5\textwidth}
\centering
\includegraphics[scale=0.4]{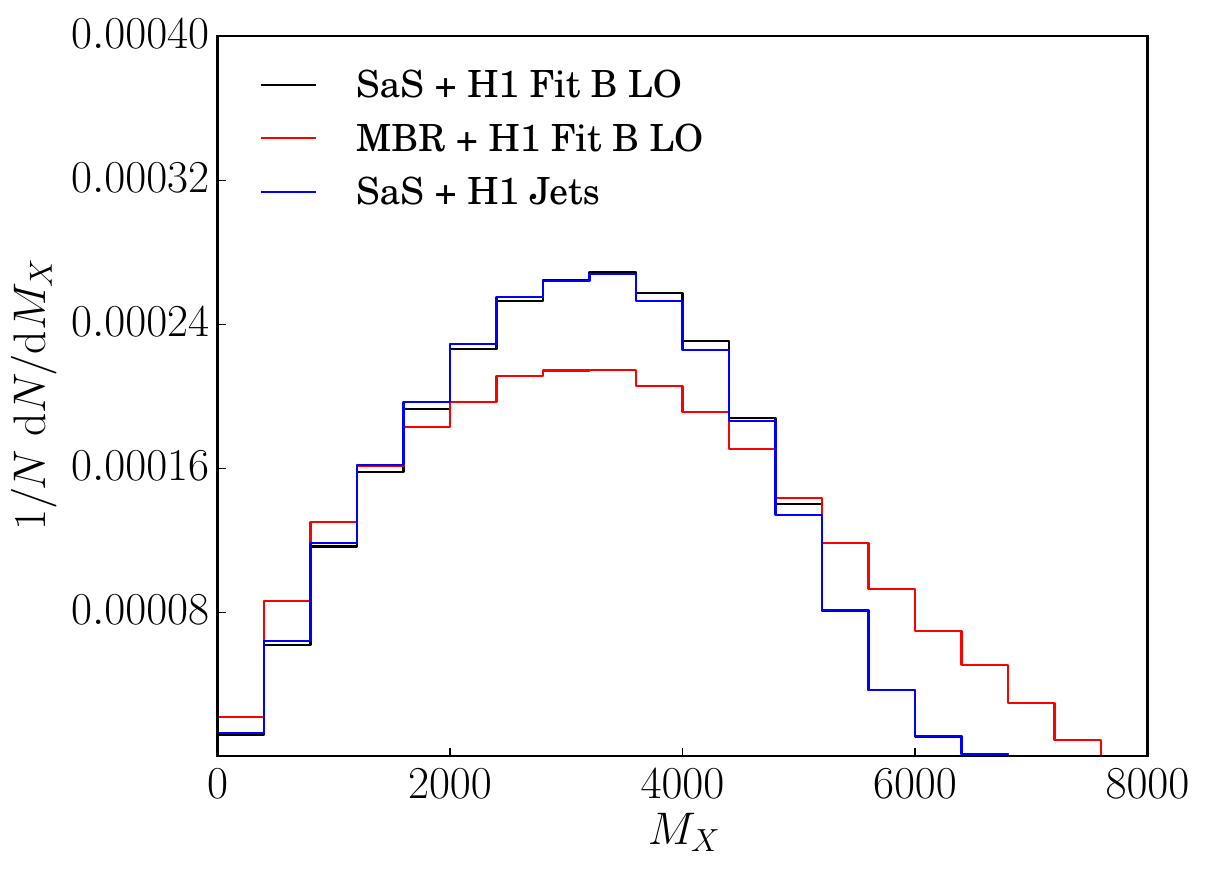}\\
(b)
\end{minipage}\\[2mm]
\begin{minipage}[t]{0.5\textwidth}
\centering
\includegraphics[scale=0.4]{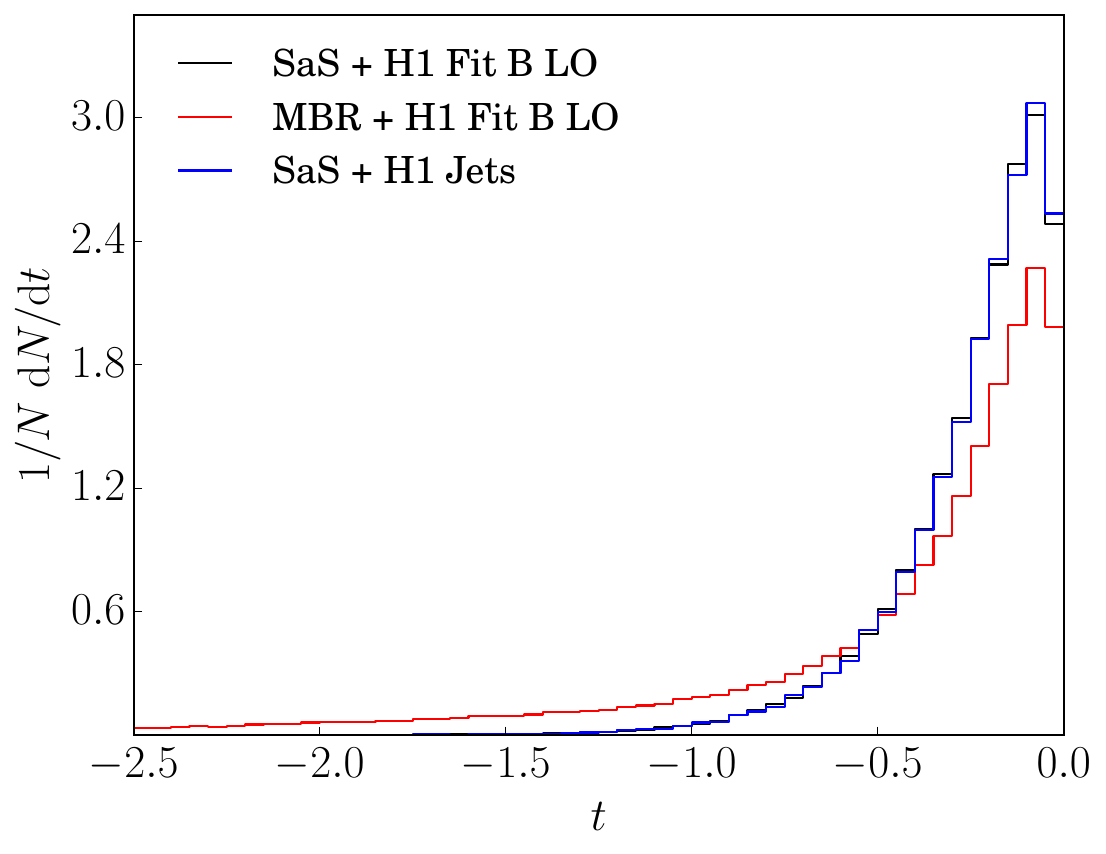}\\
(c)
\end{minipage}
\hfill
\begin{minipage}[t]{0.5\textwidth}
\centering
\includegraphics[scale=0.4]{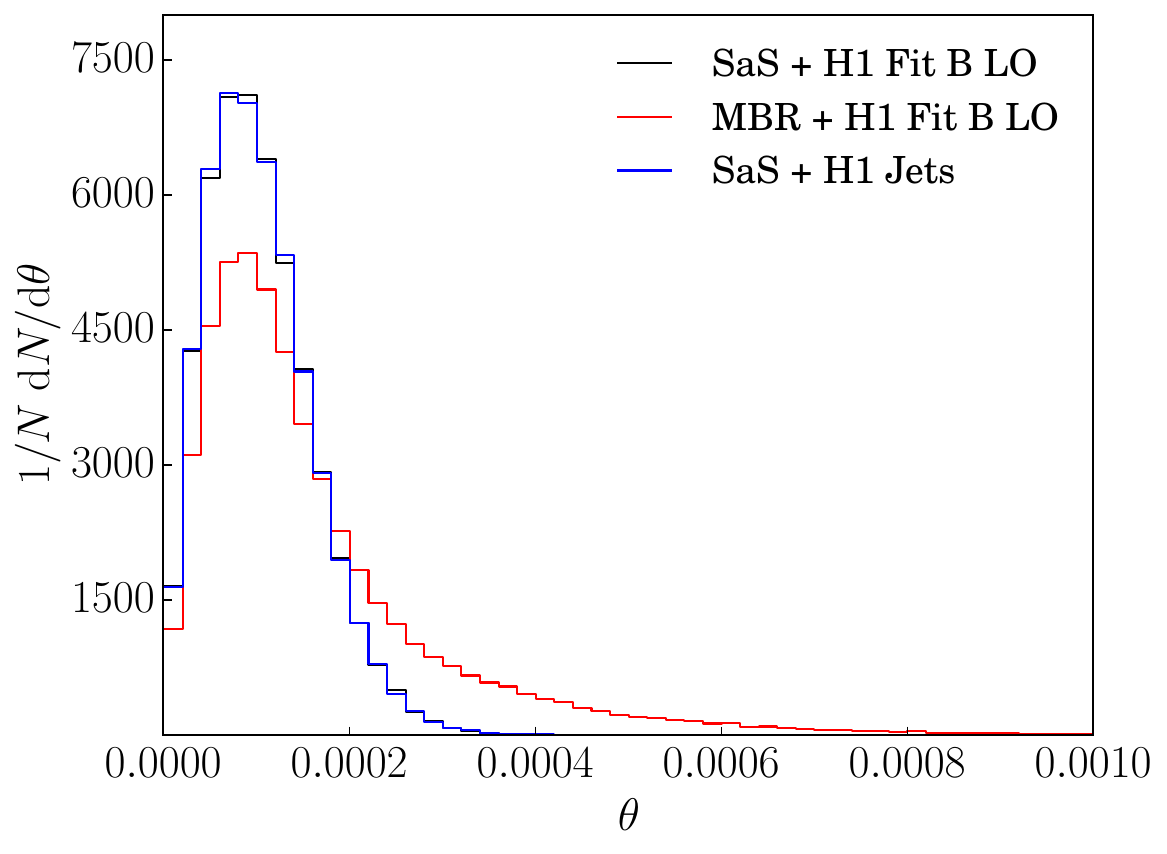}\\
(d)
\end{minipage}
\caption{\label{Fig:PomDistVal}
Some kinematics distributions obtained with variations of the Pomeron 
parametrizations: (a) $x_{\Pom}$, (b) $M_X$, (c) $t$ and (d) $\theta$.}
\end{figure}

Differential distributions of the diffractive events are also affected, since 
the kinematics of the $\Pom\p$ system is set up using the Pomeron flux 
parametrizations. A subset of these distributions is shown in 
figure~\ref{Fig:PomDistVal}, for some of the same combinations as in 
table~\ref{Tab:DiffractiveFractions}. As expected, $\Pom$ PDF 
variations do not have a large impact on the shapes 
(cf.\ figure~\ref{Fig:PomDistVal}), while the $\Pom$ flux gives 
rise to large effects in $x_{\Pom}$, hence on the broadening of the 
mass spectrum and on the tails of the $t$ and $\theta$ distributions. 
In view of these observations, we do not expect to be able to 
discrimate between the available Pomeron PDFs when comparing to data. 
Thus we will leave out this variation from now on, and focus on 
variations in the Pomeron flux. 

The diffractive event fraction is not process-independent. 
One reason is that processes may be dominated by different initial 
states, another that different $x$ and $Q^2$ scales are probed. 
In table~\ref{Tab:DiffFracProcesses} we show the fraction of events 
passing either selection for various hard processes available in 
\textsc{Pythia}~8 using the SaS flux and 
the H1 Fit B LO PDF. Firstly we note that a smaller fraction of events
pass the PDF selection than in table~\ref{Tab:DiffractiveFractions}, 
owing to the larger $x$ needed to produce these particles, 
cf.\ figures~\ref{Fig:PomeronFluxes},\ref{Fig:PomeronPDFs}. This is why top, being the
heaviest, has the smallest diffractive fraction. In addition there 
is a notable difference between the gluon-dominated Higgs production 
and the quark-induced production of $\W^{\pm}/\gamma^*/\Z^0$,
owing to the hard gluon PDF in the $\Pom$. If top production is
considered separately for $\q\qbar \to \t\tbar$ and 
$\g\g \to \t\tbar$, the PDF survival rate is (9.74 $\pm$ 0.09)\% 
and (10.55 $\pm$ 0.10)\%, respectively, displaying the difference 
between the two production channels. 

\begin{table}[tbp]
\begin{center}
\begin{tabular}{|c|c|c|}
\hline
\multicolumn{3}{|c|}{Diffractive fractions}\\
\multicolumn{3}{|c|}{$\p\p$ collisions at $\sqrt{s}=8$ TeV}\\
\hline
& PDF selection & MPI selection\\
\hline
$\q\qbar\rightarrow \W^{\pm}$      & (11.16 $\pm$ 0.10) \% & (0.70 $\pm$ 0.03) \% \\
$\q\qbar\rightarrow \gamma^*/\Z^0$ & (10.69 $\pm$ 0.10) \% & (0.76 $\pm$ 0.03) \% \\
Single top and top pair production & ( 8.51 $\pm$ 0.09) \% & (0.62 $\pm$ 0.02) \% \\
SM Higgs production                & (12.37 $\pm$ 0.10) \% & (0.86 $\pm$ 0.03) \% \\
\hline
\end{tabular}
\caption{\label{Tab:DiffFracProcesses}
Diffractive fractions obtained with \textsc{Pythia} without any 
phasespace cuts at $\sqrt{s} = 8$ TeV for various hard processes. 
\textsc{Pythia} is run with the SaS flux and 
the H1 Fit B LO PDF.}
\end{center}
\end{table}

\begin{figure}[tbp]
\centering
\includegraphics[scale=0.4]{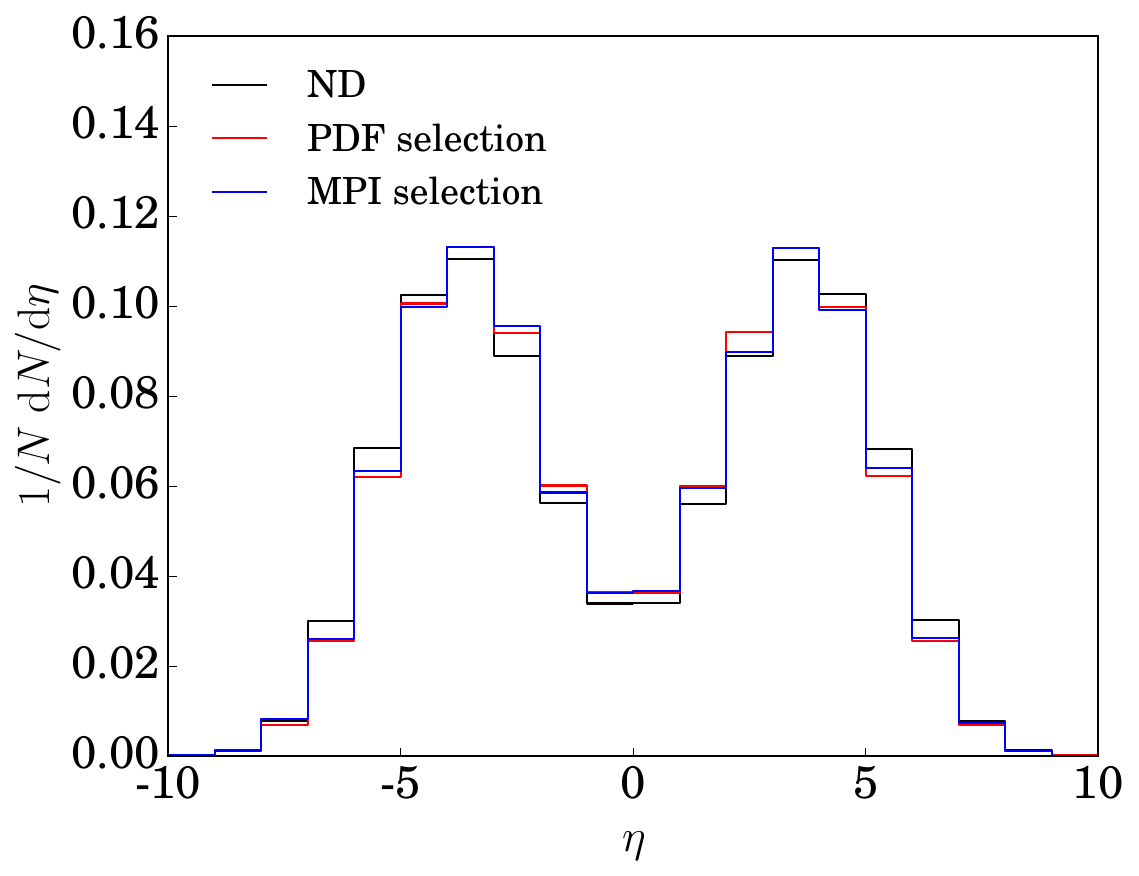}\\
\caption{\label{Fig:Wrapidity}
The rapidity of the $\W$-boson produced in $\q\qbar\to \W^{\pm}$ at 
$\sqrt{s}=8$ TeV.}
\end{figure}

In figure~\ref{Fig:Wrapidity} we show the rapidity of the $\W$-boson produced 
in the process $\q\qbar\to \W^{\pm}$ at an $8$ TeV $\p\p$ collision, comparing 
three samples; nondiffractive, PDF selected and MPI selected. It is 
observed that the diffractive $\W$'s are slighly more central than the 
nondiffractive in the CM frame, as expected from 
figure~\ref{Fig:PomeronConvolution}. The differences are small, however, 
being on the order of (5-10)\%, and might reduce when 
phase-space cuts are introduced. We will study this process further 
in section~\ref{Sec:WZcomp}. 

\subsection{The dynamical gap survival and MPI models}

In the above section we studied how the parametrization of the 
Pomeron flux and PDF affected the diffractive fractions and 
distributions, and notably by the choice of $\Pom$ flux. 
By contrast, we saw that the survival fraction in the MPI selection 
step was not significantly affected by these choices. A dependence 
does enter both via the $x$ and the $\pT$ distributions of a 
process: larger $x$ scales leaves less energy for MPIs and thereby 
gives a higher MPI survival probability, whereas larger $\pT$ values
gives a longer MPI evolution range and thereby a lower MPI survival.
Such effects are not too prominent, however, and tend to be 
overshadowed by the sensitivity to the parameters of the MPI model. 
These enter twice. Firstly, for the MPI selection, since the 
dynamical gap survival is tied to the number of MPIs in the 
$\p\p$ system. Secondly, for the properties of the diffractive system,
where the number of MPIs affects e.g.\ charged multiplicities.

The probability for obtaining MPIs is given by eq.~(\ref{eq:probMPI}), 
and hence depends on both the overlap function and the regulator 
$\pTo^{\mathrm{ref}}$. The related parameters are primarily tuned to 
minimum bias and underlying event data, e.g.\ charged particle 
pseudorapidity $\d n / \d \eta$, multiplicity $P(n)$ and transverse 
momenta $\d n / \d \pT$ and $\langle \pT \rangle(n_{\mathrm{ch}})$ 
spectra of charged particles.
This means that a change of MPI parameters for the diffractive studies
would spoil agreement with nondiffractive data. Nevertheless, 
it is interesting to study how the survival rate changes with these 
parameters for the $\p\p$ collision itself.

The MPI modelling of the $\Pom \p$ collision can be decoupled from 
that of the $\p\p$ one. Then the MPI survival rate would not be 
affected by changes, but only the particle distributions in the 
diffractive system. One inevitable free parameter is the effective 
$\Pom\p$ total cross section. It is currently set always to be 10~mb,
but could be made to depend on the mass of the diffractive system.
Also the relative normalization of $\Pom$ flux and PDFs can influence
the event activity. We will study the normalization dependence in the 
last part of this section.

\begin{figure}[tbp]
\begin{minipage}[t]{0.5\textwidth}
\centering
\includegraphics[scale=0.4]{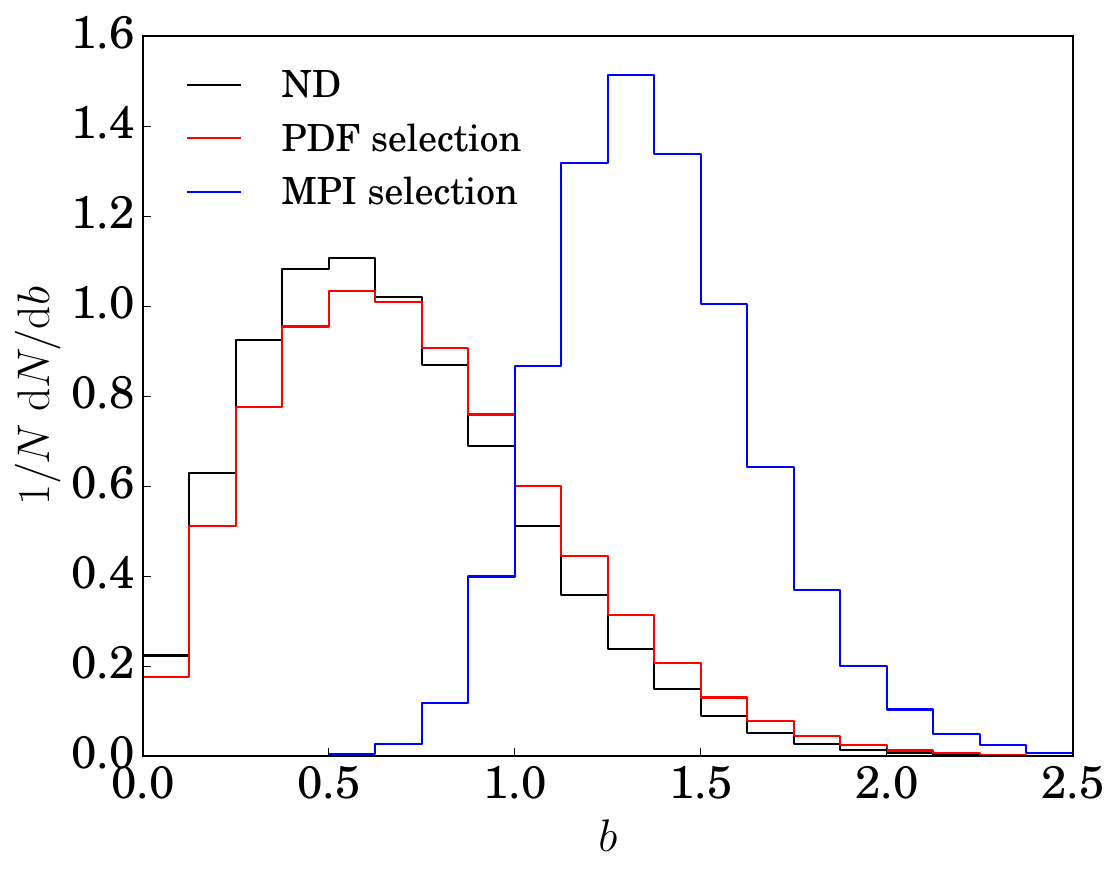}
(a)
\end{minipage}
\hfill
\begin{minipage}[t]{0.5\textwidth}
\centering
\includegraphics[scale=0.4]{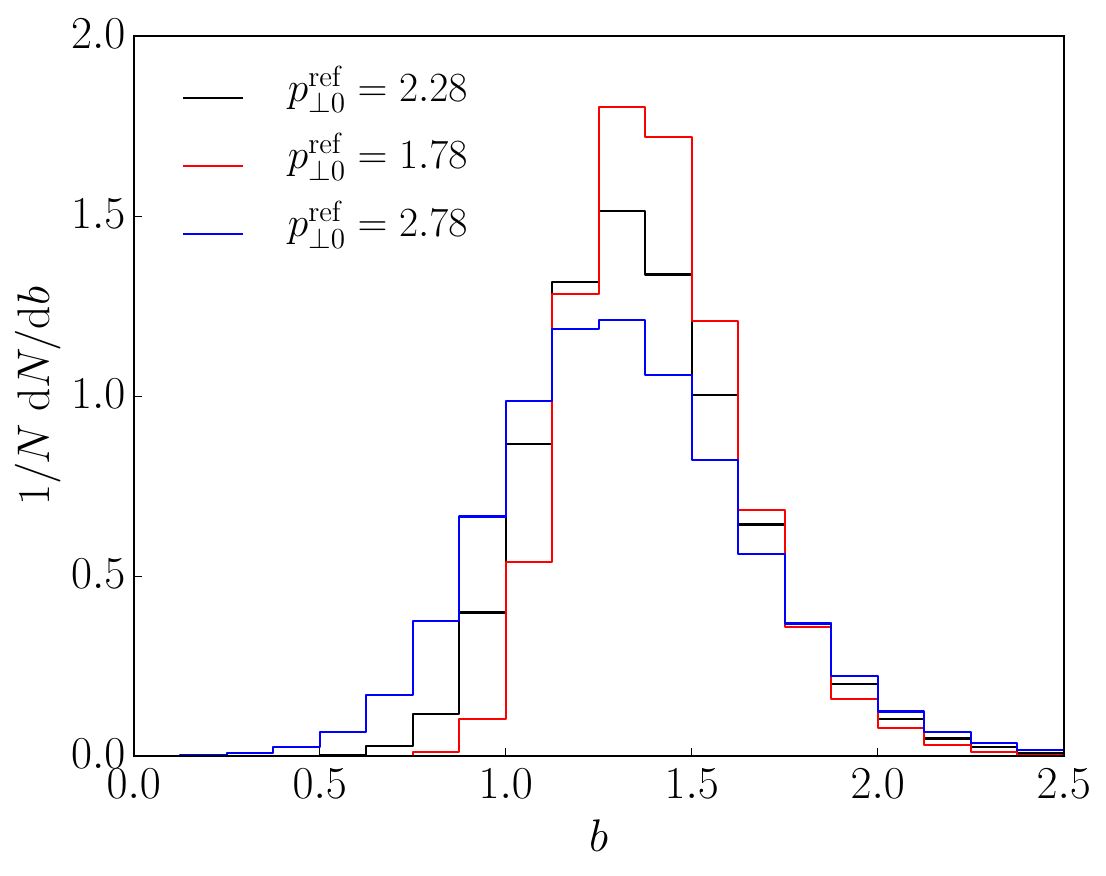}\\
(b)
\end{minipage}\\[2mm]
\begin{minipage}[t]{0.5\textwidth}
\centering
\includegraphics[scale=0.4]{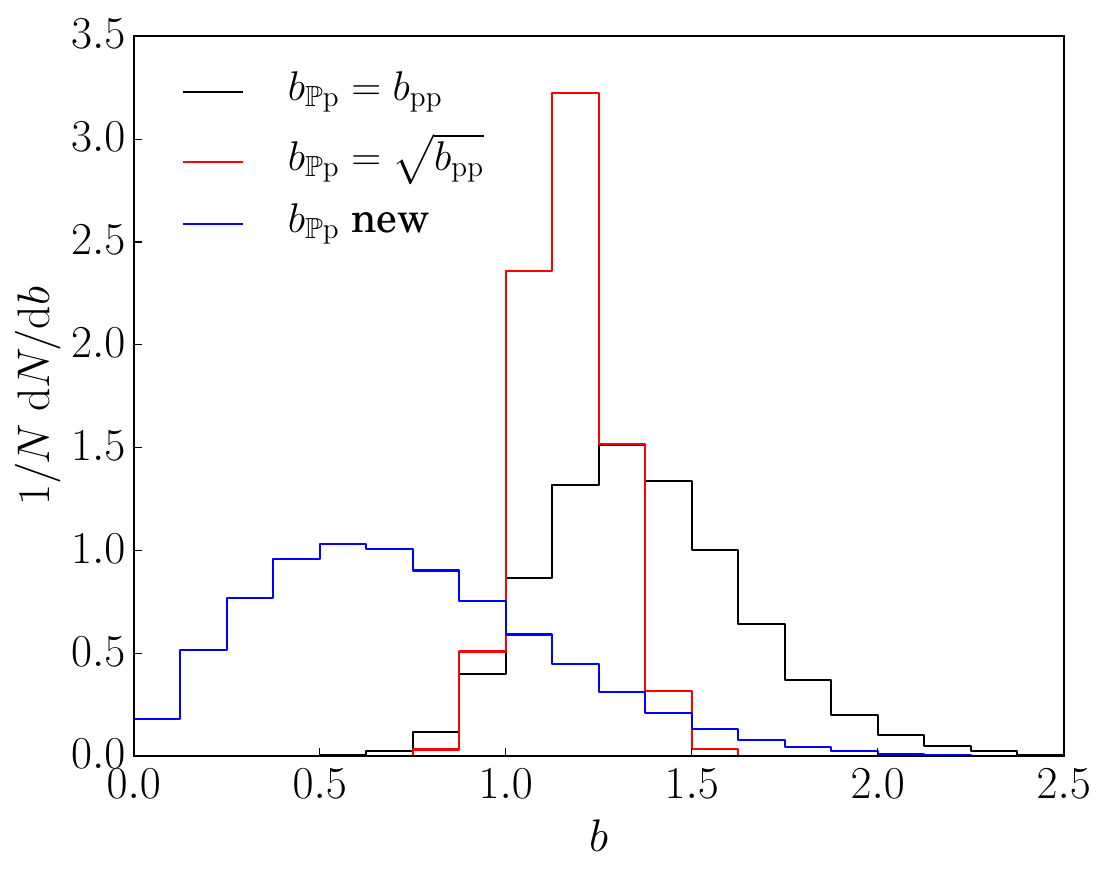}
(c)
\end{minipage}
\hfill
\begin{minipage}[t]{0.5\textwidth}
\centering
\includegraphics[scale=0.4]{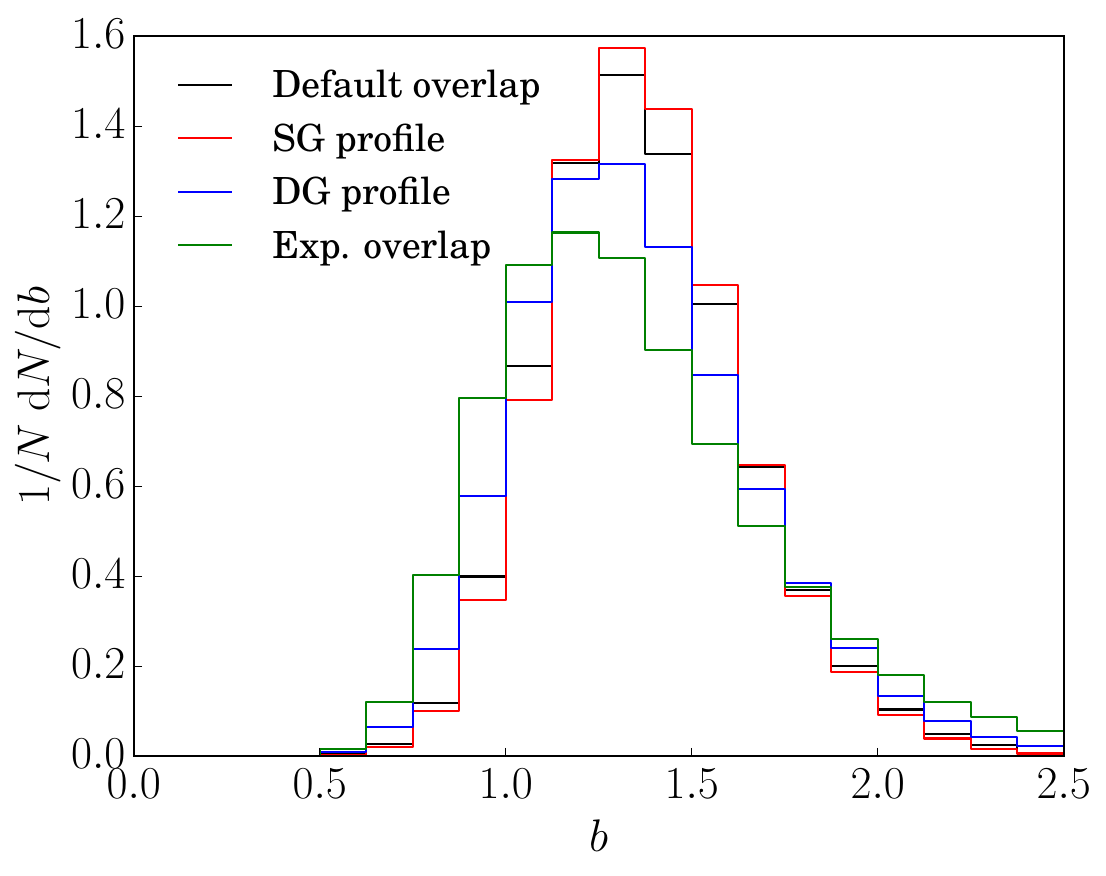}
(d)
\end{minipage}
\caption{\label{Fig:impactDist}
Impact-parameter distribution of $2 \rightarrow 2$ QCD processes with 
$\pT>20$ GeV in $\sqrt{s}=8$ TeV $\p\p$ collisions. 
(a) The change during the selection steps.
(b) The dependence on $\pTo^{\mathrm{ref}}$.
(c) The distribution in the $\Pom\p$ subcollision.
(d) The dependence on impact-parameter profile.}
\end{figure}

To begin with, consider the impact-parameter picture associated with 
hard collisions in our model, figure~\ref{Fig:impactDist}a. The $b$ scale is 
normalized such that $\langle b \rangle = 1$ for inclusive minimum-bias
events. Events with a hard interaction tend to be more central than that,
since central events have more MPIs in general and thereby a bigger
likelihood that at least one of them is at large $\pT$. The PDF selection
step does not have a significant impact, but the MPI one kills most 
low-$b$ events and pushes $\langle b \rangle$ above unity. The reason 
is obvious: for central events the average number of MPIs is high,
and thus the likelihood of only having the trigger hard process and no
further MPIs is small, while more peripheral collisions give fewer MPIs 
and thereby a higher surviving fraction. Ultimately, when 
$\langle n_{\mathrm{MPI}}(b) \rangle \ll 1$, most protons pass by 
each other without colliding at all. Thus the interesting region 
for diffraction is where $\langle n_{\mathrm{MPI}}(b) \rangle \sim 1$.

\begin{figure}[tbp]
\begin{minipage}[t]{0.5\textwidth}
\centering
\includegraphics[scale=0.4]{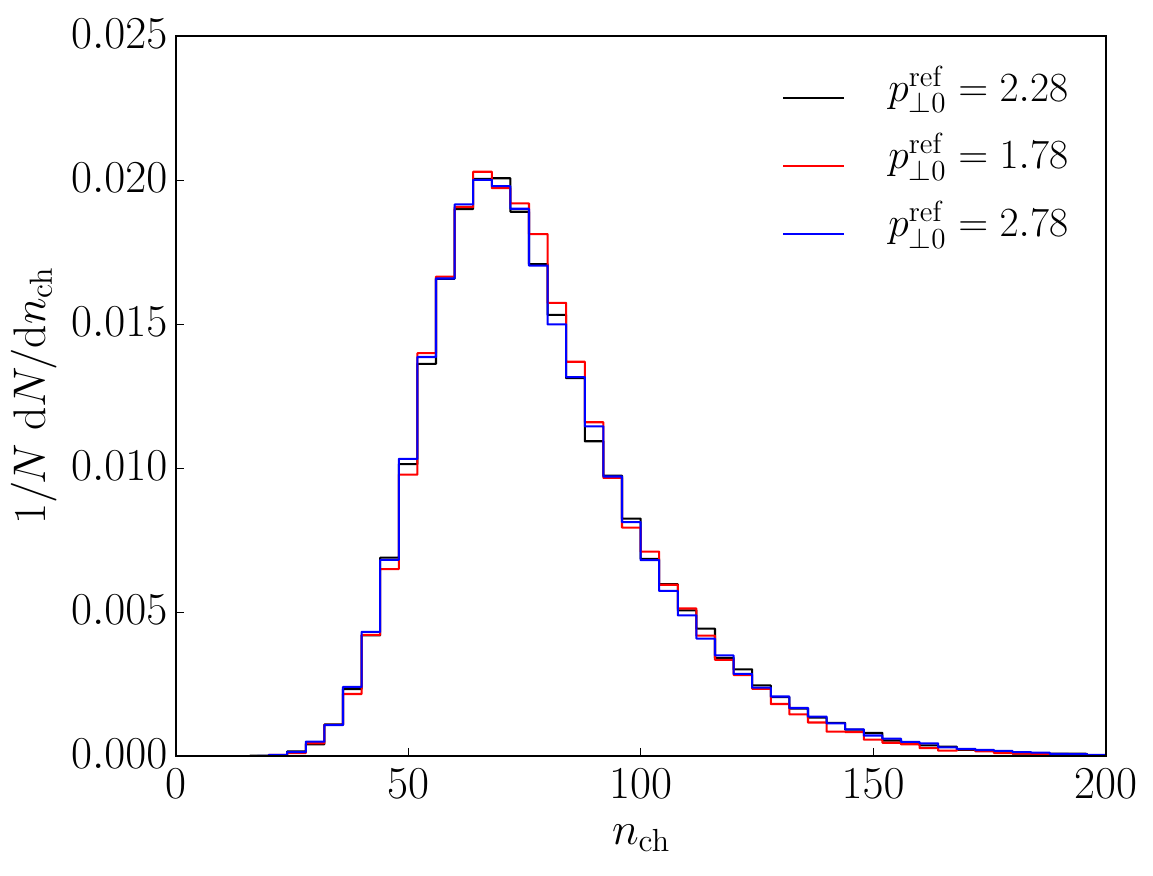}\\
(a)
\end{minipage}
\hfill
\begin{minipage}[t]{0.5\textwidth}
\centering
\includegraphics[scale=0.4]{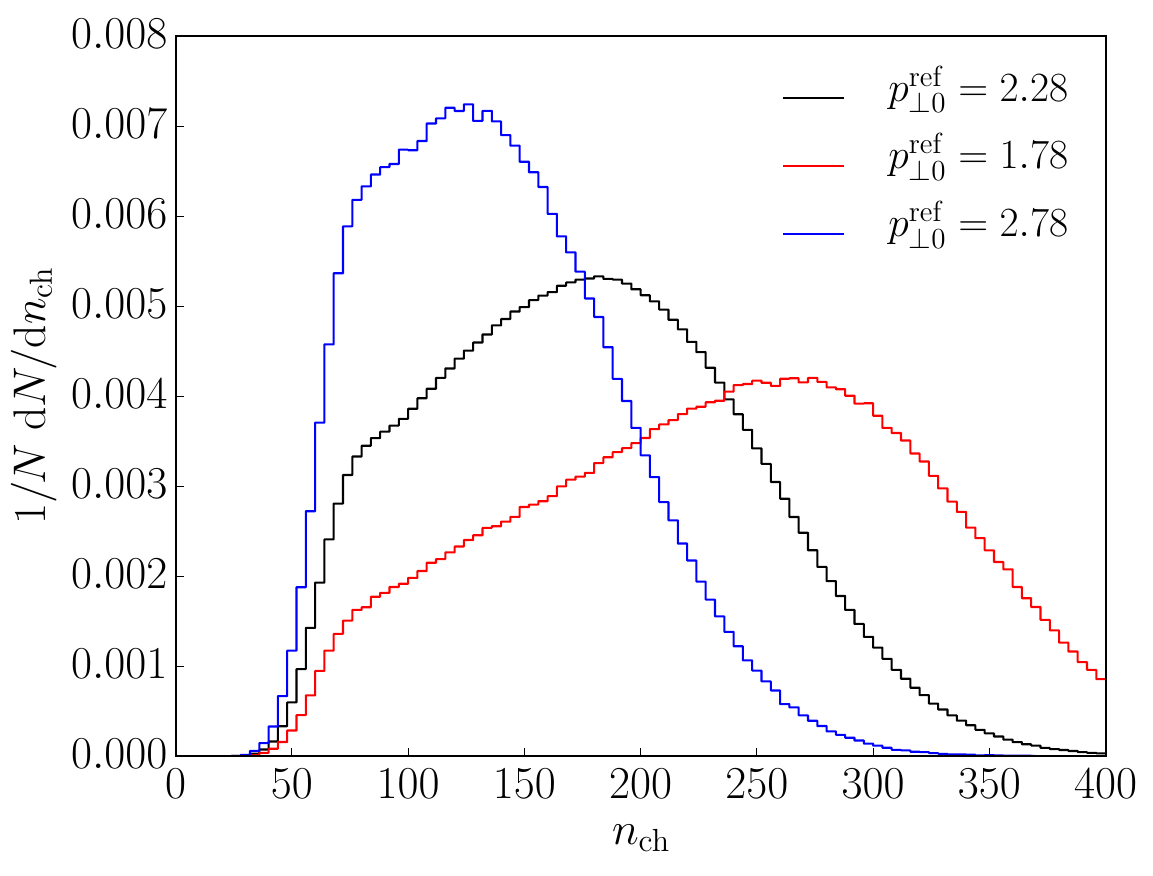}\\
(b)
\end{minipage}
\caption{\label{Fig:nch_pTref0}
Charged multiplicity distributions in the 
(a) $\Pom\p$ subsystem for diffractive events,
(b) $\p\p$ system for nondiffractive events, 
in $2\rightarrow2$ QCD processes with $\pT>20$ GeV as before.
}
\end{figure}

The $\pTo^{\mathrm{ref}}$ regulator is by default 2.28~GeV. Since an 
increase in this parameter gives less MPI in the $\p\p$ system, we 
expect an increase in the diffractive fractions, and vice versa. 
table~\ref{Tab:DiffFracpT0} confirms that this is indeed the case: 
variations of $\pm 0.5$~GeV around the default $\pTo^{\mathrm{ref}}$ 
value gives about a factor of two in the MPI selection rate. This 
major $\pTo^{\mathrm{ref}}$ dependence holds also for many other 
nondiffractive event properties, however; keeping everything else 
fixed even a variation of $\pm 0.1$~GeV would be unacceptable. 
In figure~\ref{Fig:nch_pTref0} we show the charged multiplicity 
distribution, when we change the regulator $\pTo^{\mathrm{ref}}$ for both 
diffractive and nondiffractive events, with
minor/major effects for the former/latter. The stability in the
diffractive case is because a change in the regulator also affects 
the impact parameter picture. Specifically, in this case 
$b_{\Pom\p} = b_{\p\p}$ has been assumed. A lower value of the regulator, 
giving rise to a larger number of MPIs in the $\p\p$ system, pushes 
$\langle b_{\p\p}\rangle$ to larger values for those events that
survive the diffractive MPI criterion, figure~\ref{Fig:impactDist}b.
More precisely, the change is to $b$ values where the average $\p\p$ 
MPI activity is restored to its original level. 
With $b_{\Pom\p} = b_{\p\p}$ the same then holds when MPI activity
is generated in the diffractive system, such that the effects of a 
smaller regulator and a larger impact parameter almost completely 
cancel.

\begin{table}[tbp]
\begin{center}
\begin{tabular}{|c|c|c|}
\hline
\multicolumn{3}{|c|}{Diffractive fractions}\\
\multicolumn{3}{|c|}{$\p\p$ collisions at $\sqrt{s}=8$ TeV}\\
\hline
& PDF selection & MPI selection\\
\hline
$\pTo=1.78$ & (14.50 $\pm$ 0.11) \% & (0.39 $\pm$ 0.02) \% \\
$\pTo=2.28$ & (14.33 $\pm$ 0.11) \% & (0.98 $\pm$ 0.03) \%\\ 
$\pTo=2.78$ & (14.19 $\pm$ 0.11) \% & (2.00 $\pm$ 0.04) \% \\
\hline
\end{tabular}
\caption{\label{Tab:DiffFracpT0}
Diffractive fractions for the $2\rightarrow2$ QCD processes with 
$\pT>20$ GeV in $\sqrt{s}=8$ TeV $\p\p$ collisions. 
\textsc{Pythia} is run with the SaS flux and 
the H1 Fit B LO PDF.}
\end{center}
\end{table}

\begin{figure}[tbp]
\begin{minipage}[t]{0.5\textwidth}
\centering
\includegraphics[scale=0.4]{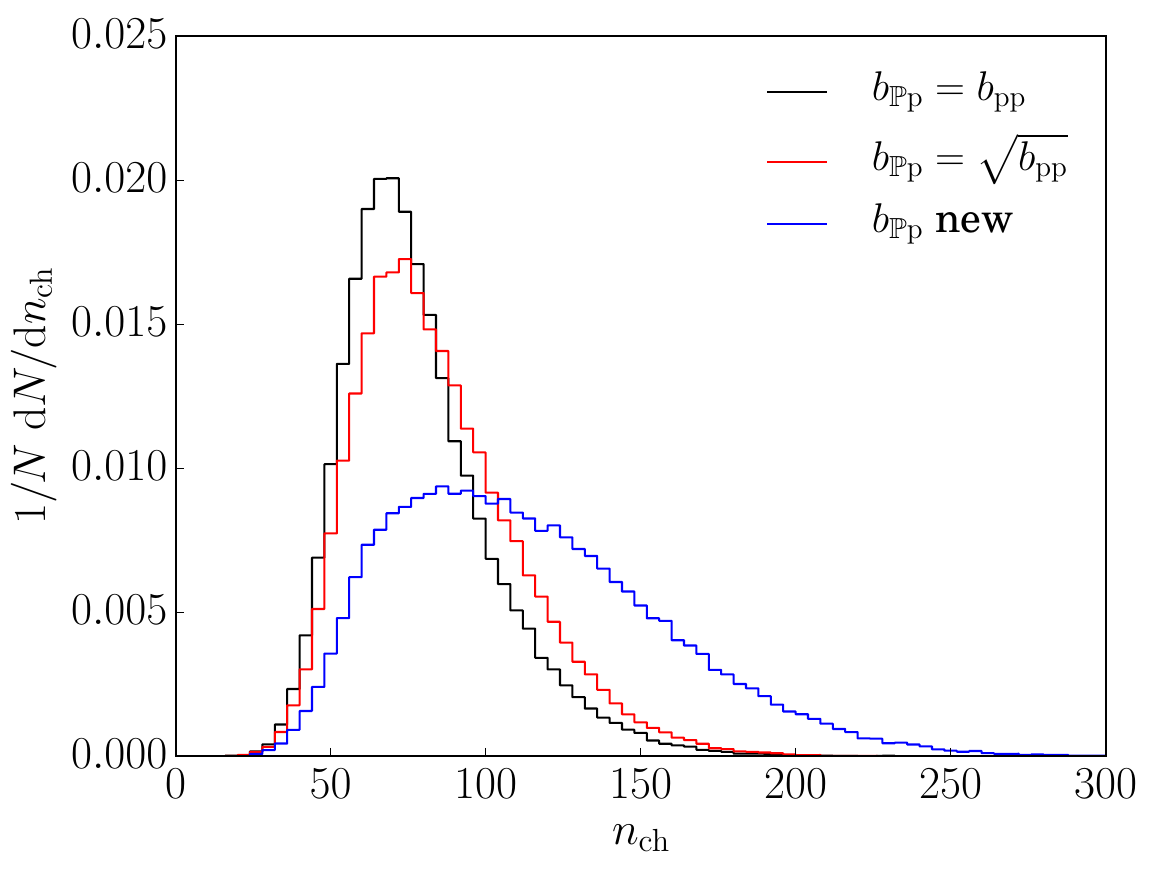}\\
(a)
\end{minipage}
\hfill
\begin{minipage}[t]{0.5\textwidth}
\centering
\includegraphics[scale=0.4]{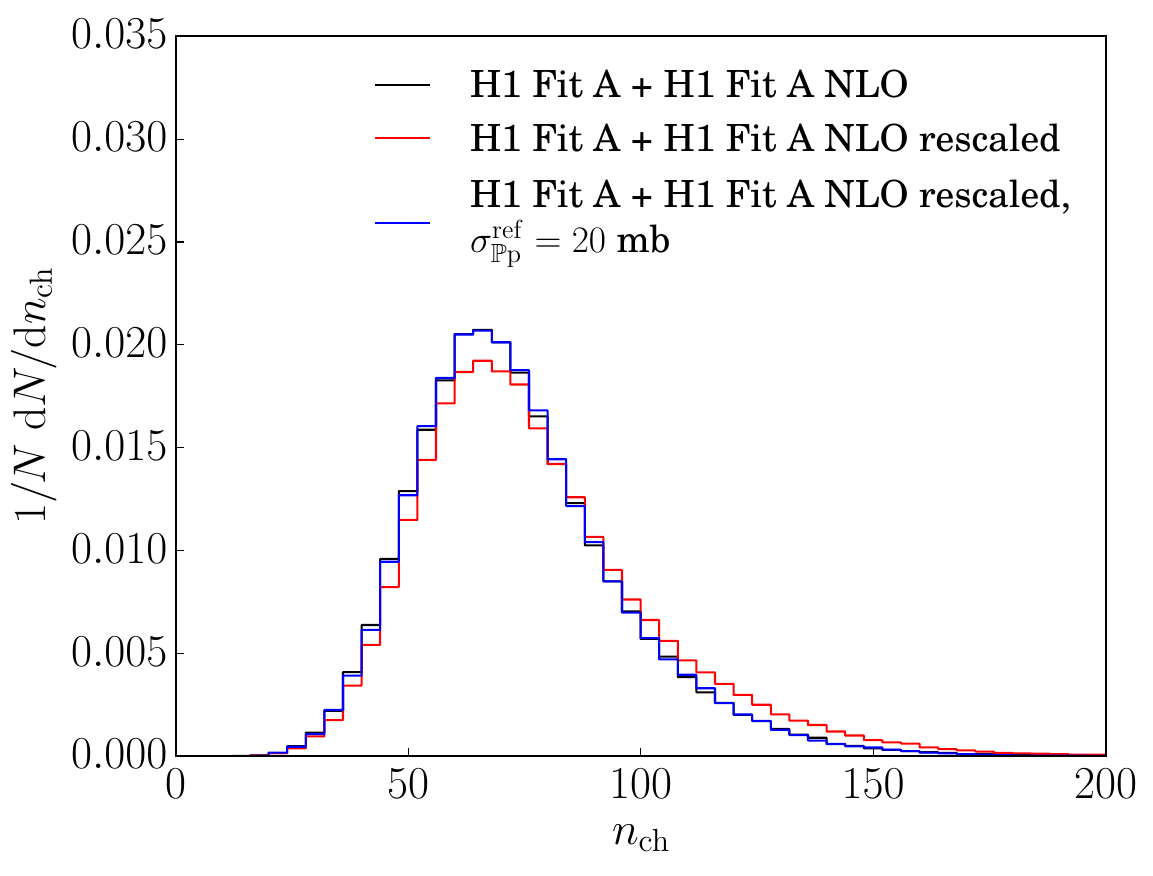}\\
(b)
\end{minipage}
\caption{\label{Fig:nchPomp}
Charged multiplicity distribution distributions for the 
$\Pom\p$ diffractive subsystem, for events with $2\rightarrow2$ QCD 
processes with $\pT>20$ GeV as before. 
(a) For three different $b_{\Pom\p}$ impact-parameter profiles.
(b) With or without rescaled $\Pom$ flux and PDFs, see text.}
\end{figure}

As we have already discussed, the modelling of the $\Pom$ size could 
also affect the MPI machinery for the $\Pom\p$ subcollision via the 
impact parameter $b_{\Pom\p}$. The currently 
implemented three alternatives are compared in 
figure~\ref{Fig:impactDist}c. The maybe less realistic last option of 
picking a new $b_{\Pom\p}$ value at random implies a significant fraction 
of events with small $b_{\Pom\p}$ and thereby the possibility of many MPIs.
The average $\langle n_{\mathrm{MPI}}\rangle$ for the three options is 
1.66, 2.04 and 4.09, respectively, thus giving rise to 0.66, 1.04 
and 3.09 additional MPIs besides the hardest process.
This is reflected notably in the charged multiplicity distribution,
figure~\ref{Fig:nchPomp}a. 


\begin{table}[tbp]
\begin{center}
\begin{tabular}{|c|c|c|}
\hline
\multicolumn{3}{|c|}{Diffractive fractions}\\
\multicolumn{3}{|c|}{$\p\p$ collisions at $\sqrt{s}=8$ TeV}\\
\hline
& PDF selection & MPI selection\\
\hline
No impact parameter dependence & (14.36 $\pm$ 0.11) \% & (0.38 $\pm$ 0.02) \% \\
Single gaussian matter profile & (14.25 $\pm$ 0.11) \% & (0.93 $\pm$ 0.03) \% \\
Double gaussian matter profile & (14.24 $\pm$ 0.11) \% & (1.04 $\pm$ 0.03) \% \\
Default overlap                & (14.33 $\pm$ 0.11) \% & (0.98 $\pm$ 0.03) \%\\
Exponential overlap            & (14.50 $\pm$ 0.11) \% & (1.28 $\pm$ 0.04) \%\\ 
\hline
\end{tabular}
\caption{\label{Tab:DiffFracImpact}
Diffractive fractions for the $2\rightarrow2$ QCD processes with 
$\pT>20$ GeV in $\sqrt{s}=8$ TeV $\p\p$ collisions. 
\textsc{Pythia} is run with the SaS flux and 
the H1 Fit B LO PDF.}
\end{center}
\end{table}

The MPI survival rate is highly dependent on the proton matter profile,
table~\ref{Tab:DiffFracImpact} and figure~\ref{Fig:impactDist}d.  
Diffraction thrives when $\langle n_{\mathrm{MPI}}(b) \rangle \sim 1$,
so this $b$ region should be as broad as possible for a large
diffractive 
rate. Conversely, a sharp proton edge implies less diffraction. The 
default overlap function $\exp(-b^{1.85})$ is close to a Gaussian, and 
the two have about the same MPI selection rate. The double Gaussian and 
the exponential overlap are examples of broader distributions, thus with 
more diffraction, whereas the option without any $b$ dependence 
represents the other extreme (not shown in figure~\ref{Fig:impactDist}d), 
with less diffraction. Overall the variation is not so dramatic, 
however, if only experimentally acceptable variations are considered.

Finally we turn to the relative normalization of the $\Pom$ 
PDF and flux. From eq.~(\ref{eq:fipD}) we know that the PDF 
selection step depends on the convolution of the $\Pom$
flux and PDFs. Thus it has no net effect if the flux is scaled down 
by a factor of two and the PDFs are scaled up by the same amount,
so as to bring the H1 PDFs to be approximately normalized to unit 
momentum sum. It does have consequences for the MPI selection step,
however, since the average MPI rate comes up in the $\Pom\p$ system. 

Compared with the (1.35 $\pm$ 0.04) \% MPI selection rate in 
table~\ref{Tab:DiffractiveFractions} for the H1 Fit A flux+PDF combination, 
such a rescaling changes the rate to (1.40 $\pm$ 0.04) \%, ie.\
no effects are seen. The rescaling 
however, does change the multiplicity distribution, figure~\ref{Fig:nchPomp}b, 
as a consequence of the increased $\d\sigma_{\mathrm{MPI}}$ in 
eq.~(\ref{eq:probMPI}). This could be compensated by a corresponding 
increase of $\sigma_{\mathrm{ref}}$ from the default 10~mb to 20~mb, 
thereby restoring both the MPI selection rate and the multiplicity 
distribution, cf.\ the blue line in figure~\ref{Fig:nchPomp}b.

\subsection{Energy and scale dependence}

Here we study the model dependence on the scales in the hard process 
and the energy of the collision.

\begin{figure}[tbp]
\begin{minipage}[t]{0.5\textwidth}
\centering
\includegraphics[scale=0.4]{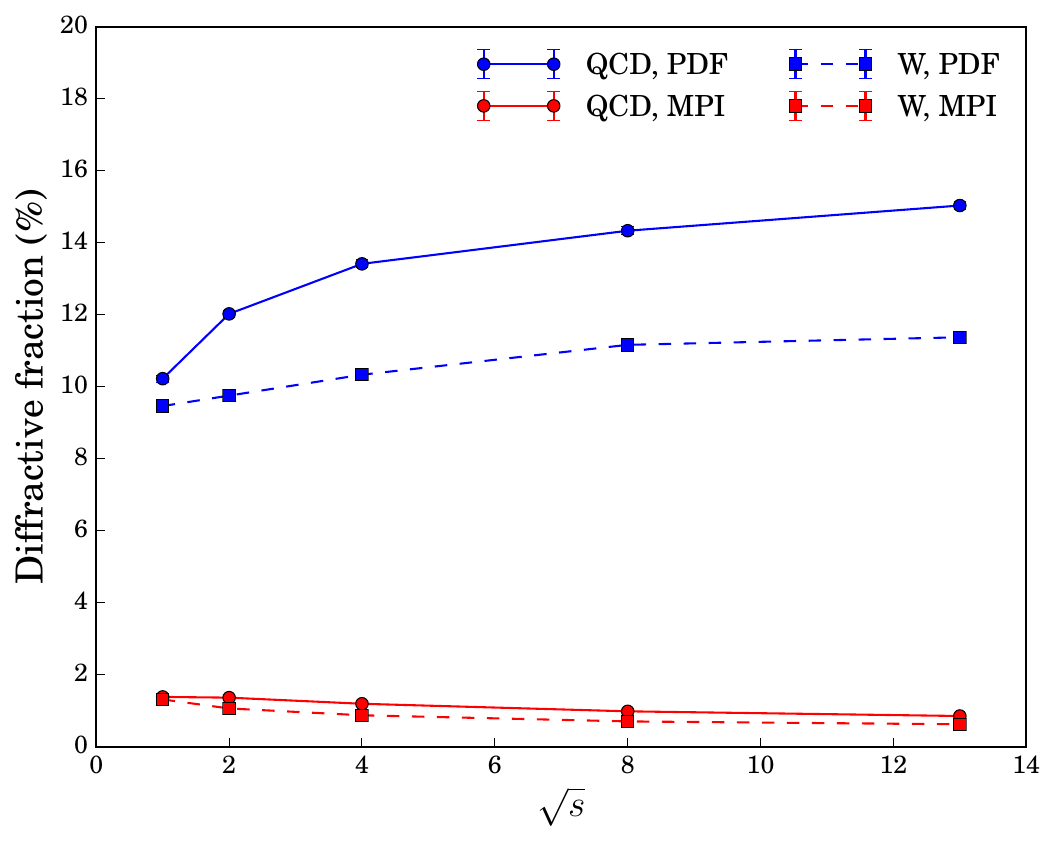}\\
(a)
\end{minipage}
\hfill
\begin{minipage}[t]{0.5\textwidth}
\centering
\includegraphics[scale=0.4]{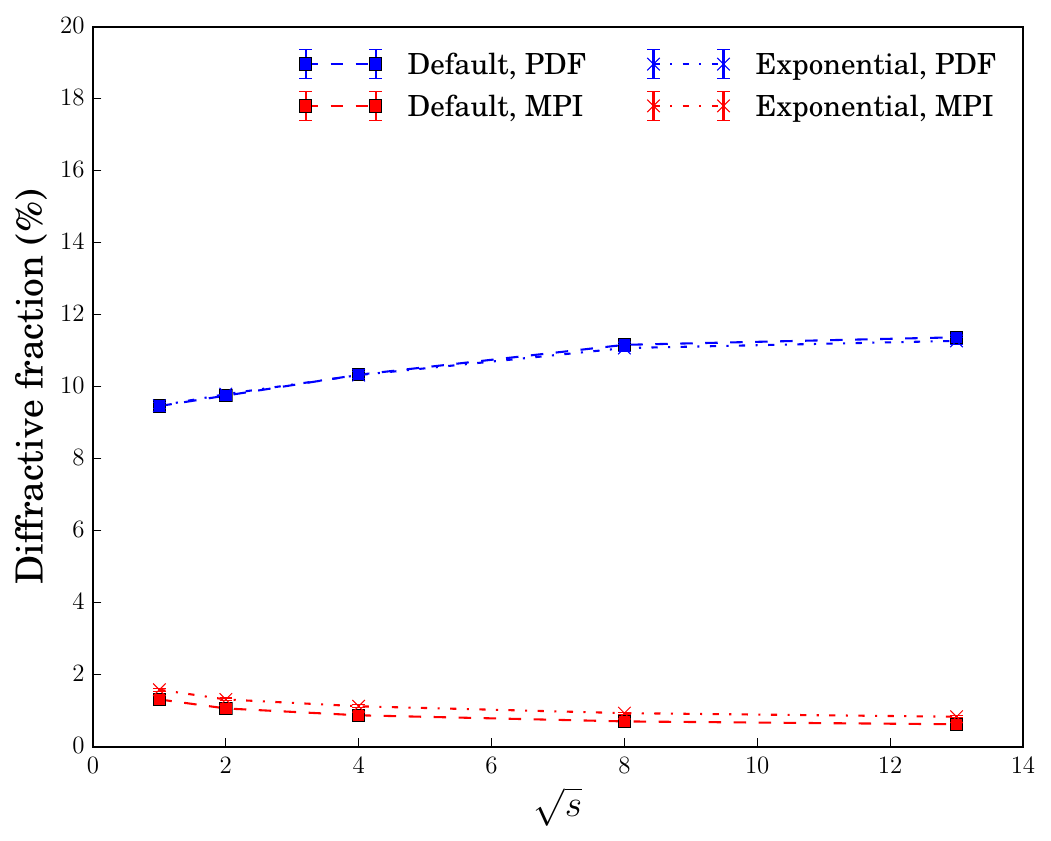}\\
(b)
\end{minipage}
\caption{\label{Fig:EnergyVariation}
(a) The diffractive fractions obtained in 
$2\rightarrow2$ QCD processes with $\pT>20$ GeV 
(circles and solid lines) and $\q\qbar\to W^{\pm}$ 
(squares and dashed lines) in pp collisions at 
different energies. 
(b) The diffractive fractions obtained in 
$\q\qbar\to W^{\pm}$ with the default overlap 
function (squares and dashed lines) and the 
exponential overlap function (crosses and dashed-dotted lines).
\textsc{Pythia} is 
run with the SaS flux and 
the H1 Fit B LO PDF.}
\end{figure}

In figure~\ref{Fig:EnergyVariation} the diffractive fractions are 
compared at different collision energies, $\sqrt{s}$, for 
$2\rightarrow2$ QCD processes with $\pT>20$ GeV, and for $\W^{\pm}$
production. In the PDF selection step the diffractive rate increases 
with energy. The difference between the two processes indicates
that this rise can depend on the incoming flavours and the relevant
ranges of $x$ values. Depending on the $\Pom$ flux and PDF, 
such as a freezing of the latter at small $x$, the 
fraction might even decrease with energy.

A larger collision energies also implies a higher average number 
of MPIs, in addition to the hardest collision, thus implying a 
reduced fraction of events passing the MPI criterion, see 
figure~\ref{Fig:EnergyVariation}. There is a compensatory effect of 
diffraction shifting to larger impact parameters, as already 
discussed for the $\pTo^{\mathrm{ref}}$ variations. For the 
close-to-Gaussian default overlap the relative size of the 
$\langle n_{\mathrm{MPI}} \rangle \approx 1$ region decreases
with energy, however, resulting in the trend shown. By comparison
an exponential overlap decreases slower than the close-to-Gaussian, 
hence resulting in less suppression with increasing energy.

Finally, table~\ref{Tab:DiffFracW} shows the number of events 
passing the PDF and MPI selections when the mass of the produced 
particle is changed. In the PDF selection step heavier particles 
are less likely to be produced diffractively, as they require 
larger $x$-values, where the probability for diffraction is lower 
(cf.\ figure~\ref{Fig:PomeronConvolution}). The same trend was observed 
in table~\ref{Tab:DiffFracProcesses}, but was there mixed up by 
the use of different production channels. After the MPI selection 
step the mass dependence is not as clearly visible. A partial 
compensation can indeed occur, since a higher subcollision mass
implies more momentum taken out of the incoming protons and thereby
less left for subsequent collisions. 

\begin{table}[tbp]
\begin{center}
\begin{tabular}{|c|c|c|}
\hline
\multicolumn{3}{|c|}{Diffractive fractions}\\
\multicolumn{3}{|c|}{$\p\p$ collisions at $\sqrt{s}=8$ TeV}\\
\hline
& PDF selection & MPI selection\\
\hline
$M_W = 50$ GeV      & (11.52 $\pm$ 0.10) \% & (0.72 $\pm$ 0.03) \% \\
$M_W = 80.385$ GeV  & (10.69 $\pm$ 0.10) \% & (0.70 $\pm$ 0.03) \% \\
$M_W = 150$ GeV     & (10.46 $\pm$ 0.10) \% & (0.72 $\pm$ 0.03) \% \\
$M_W = 500$ GeV     & ( 9.47 $\pm$ 0.09) \% & (0.65 $\pm$ 0.03) \% \\
\hline
\end{tabular}
\caption{\label{Tab:DiffFracW}
Diffractive fractions for the process $\q\qbar\rightarrow Z^0$ 
in $\sqrt{s}=8$ TeV $\p\p$ collisions. 
\textsc{Pythia} is run with the SaS flux and 
the H1 Fit B LO PDF.}
\end{center}
\end{table}

\subsection{\label{Sec:SoftHard}Comparison with soft diffraction}

The new model for hard diffraction complements
the existing one for soft (or rather inclusive) diffraction. 
The latter already has a hard component arising from the MPI 
model, which is used to pick the hardest process and all subsequent 
scatterings in the $\Pom\p$ system, except for low-mass diffractive 
systems where no perturbative framework can be applied. The soft
diffractive model only allows for $2\to2$ QCD processes, unlike the 
new hard one, but for QCD processes a comparison between the two is 
meaningful. To this end, the $\pT$ of the hardest process in an event 
will be used. This is not a physically measurable observable, unlike 
e.g.\ the closely related $\pT$ of the hardest jet in an event, but 
for the relative comparison of hard and soft diffraction it is cleaner. 
\begin{figure}[tbp]
\begin{minipage}[t]{0.5\textwidth}
\centering
\includegraphics[scale=0.4]{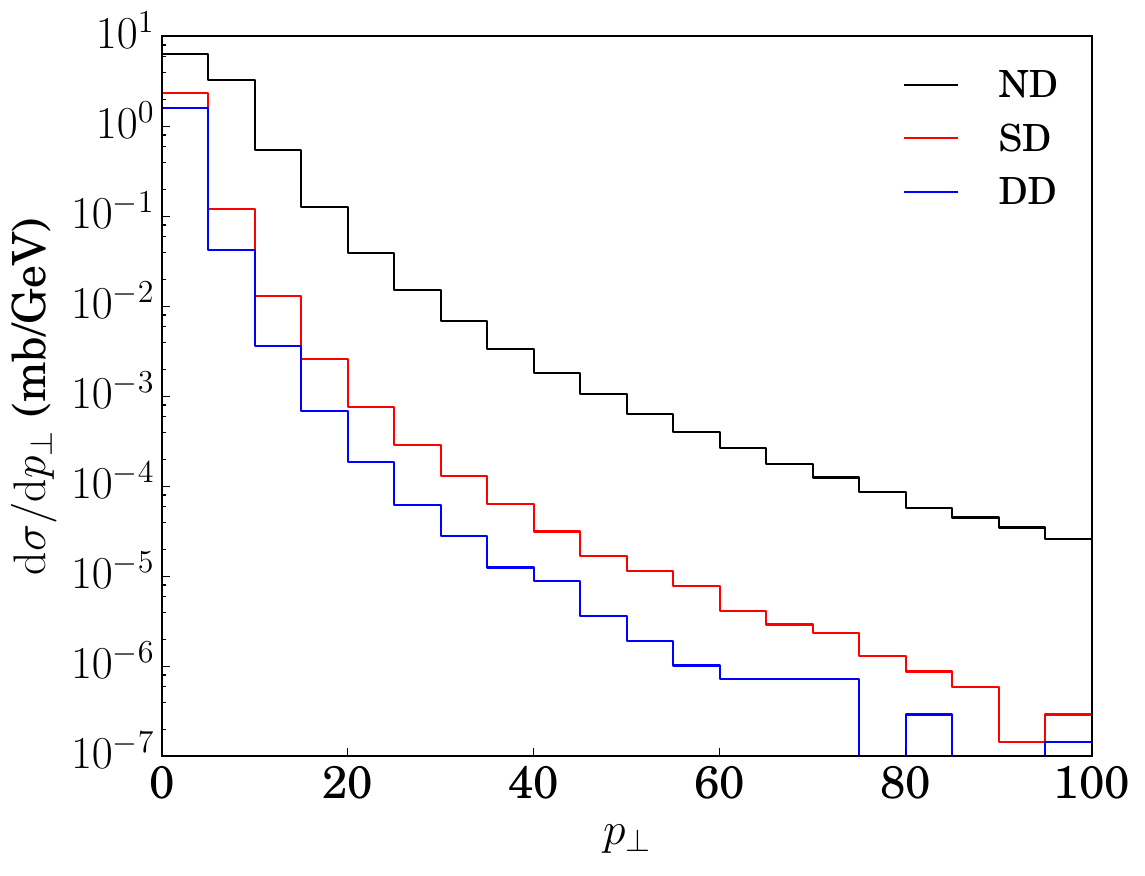}\\
(a)
\end{minipage}
\hfill
\begin{minipage}[t]{0.5\textwidth}
\centering
\includegraphics[scale=0.4]{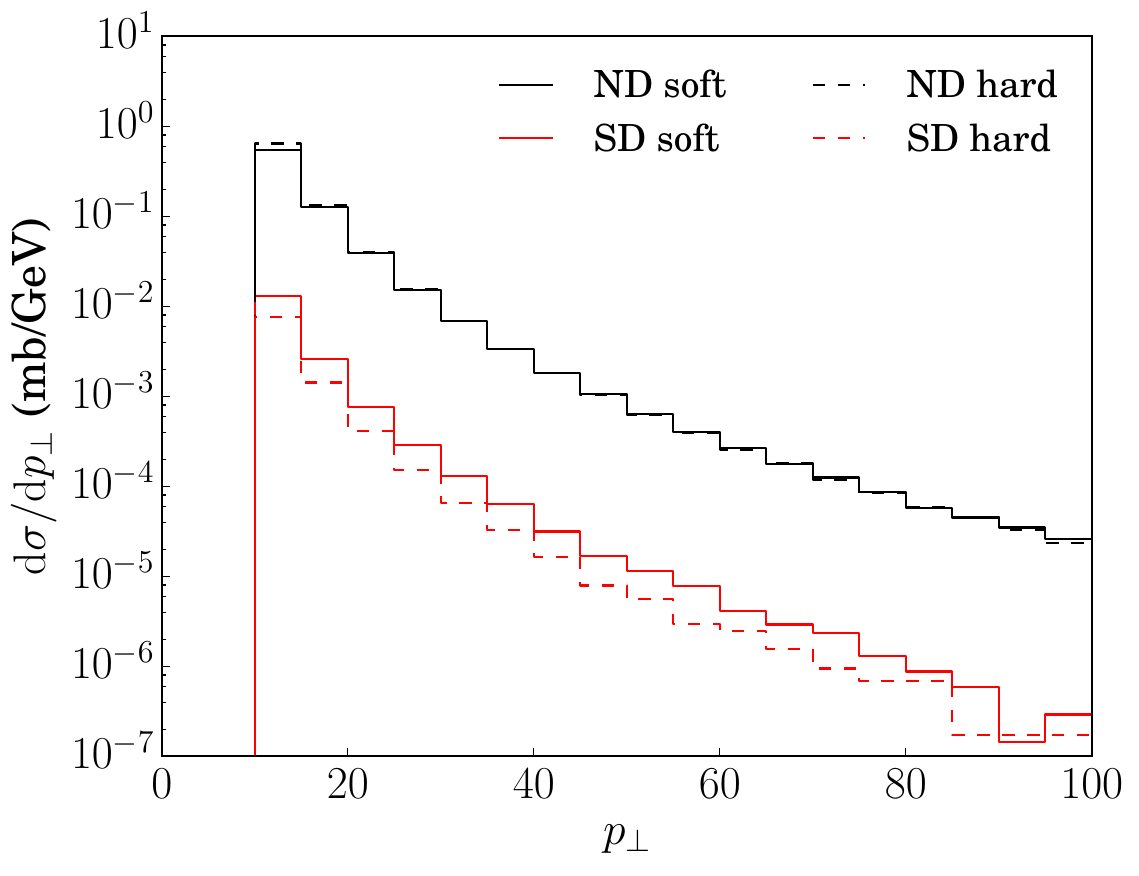}\\
(b)
\end{minipage}
\caption{\label{Fig:pThard}
The $\pT$ of the hardest process obtained with (a) the soft (or inclusive)
diffraction framework, and (b) both the soft and hard diffraction frameworks 
for events with $\pT>10$ GeV.}
\end{figure}

The MPI framework predominantly gives low-$\pT$ interactions, be it 
for diffractive or nondiffractive events. Thus only a small fraction 
of the events will have $\pT$ values at $\sim10$ GeV or more, see
figure~\ref{Fig:pThard}a. Note that the $\pT$ spectrum falls faster 
for diffractive than nondiffractive events, mainly as a consequence
of the former having a $\Pom\p$ invariant mass spectrum peaked towards
lower values.  

\begin{table}[tbp]
\begin{center}
\begin{tabular}{|c|c|c|}
\hline
\multicolumn{3}{|c|}{Cross sections}\\
\multicolumn{3}{|c|}{$\p\p$ collisions at $\sqrt{s}=8$ TeV}\\
\hline
& Soft diffraction & Hard diffraction\\
\hline
ND sample, $\pT>10$ GeV (mb) & 3.730 & 4.239\\
ND sample, $\pT>20$ GeV (mb) & 0.348 & 0.353\\
SD sample, $\pT>10$ GeV (mb) & 0.084 & 0.048\\
SD sample, $\pT>20$ GeV (mb) & 0.0066 & 0.0035\\
\hline
\end{tabular}
\caption{\label{Tab:Crosssections}
Cross sections obtained with the two diffractive frameworks.
Extracted from figure~\ref{Fig:pThard} by integration.}
\end{center}
\end{table}

In figure~\ref{Fig:pThard}b the $\pT$ of the hardest process for the
two samples is compared. One is obtained by generating inclusive (soft)
events and keeping only those with large enough $\pT$, the other by
generating only hard events above 10~GeV. Nondiffractive events are 
shown as a sanity check, as for them the two approaches should give 
the same results. A closer look at integrated cross sections, 
table~\ref{Tab:Crosssections}, shows a small discrepancy for the 
$\pT > 10$~GeV case, while the $\pT > 20$~GeV agree much better. 
This discrepancy is caused by not having a ``Sudakov factor'' 
in the hard model. That is, in the soft model the rate at lower 
$\pT$ scales is reduced by the requirement of not having an
interaction at a higher $\pT$ scale, whereas no such reduction is
implemented in the hard framework, which only uses pure matrix elements. 

The single diffractive events show differences in the normalisation, 
while the shape of the $\pT$ distributions agree between the two 
frameworks. The normalisation
differences arise from the two different ways of handling the
survival rate. The soft diffractive framework assumes an
effective flux of $\Pom$'s inside the proton, rescaled to get the 
desired total diffractive cross section, and thereby implicitly 
includes an average rapidity gap survival factor. The hard diffractive
framework has a higher initial $\Pom$ flux but then explicitly 
implements a dynamical event-by-event survival factor. As it works
out, single diffractive high-$\pT$ events are somewhat more suppressed
in the latter case. This is indeed what we would expect: there should
be more MPIs in high-$\pT$ (and high-mass) events than in low-$\pT$
ones, and thus more MPI survival suppression. Put another way, the 
soft implementation overestimates the suppression at low $\pT$ and 
underestimates it at high $\pT$. (Assuming our new model is the right 
way to view the matter.)

In the future it would be desirable to include such dynamical effects
also in the soft framework, so that the two descriptions can be made
to agree in the high-$\pT$ region. This is not a trivial task, however.

\section{Comparisons with data}

In this section we compare the new model for hard diffraction with
some available data. While many results have been presented for soft 
diffractive processes, less is available on hard ones. 

At the Tevatron, both the CDF and D0 collaborations studied hard 
diffractive events. We have chosen here to compare with two analyses, 
one in which only the diffractive fractions are measured, the other 
in which also the distributions of the hard collisions are reported. 

At the LHC, diffraction has been studied both by ATLAS 
\cite{Aad:2011eu, Aad:2012pw, Aad:2015xis} 
and CMS \cite{Chatrchyan:2011wb, Chatrchyan:2012vc,
Khachatryan:2015gka}.
One key observation there is that the \textsc{Pythia} default $\Pom$ 
flux shape does not describe the rapidity gap distribution so well,
suggesting that a new parametrization may be needed. In other respects
the model seems to do a reasonable job. For hard diffraction we
will compare to the latest ATLAS study, \cite{Aad:2015xis}, and a 
similar CMS study, \cite{Chatrchyan:2012vc}.
 
Unfortunately, neither of the studies at hand are implemented as 
Rivet \cite{Buckley:2010ar} analyses, so we have tried to apply 
the relevant experimental cuts as best as we can. This makes 
comparisons with data less than reliable, and results should therefore 
be taken as a first indication only. At least for LHC the intention is 
that the new \textsc{Pythia} options can be directly tested by the
experimental community, to allow more precise comparisons in  the 
future.  

\subsection{\label{Sec:WZcomp}Diffractive $\W/\Z$ production at the Tevatron}

CDF has measured the fraction of events with a diffractively produced 
$\W/\Z$ boson at $\sqrt{s}=1.96$~TeV \cite{Aaltonen:2010qe}. 
The surviving antiproton was measured in a Roman Pot forward 
spectrometer, and the boson decay products in the central detector. 
The observed fraction of events with forward antiprotons was doubled,
to compensate for there being no Roman Pots on the proton side.
Only the $\e$ and $\mu$ leptonic decays of the bosons were taken into 
account. The cuts used in the analysis are listed in 
table~\ref{Tab:CDFcutsWZ}, along with the number of events that 
survive after each step. To this end, the internal $\W$- and $\Z$-finder 
projections available in Rivet \cite{Buckley:2010ar} have been used as 
a starting point; these have previously been validated for other CDF 
analyses. In addition the diffractive properties are derived from
the measured antiproton as 
\begin{align}
t &= -\pT^2\label{eq:tRPS}\\
x_{\Pom}^{\mathrm{RPS}} &= 1 -
\frac{2|p_z|}{\sqrt{s}}\label{eq:xiRPS}
\end{align}
which has been compared to Monte Carlo truth, giving good agreement.

\begin{table}
\begin{center}
\begin{tabular}{|l|c|c|c|c|c|c|}
\hline
\multirow{2}{*}{CDF cuts} & 
  \multicolumn{3}{|c|}{$W$ sample} & 
  \multicolumn{3}{|c|}{$Z$ sample}\\
 & ND & SD & $\frac{\mathrm{SD}\times2}{\mathrm{ND}}$ (\%) & ND & SD & $\frac{\mathrm{SD}\times2}{\mathrm{ND}}$(\%) \\
\hline
Lepton $E_T^e(p_T^{\mu})>$ 25 GeV & 670602 & 2827 & 0.84 & 667851 & 2466 & 0.74\\
Missing $E_T>$ 25 GeV & 595236 & 2490 & 0.84 & - & - & - \\
One electron in $|\eta|<$ 2.8  & - & - & - & 642250 & 2366 & 0.74\\
One lepton in $|\eta|<$ 1.1  & 331316 & 1374 & 0.83 & 366566 & 1397 & 0.76\\
$M_T^W = $ [40, 120] GeV & 327671 & 1361 & 0.83 & - & - & - \\
$M_Z = $ [66,116] GeV  & - & - & - & 36814 & 1397 & 0.76\\ 
$|t|<$ 1 GeV$^2$ & - & 1348 & 0.82 & - &  1383 & 0.75\\
$x_{\Pom} = $ [0.03,0.1] & - & 366 & 0.23 & - & 346 & 0.19 \\
\hline
\end{tabular}
\caption{\label{Tab:CDFcutsWZ}
Cuts used in \cite{Aaltonen:2010qe}. Number of events listed in
each of the samples are based on Monte Carlo truth obtained when
generating $10^6$ inclusive events. A blank means that a specific
cut was not relevant.}
\end{center}
\end{table}

The results in table~\ref{Tab:CDFcutsWZ} are obtained with 
\textsc{Pythia}~8 using the SaS flux and the 
H1 Fit B LO PDF, starting out from an inclusive MPI-selected sample. 
We note that a large fraction of the diffractive events do not pass 
the experimental $x_{\Pom}$ cut. Therefore, although we begin with 
a ``Monte Carlo truth'' fraction of $\sim 1\%$ diffractive $\W/\Z$, 
this is reduced to $\sim 0.2\%$ by the $x_{\Pom}$ cut. Results look 
better for other choices of $\Pom$ flux, see table~\ref{Tab:WZfrac}, 
but even at best still with a factor two discrepancy. Note that
it is the fluxes that rise fastest in the low-$x_{\Pom}$ region that
gives fractions closer to data. 

\begin{table}[tbp]
\begin{centering}
\begin{tabular}{|l | c | c|}
\hline
$\Pom$ PDF & & \\
$\Pom$ flux 
  & ($\p\pbar\rightarrow \pbar' + \W$) $\times$ 2
  & ($\p\pbar\rightarrow \pbar' + \Z$) $\times$ 2 \\
\hline
CDF & (1.0$\pm$0.11) \% & (0.88$\pm$0.22) \%\\
\hline
H1 Fit B LO & & \\
SaS & (0.19 $\pm$ 0.03) \% & (0.24 $\pm$ 0.04) \%\\ 
\hline
H1 Fit B LO & & \\
MBR                 & (0.29 $\pm$ 0.04) \% & (0.20 $\pm$ 0.03) \%\\ 
\hline
H1 Jets & & \\
SaS & (0.29 $\pm$ 0.04) \% & (0.24 $\pm$ 0.04) \%\\ 
\hline
H1 Fit A NLO & & \\
H1 Fit A            & (0.46 $\pm$ 0.05) \% & (0.35 $\pm$ 0.04) \%\\ 
\hline
H1 Fit B LO & & \\
H1 Fit A            & (0.44 $\pm$ 0.05) \% & (0.29 $\pm$ 0.04) \%\\ 
\hline
\end{tabular}
\caption{\label{Tab:WZfrac}Diffractive fractions for 
the $\W\rightarrow l\nu$ and 
$\Z\rightarrow l^+l^-$, $l=e,\mu$ in $\sqrt{s}=1.96$ TeV 
$\p\pbar$ collisions.}
\end{centering}
\end{table}

We can compare these values to the results from \cite{Alvero:1998ta}, where no
gap survival factor is included. The authors only show results on
$\W$ production and use different integration limits on $x_{\Pom}$. 
A subset of the results are listed in Table~\ref{Tab:WZfracAlvero}.
\begin{table}[tbp]
\begin{centering}
\begin{tabular}{|l | c | c|}
\hline
$\Pom$ PDF & & \\
$\Pom$ flux 
  & $x_{\Pom}=0.01$ & $x_{\Pom}=0.1$ \\
\hline
CDF & - & (1.0$\pm$0.11) \%\\
\hline
Fit B & & \\
DL, $\epsilon=0.14$ & 0.14 \% & 5.1 \%\\ 
\hline
Fit D & & \\
DL, $\epsilon=0.14$ & 0.18 \% & 6.9 \%\\ 
\hline
Fit SG & & \\
DL, $\epsilon=0.14$ & 0.14 \% & 4.1 \%\\ 
\hline
\end{tabular}
\caption{\label{Tab:WZfracAlvero}Diffractive fractions for 
the $\W$ production from \cite{Alvero:1998ta}.}
\end{centering}
\end{table}
It is worth noting that the results using the lower integration
limit are of the same order as the default settings of
\textsc{Pythia}~8, whereas the high integration limit (which is
that of CDF) are higher than both data and our model. This we
interpret as being  due to the lack of suppression factor, as their 
calculations do not take MPIs into account. 

The diffractive fraction can also be increased by changing the free 
parameters of the MPI framework, with the caveat that nondiffractive 
events will then no longer describe data as well. Table~\ref{Tab:WZfracMPI}
shows the diffractive fractions obtained when varying some of 
the MPI parameters. This variation is still not sufficient when combined
with the default flux and PDF in \textsc{Pythia}~8. If combined with
some of the fluxes in table~\ref{Tab:WZfrac} it would be possible to obtain 
fractions close to the experimentally observed values, however. 

\begin{table}[tbp]
\begin{centering}
\begin{tabular}{|l | c | c|}
\hline
Parameter 
  & ($\p\pbar\rightarrow \pbar' + \W$) $\times$ 2
  & ($\p\pbar\rightarrow \pbar' + \Z$) $\times$ 2 \\
\hline
CDF & (1.0$\pm$0.11) \% & (0.88$\pm$0.22) \%\\
\hline
$\pTo^{\mathrm{ref}}=2.78$ GeV & (0.59 $\pm$ 0.06) \% 
& (0.49 $\pm$ 0.05) \%\\ 
\hline
Exponential overlap            & (0.25 $\pm$ 0.04) \% 
& (0.24 $\pm$ 0.04) \%\\ 
\hline
\end{tabular}
\caption{\label{Tab:WZfracMPI}Diffractive fractions for 
the $W\rightarrow l\nu$ and 
$Z\rightarrow l^+l^-$, $l=\e,\mu$ in $\sqrt{s}=1.96$ TeV 
$\p\pbar$ collisions.}
\end{centering}
\end{table}

\subsection{Diffractive dijets at the Tevatron}

Another interesting measurement performed at CDF was 
the process $\pbar\p \rightarrow \pbar + X_{\p},\, 
X_{\p} \rightarrow X + J + J$, ie. SD dijet production with 
a leading antiproton. CDF measured this at three 
different energies, $\sqrt{s} = 630$, 1800 and 1960~GeV 
\cite{Affolder:2001zn, Affolder:2000vb, Aaltonen:2012tha}. 
Here not only the diffractive fractions were measured,
but a number of differential distributions. Large 
discrepancies were found between the diffractive structure functions 
determined from CDF data and those extracted by the H1 Collaboration 
from diffractive deep inelastic scattering data at HERA. 
The discrepancies 
are both in normalisation and shape and were interpreted as a 
breakdown of factorization. 

\begin{table}
\begin{center}
\begin{tabular}{|c|c|}
\hline
\multicolumn{2}{|c|}{CDF cuts}\\
\hline
Jet $E_T^{1,2}$ & $>$ 7 GeV \\
Jet $E_T^{3}$ & $>$ 5 GeV \\
Jet $|\eta^{1,2,3}|$ & $<$ 4.2 \\
$\Delta R$ & 0.7 \\
$|t|$ & $<$ 1 GeV$^2$ \\
$x_{\Pom}^{\mathrm{RPS}}$ & [0.035,0.095]\\
\hline
\end{tabular}
\caption{\label{Tab:CDFcutsJets}
Cuts used in \cite{Affolder:2000vb}.}
\end{center}
\end{table}

Our comparison focuses on the 1800~GeV data
(\cite{Affolder:2000vb}), since this 
also includes a measurement of the 
diffractive structure function. The cuts used in the analysis are
listed in table~\ref{Tab:CDFcutsJets}. 
The jets are identified with the CDF cone algorithm as implemented 
in Rivet \cite{Buckley:2010ar}, with a cone radius of $0.7$. Jet
energy scale corrections for underying-event activity are done 
separately for diffractive and nondiffractive events, as outlined
in the CDF article, but only has a minor impact on relative rates.
The momentum transfer of the antiproton is evaluated using 
eq.\ (\ref{eq:tRPS}) and the momentum loss of the antiproton using 
eq.\ (\ref{eq:xiRPS}).

We begin by evaluating the suppression factor introduced by the
MPI framework. This is evaluated by running two samples of
$10^{6}$ events, one with and one without the MPI criterion, both
using the cuts of table~\ref{Tab:CDFcutsJets} and the
SaS flux and the H1 Fit B LO PDF. We obtain a
suppression factor of 0.11, to be compared with the quoted
discrepancies from CDF of $0.06\pm0.02 \, (0.05\pm0.02)$ when using
the H1 Fit 2 (Fit 3), respectively \cite{Affolder:2000vb}. A
similar suppression factor as for SaS is obtained when using the H1 Fit B
flux, based on the same parametrization as the H1 Fit 2 and 3
fluxes, although with different values for the free parameters of
the model. Using this flux, however, allows for approximately two
times more events passing the experimental cuts. This is due to
the fact that the H1 Fit B flux is less restrictive in the
low-$x_{\Pom}$ region, where the experiment is performed. Hence
we expect better agreement with data when using the H1 Fit B
flux, as compared to SaS. We are not able to directly compare to
the suppression factors obtained in \cite{Alvero:1998ta}, as these have been
calculated with different kinematical cuts (eg.\ $E_T>10$ GeV and
$0.05<x_{\Pom}<0.1$), but the numbers obtained are
still interesting in their own right. Alvero et. al. obtain
suppression factors of 0.061 (fit B and DL flux, $\epsilon=0.14$), 
0.029 (fit D, same flux) and 0.12 (fit SG, same flux), thus ranging 
from the measured suppression factor to our one.

\begin{figure}[tbp]
\begin{minipage}[t]{0.5\textwidth}
\centering
\includegraphics[scale=0.4]{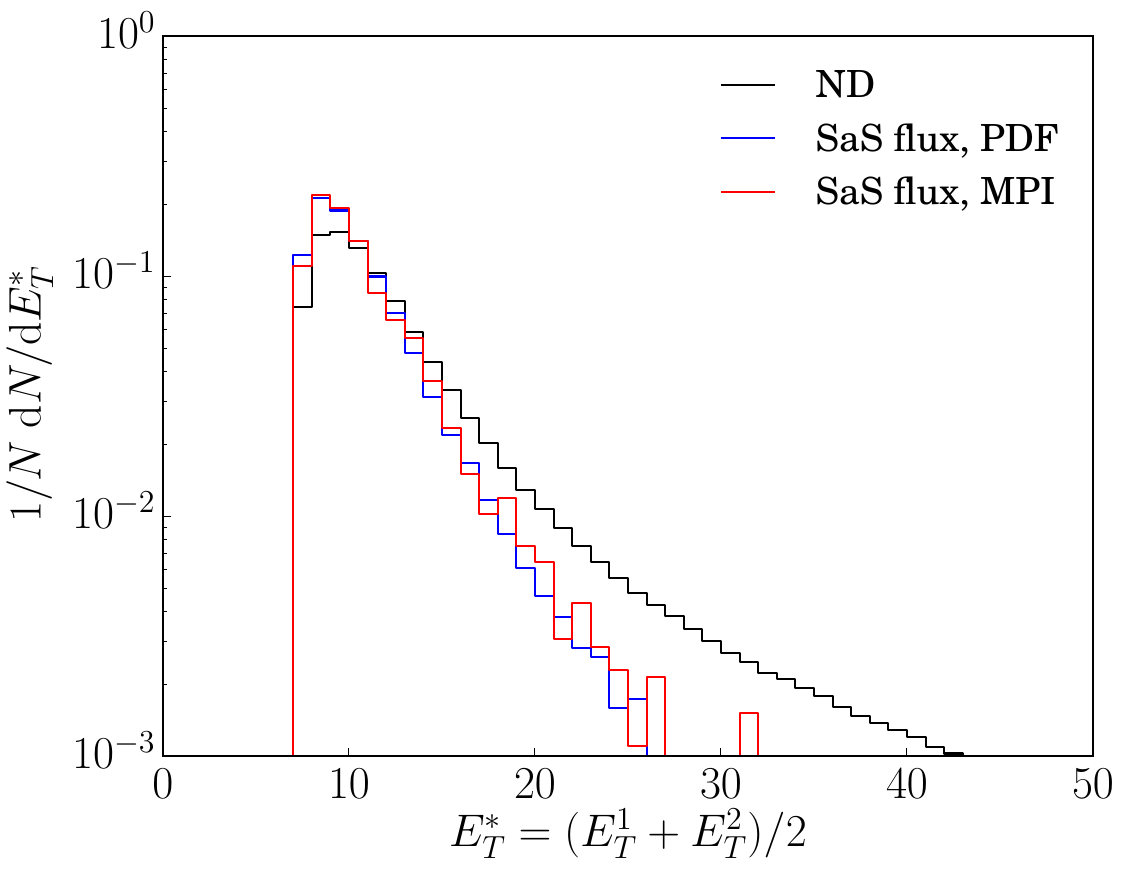}\\
(a)
\end{minipage}
\hfill
\begin{minipage}[t]{0.5\textwidth}
\centering
\includegraphics[scale=0.4]{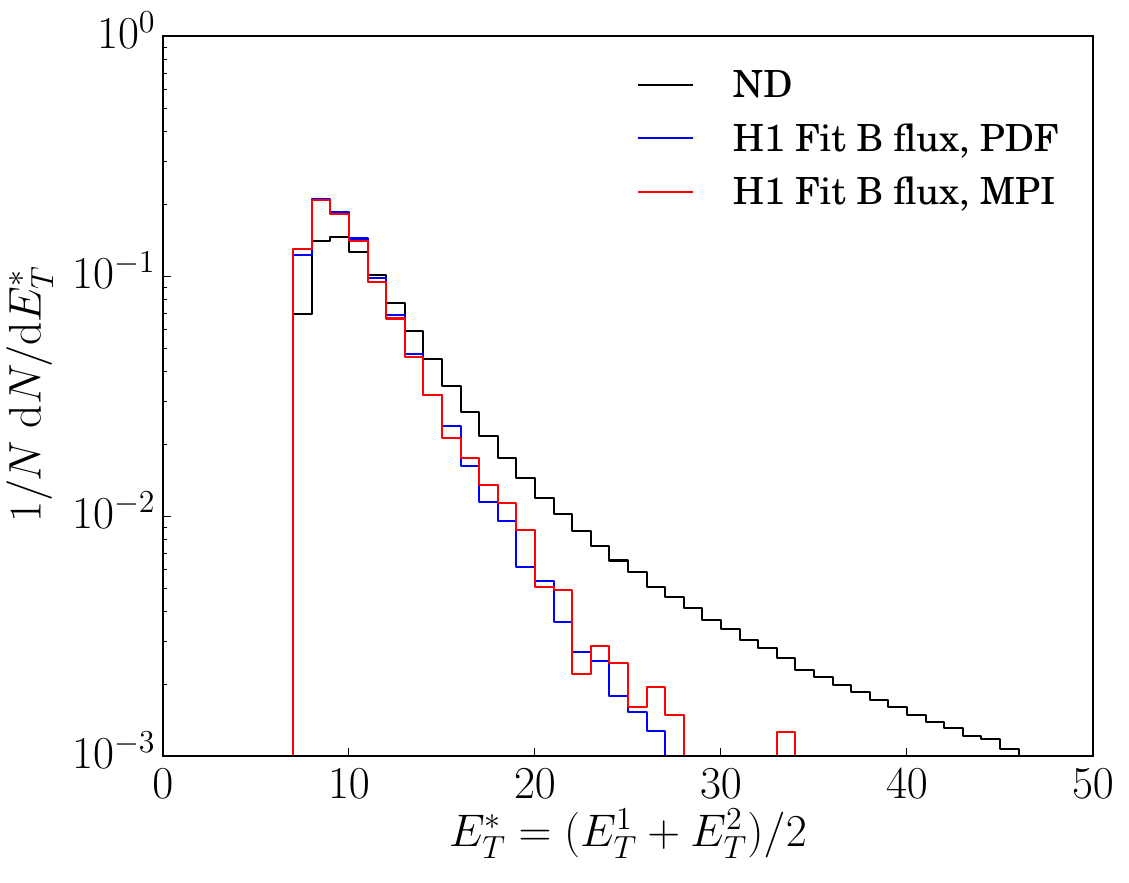}\\
(b)
\end{minipage}
\hfill
\begin{minipage}[t]{0.5\textwidth}
\centering
\includegraphics[scale=0.4]{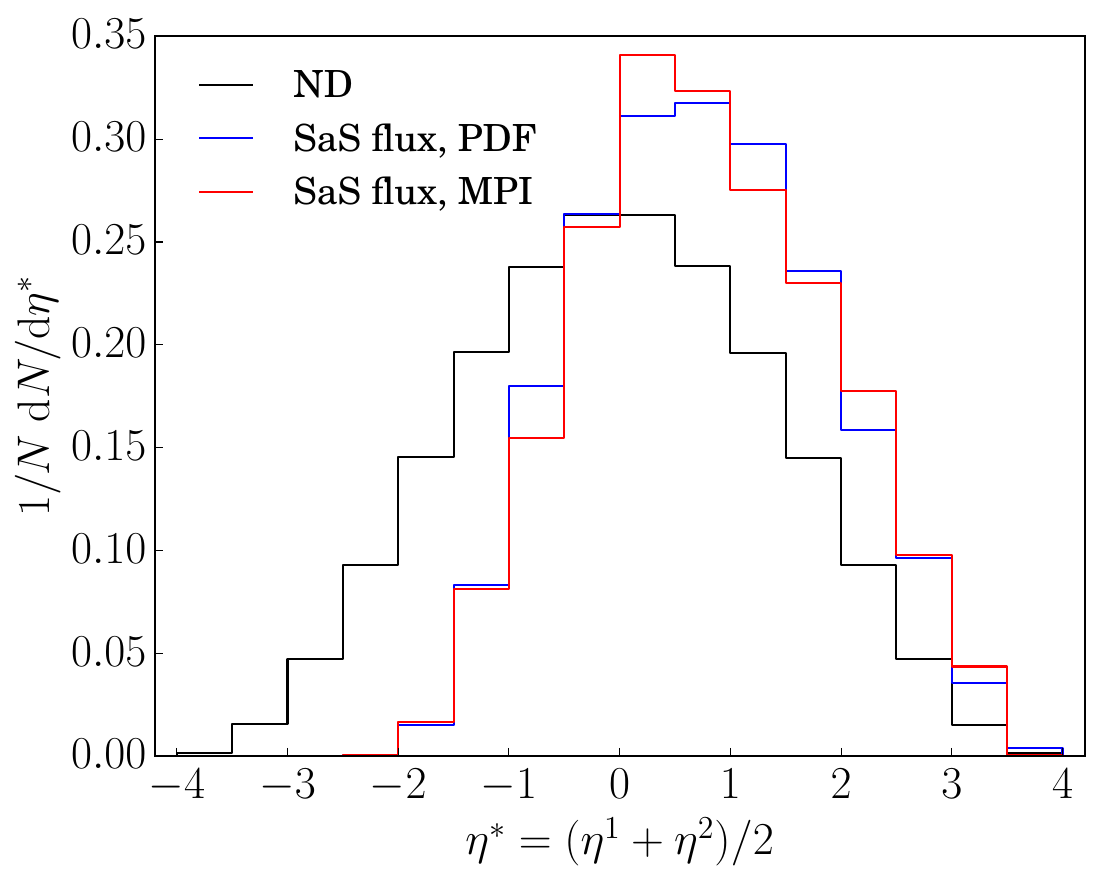}\\
(c)
\end{minipage}
\begin{minipage}[t]{0.5\textwidth}
\centering
\includegraphics[scale=0.4]{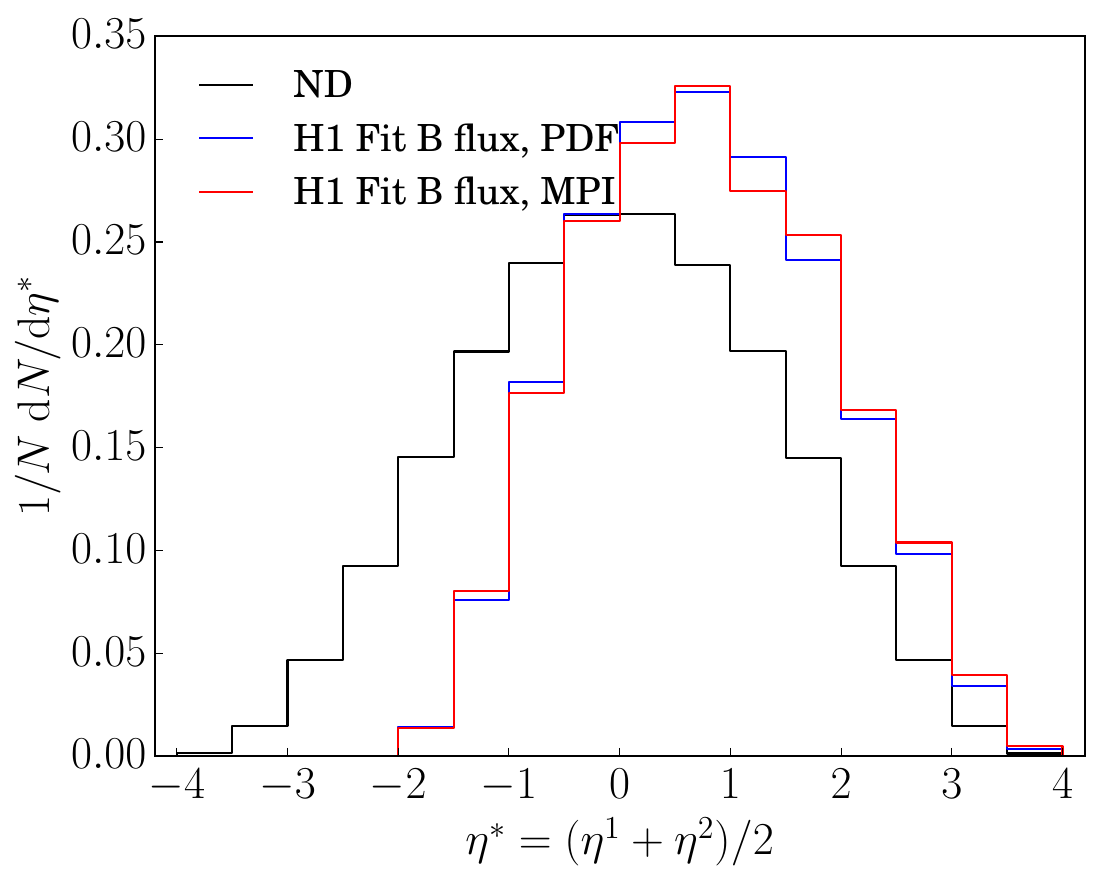}\\
(d)
\end{minipage}
\hfill
\begin{minipage}[t]{0.5\textwidth}
\centering
\includegraphics[scale=0.4]{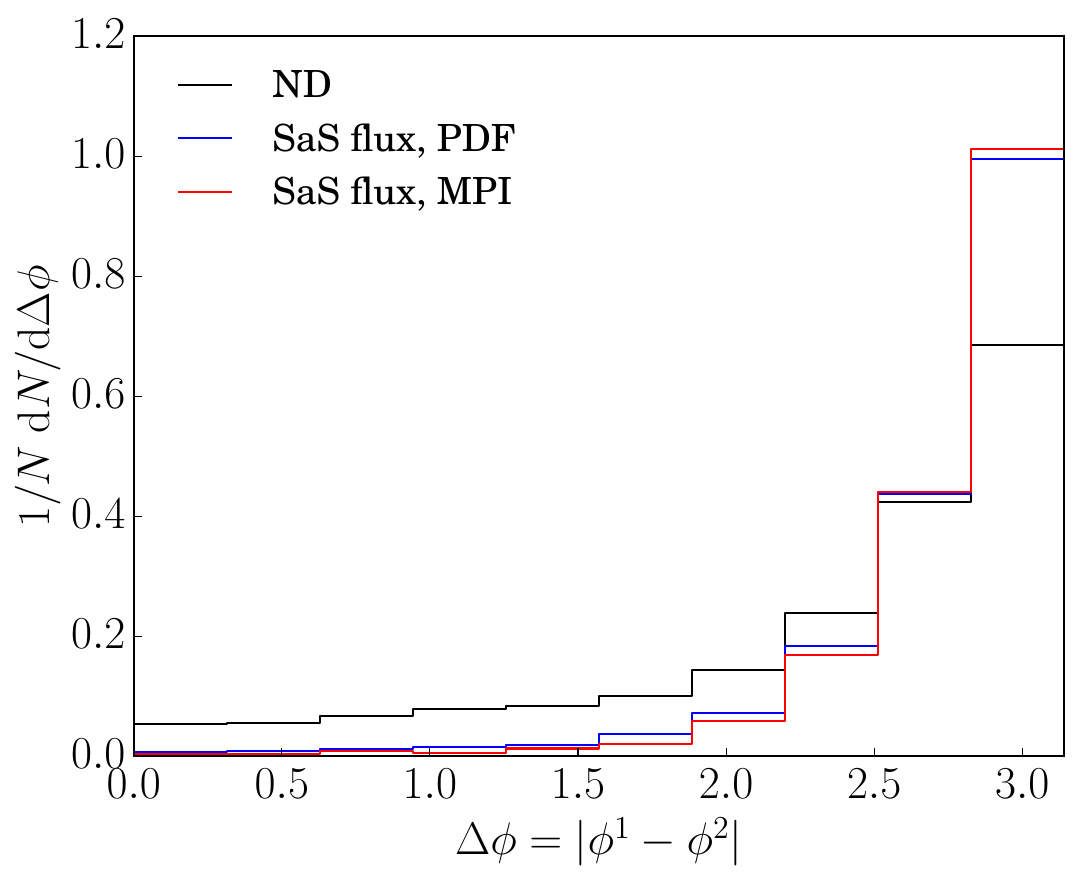}\\
(e)
\end{minipage}
\begin{minipage}[t]{0.5\textwidth}
\centering
\includegraphics[scale=0.4]{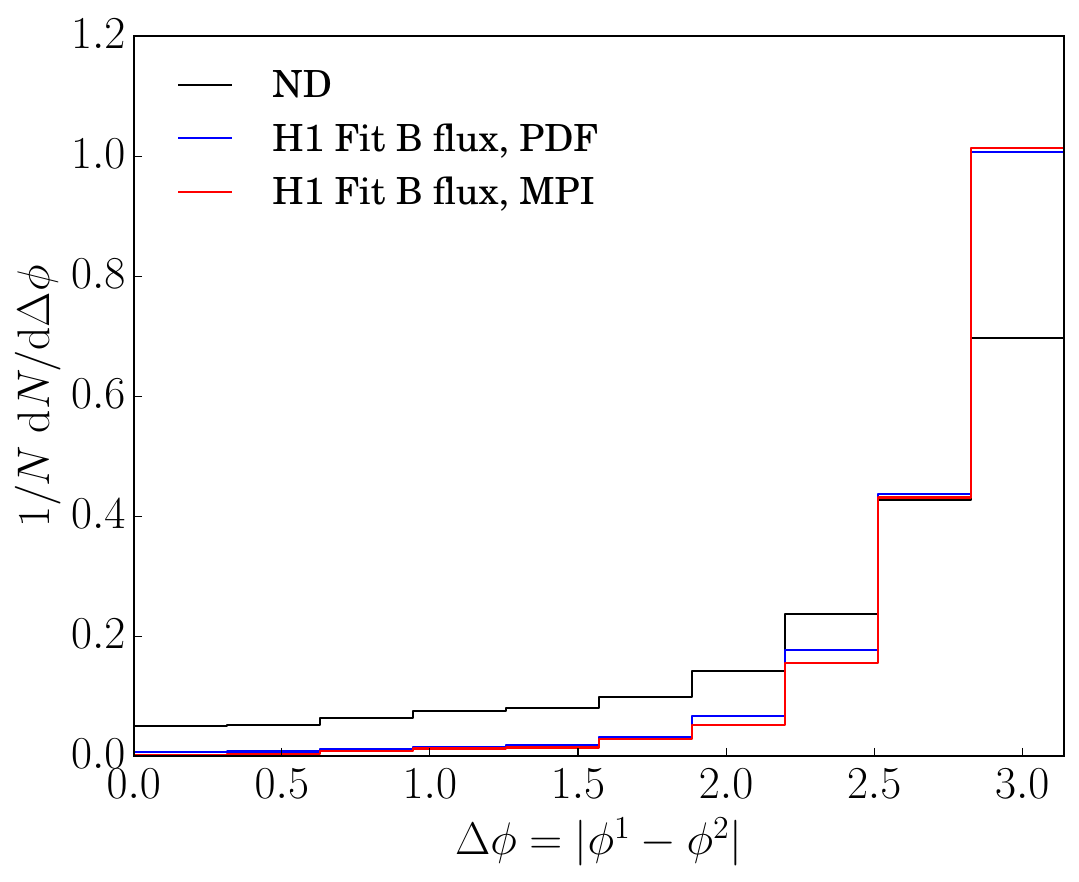}\\
(f)
\end{minipage}
\caption{\label{Fig:DiJetKin}
The mean $E_T$ of the leading jets in both SD and ND
events using (a) the SaS and (b) the H1 Fit B
flux. 
The mean $\eta$ of the leading jets in both SD and ND
events using (c) the SaS and (d) the H1 Fit B
flux. 
}
\end{figure}

Results on kinematical distributions using both the
SaS and the H1 Fit B flux 
are shown in figure~\ref{Fig:DiJetKin}. 
The SD $E_T^*$ distribution has a steeper falloff than the ND
distribution, indicating a lower center-of-mass energy in the
collision. Likewise the $\eta^*$ distribution is shifted
towards positive $\eta$, the proton direction, indicating a boost
of the center-of-mass system. The final kinematical distribution 
here is the difference in azimuthal angle between the two 
leading jets. This observable was not shown in the 1800 GeV analysis 
but in the 1960 GeV one. The SD events there show a tendency 
to be more back-to-back than the ND ones. This can 
also be attributed to the lower energy in the $\Pom\p$ 
collision than in the full $\p\pbar$ system, leaving less space 
for initial-state radiation.

\begin{figure}[tbp]
\begin{minipage}[t]{0.5\textwidth}
\centering
\includegraphics[scale=0.4]{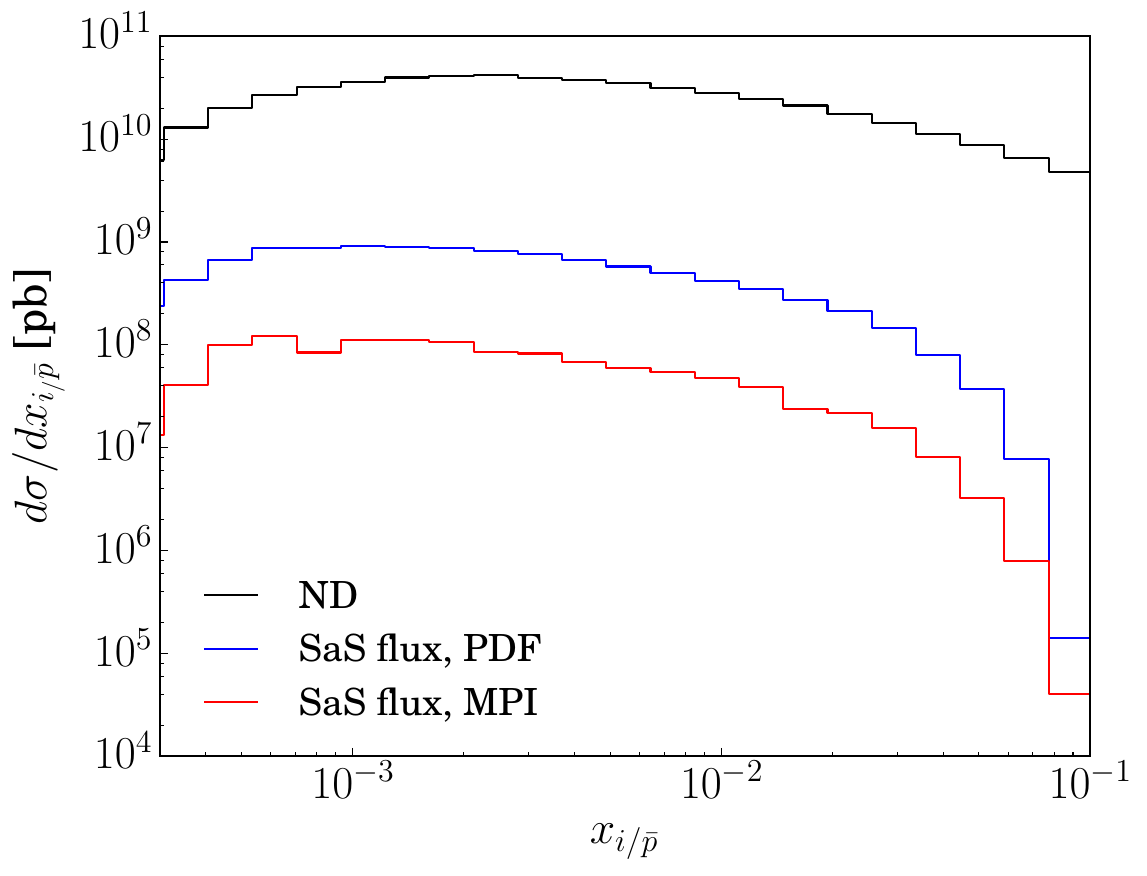}\\
(a)
\end{minipage}
\hfill
\begin{minipage}[t]{0.5\textwidth}
\centering
\includegraphics[scale=0.4]{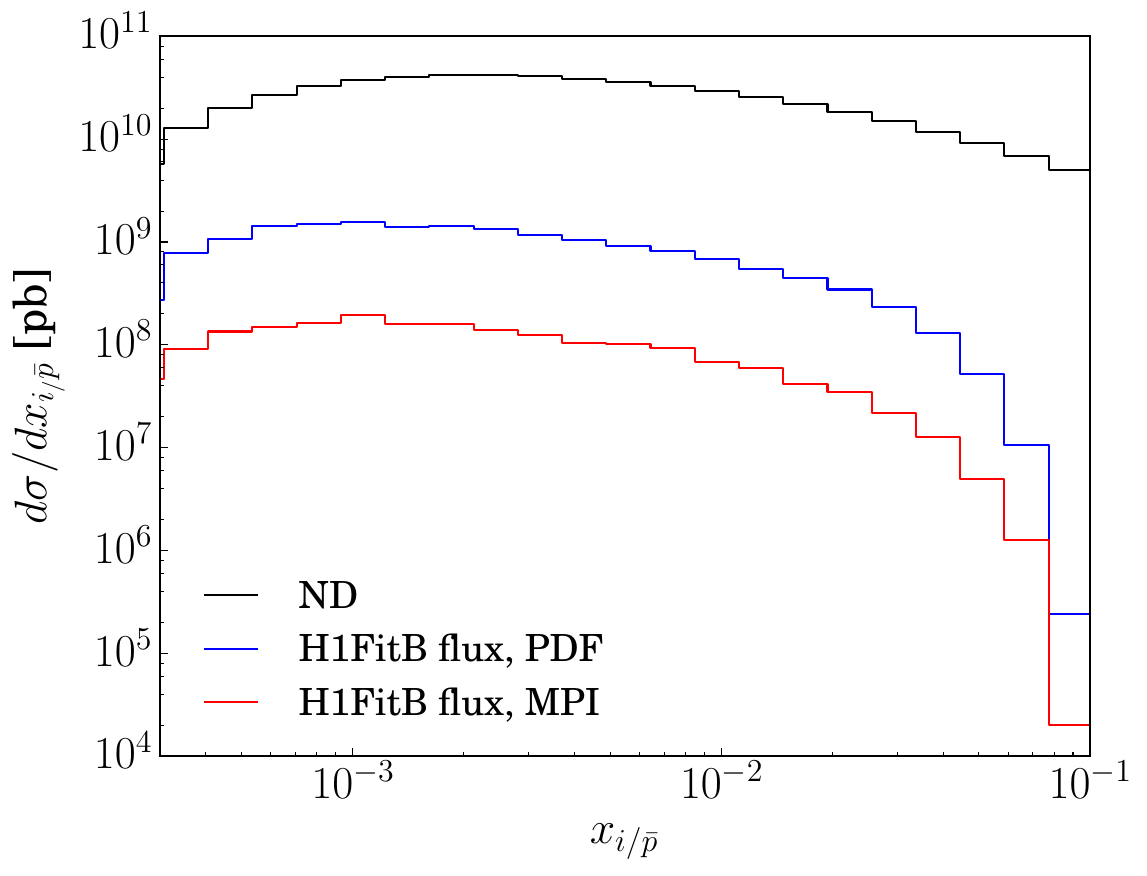}\\
(b)
\end{minipage}
\begin{minipage}[t]{0.5\textwidth}
\centering
\includegraphics[scale=0.4]{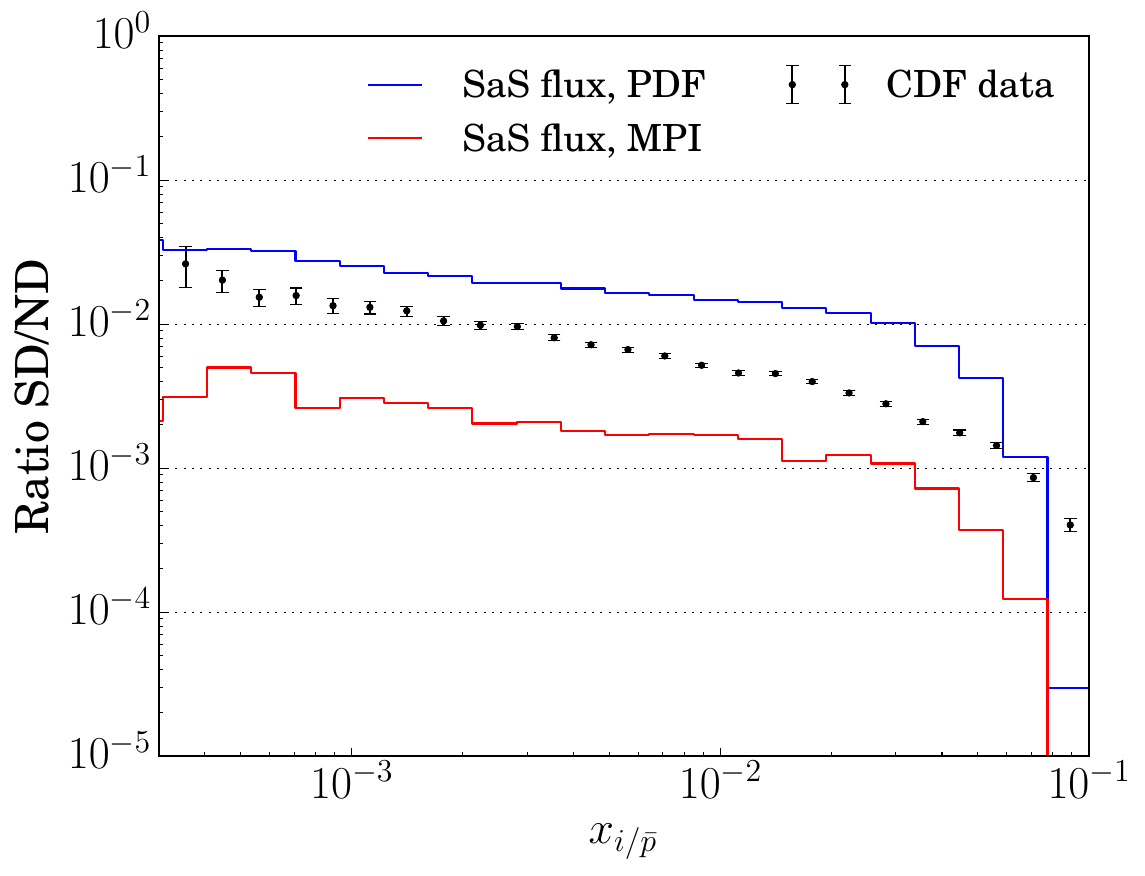}\\
(c)
\end{minipage}
\hfill
\begin{minipage}[t]{0.5\textwidth}
\centering
\includegraphics[scale=0.4]{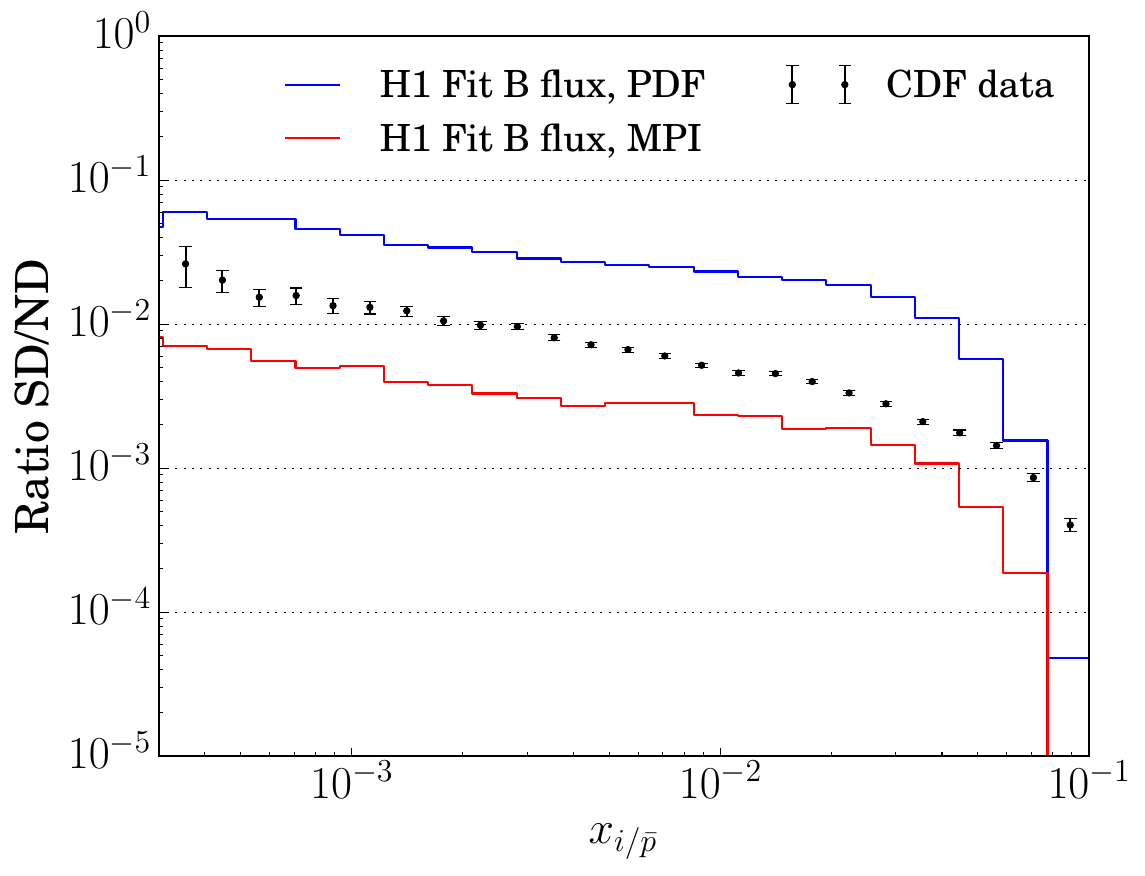}\\
(d)
\end{minipage}
\caption{\label{Fig:DiJetX}
The antiproton momentum fraction carried by the parton entering
the hard collision, for \textsc{Pythia}~8 compared with
CDF data. \textsc{Pythia} is run with the H1 Fit B LO PDF and 
(a) the SaS or (b) H1 Fit B flux. (c) and (d)
shows the ratio SD to ND using (a) and (b).}
\end{figure}

The momentum fraction of the antiproton carried by the subcollision
parton can be evaluated from the jets using
\begin{equation}
x = \frac{1}{\sqrt{s}} \sum_{i=1}^3 E_T^i e^{-\eta^i},
\end{equation}
where the sum is over the two leading jets,  
plus a third if it has $E_T > 5$ GeV. The result is shown in
figure~\ref{Fig:DiJetX}, for the two $\Pom$ fluxes used in
figure~\ref{Fig:DiJetKin}. As expected the SaS
flux, figure~\ref{Fig:DiJetX}a,
suppress the diffractive events too much, as the suppression
factor is too large compared to experimental value from CDF. 
With this flux, the PDF selected samples lie above the CDF data, 
but then drop by an order of magnitude by the MPI selection, to lie 
well below the data, by a factor of five. There is also some discrepancy in 
shape. Changing to the H1 Fit B flux, figure~\ref{Fig:DiJetX}b,
the PDF selected sample lies above the data as expected, with the
MPI selected sample a bit below, although only by a factor of
three. The suppression is still too large, and shapes still
disagree, but not as markedly as in figure~\ref{Fig:DiJetX}a. 

There are some aspects of the CDF article that we don't understand, 
however. The key figure~4 of \cite{Affolder:2000vb} is intended to
show the H1 predictions for the diffractive structure function 
along with the experimentally measured one. The information provided
on how the former prediction is obtained is inconsistent with the
curve shown, however, in normalization and shape. In the end we 
therefore put more faith in the suppression factor between CDF and HERA, 
already presented above, than in absolute numbers. Assuming we could 
have reproduced the CDF curve intended to represent
the predictions of the H1 PDFs, that then is suppressed by an
average multiplicative factor of $0.05-0.06$ in data but $0.11$ 
in our model, we should have been a factor of $\sim 2$ above data, 
which is inconsistent with the outcome in figure~\ref{Fig:DiJetX}.

\subsection{CMS diffractive contribution to dijet
production}

CMS has studied the diffractive contribution to dijet events at 
$\sqrt{s}=7$~TeV $\p\p$ collisions \cite{Chatrchyan:2012vc}, 
The cross section is presented as a function of
$\widetilde{\xi}$, an approximation to the fractional momentum
loss of the scattered proton correspinding to the $x_{\Pom}$
variable. Dijets were selected with $\pT>20$
GeV in the $|\eta|<4.4$ range using the anti-$\kT$ algorithm with
a cone size of $R=0.5$ \cite{Cacciari:2008gp}. $\widetilde{\xi}$ 
was reconstructed using
particles in the region $|\eta|<2.4$ with $\pT>0.2$ GeV for
charged particles as well as particles in the range $3.0 < |\eta|
< 4.9$ with $E>4$ GeV. To enhance the diffractive contribution
additional requirements was introduced, such that the minimum
rapidity gap was of 1.9 units (no particles was allowed in the
region $|\eta|>3$). Finally a cut on $\widetilde{\xi} <0.01$ was
introduced.

With these cuts, rapidity gap survival probabilities are in the range
$0.08\pm0.04$ (NLO) to $0.12\pm0.05$ (LO), where the NLO gap
survival probability was found using \textsc{PomPyt} and
\textsc{PowHeg}\cite{Alioli:2010xa}+\textsc{Pythia}~8 and the 
LO gap survival probability was
found using \textsc{PomPyt} and \textsc{PomWig}.
 
Implementing the same cuts in \textsc{Pythia}~8, using the SaS
flux and the H1 Fit B LO PDF gives a rapidity gap survival 
probability of 0.06, compatible with the CMS results. Changing
from the SaS flux to the H1 Fit B flux gives the same
suppression factor, but allows for more events to pass the
experimental cuts. We thus see the same trend as in the CDF
analysis, where the SaS flux is too restrictive at low
$x_{\Pom}$.

\subsection{ATLAS dijets with large rapidity gaps}

Recently, the ATLAS collaboration published a study of dijets
with large rapidity gaps in $\sqrt{s}=7$ TeV $\p\p$ collisions
\cite{Aad:2015xis}.
Dijets were selected with $\pT > 20$ GeV in the $|\eta|< 4.4$
range, and the cross section was measured in terms of
$\Delta\eta^F$, the size of the observed rapidity gap, as well as
in $\widetilde{\xi} = \sum\pT^ie^{\pm\eta^i} / \sqrt{s}$, 
the estimate of the fractional momentum
loss deduced from charged and neutral particles in the ATLAS
detector (the sign on $\eta$ depends on where in the detector
the largest gap is located). Cuts used in the analysis are listed in
table~\ref{Tab:ATLAScuts}. 

\begin{table}
\begin{center}
\begin{tabular}{|c|c|}
\hline
\multicolumn{2}{|c|}{Jet cuts}\\
\hline
Jet $E_T^{1,2}$ & $>$ 20 GeV \\
Jet $|\eta^{1,2}|$ & $<$ 4.4 \\
Anti-$\kT$ $\Delta R$ & 0.6 \\
\hline
\multicolumn{2}{|c|}{Neutral particles}\\
\hline
$|p|$ & $>$ 200 MeV\\
$|\eta|$ & $<$ 4.8\\
\hline
\multicolumn{2}{|c|}{Charged particles}\\
\hline
$|p|$ & $>$ 500 MeV or \\
$\pT$ & $>$ 200 MeV\\
$|\eta|$ & $<$ 4.8\\
\hline
\end{tabular}
\caption{\label{Tab:ATLAScuts}
Cuts used in \cite{Aad:2015xis}.}
\end{center}
\end{table}

Experimental results were compared with the \textsc{Pythia}~8 
soft diffractive framework, which predicts both the ND, SD and DD
contributions to the dijet production. Three different flux models
were compared: SaS, Donnachie-Landshoff and MBR. All three 
predict cross sections in the range of the data, without any need
for additional gap survival probability factors. The \textsc{PomWig} 
generator \cite{Cox:2000jt}, on the other hand, needed an additional
suppression of
$S^2=0.16\pm0.04\textrm{(stat)}\pm0.08\textrm{(sys)}$ in order to
describe data. 

In this section we use the new model for hard diffraction to
study the same cross sections. The new model currently only includes 
the SD contribution, hence we will not be able to describe all
aspects of data, especially in the high-$\Delta\eta^F$ and
low-$\widetilde{\xi}$-regions, where the SD and DD contributions
are comparable in size, at least according to the soft
diffraction model available in \textsc{Pythia}~8.
We could also expect the normalisation of the SD events
obtained with the hard diffraction framework to be lower than
in the soft one and thus in data, because of the difference
in normalisation between the two frameworks (cf.
section~\ref{Sec:SoftHard}). The ND contribution should not differ from 
the ATLAS analysis, however, since no changes have been implemented in this
framework. 

\begin{figure}[tbp]
\begin{minipage}[t]{0.5\textwidth}
\centering
\includegraphics[scale=0.4]{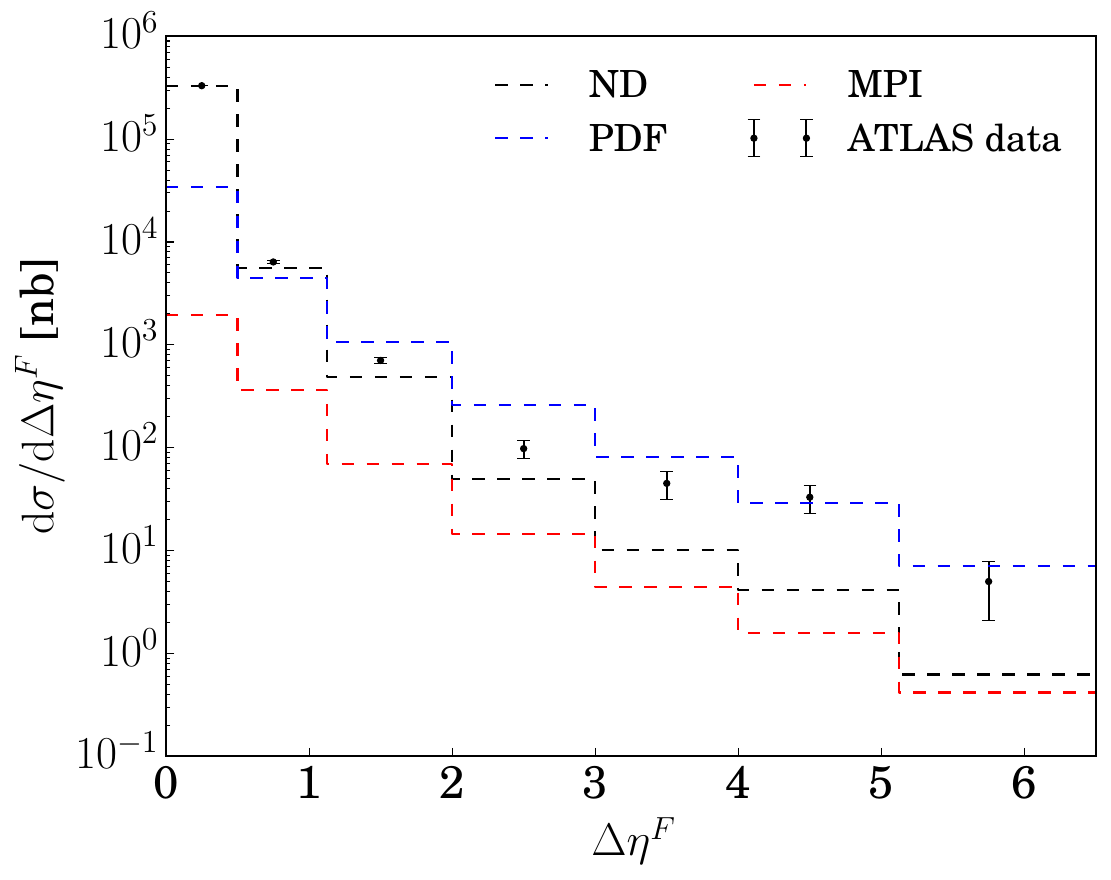}\\
(a)
\end{minipage}
\hfill
\begin{minipage}[t]{0.5\textwidth}
\centering
\includegraphics[scale=0.4]{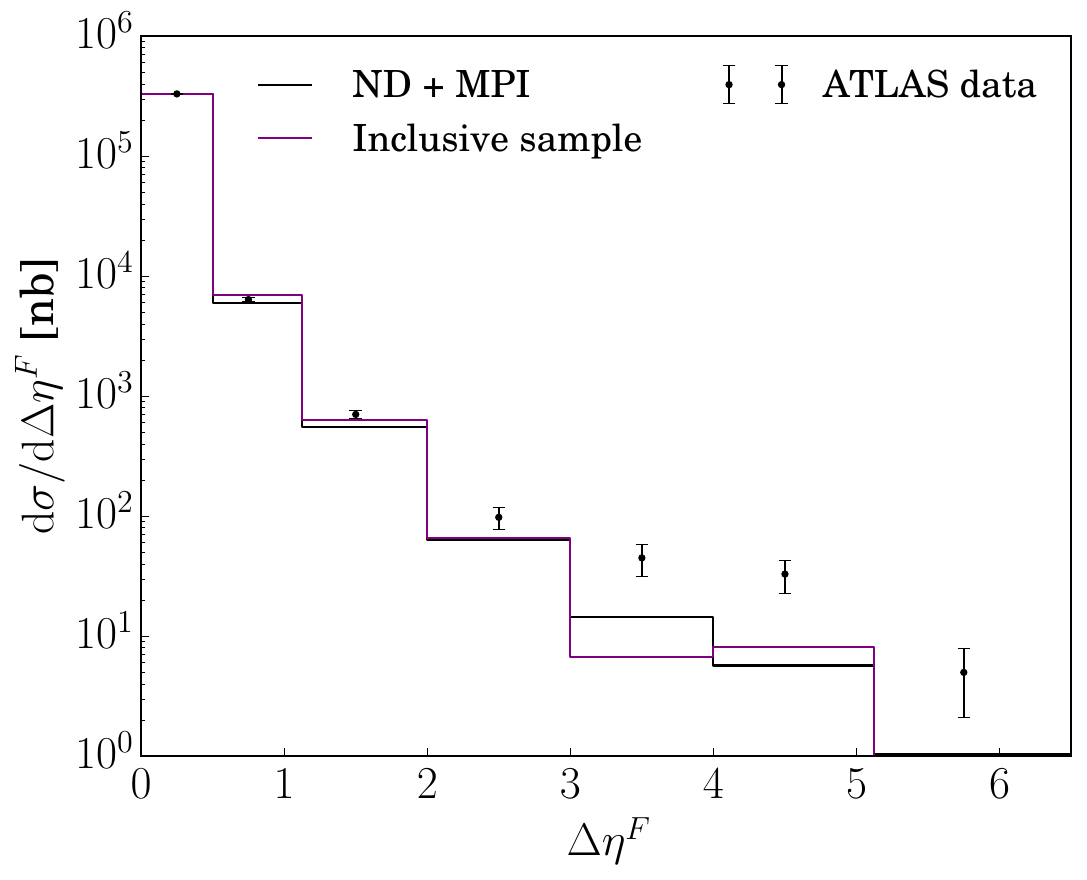}\\
(b)
\end{minipage}
\begin{minipage}[t]{0.5\textwidth}
\centering
\includegraphics[scale=0.4]{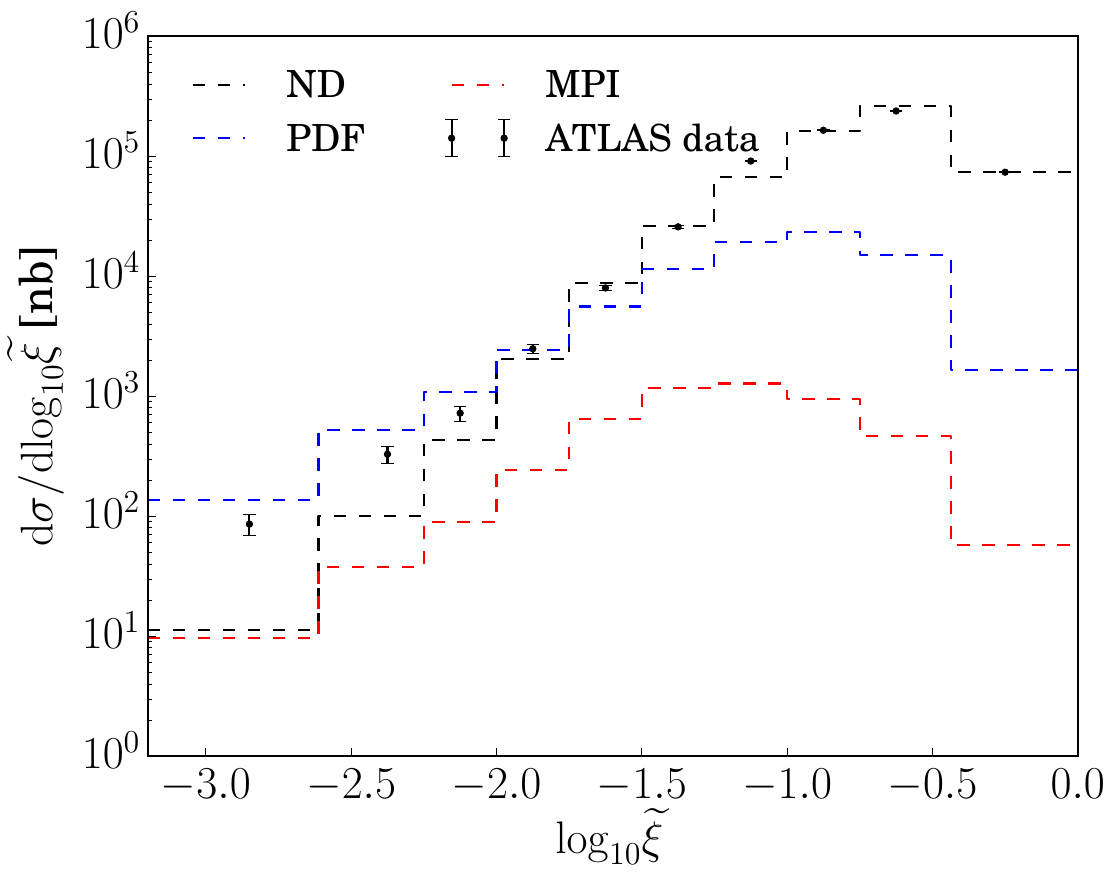}\\
(c)
\end{minipage}
\hfill
\begin{minipage}[t]{0.5\textwidth}
\centering
\includegraphics[scale=0.4]{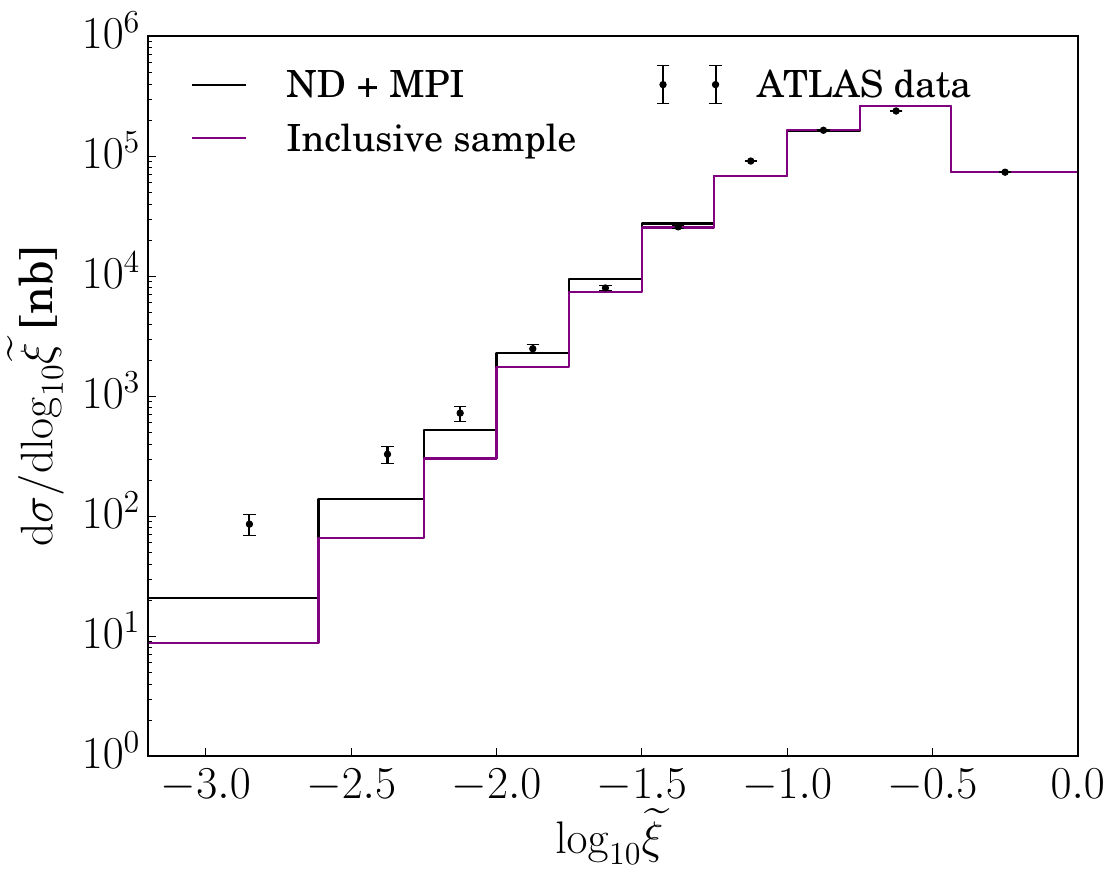}\\
(d)
\end{minipage}
\caption{\label{Fig:ATLAScomp}
The dijet cross sections as a function of the size of the
rapidity gap (a), (b) and the fractional momentum loss of the proton
(c), (d). Compared to the hard diffraction model of \textsc{Pythia}~8
using the SaS flux and H1 Fit B LO PDF. In (b)
and (d) the ND + MPI sample is a sum of the black and red dotted
lines from (a) and (c), whereas the inclusive sample are
generated directly with \textsc{Pythia}~8. Only
statistical errors are included in the ATLAS errorbars.}
\end{figure}

The ND distribution was normalized to data, where the
normalization factor was found using the
first bin of the $\Delta\eta^F$ distribution. This approach has
also been used in our analysis, although when generating an
inclusive sample (e.g.\ the purple distribution in
figures~\ref{Fig:ATLAScomp}b and \ref{Fig:ATLAScomp}d) this
normalization is applied to the full sample, unlike in the ATLAS
paper. In this sample, no classification of events occurs, hence
the normalization cannot be performed only on the ND sample. In
the exclusive samples, the distinction between ND and SD is
performed, and we can apply the normalization to only the ND
sample (cf. the black distribution in figures~\ref{Fig:ATLAScomp}b 
and \ref{Fig:ATLAScomp}d).

In figure~\ref{Fig:ATLAScomp} we show the results obtained with the
model for hard diffraction. Three samples are compared:
ND, PDF-selected SD and MPI-selected SD. Note that the
MPI-selected sample lies about a factor of 10 below the PDF-selected
one, as usual, and that the suppression due to the MPI-framework is
constant over both intervals. The new model undershoots the
data in the regions where the DD contribution is non-negligible
($\Delta\eta^F > 1$ and $\mathrm{log}_{10}\widetilde{\xi} < -0.5$). 
When this contribution is included in the framework, a better
agreement with data should be possible, and overall the picture  
should be consistent with the soft diffractive framework.

\section{Summary and outlook}

In this article we have studied hard diffraction by combining 
two concepts, the Ingelman--Schlein picture of a Pomeron and 
the \textsc{Pythia} model for multiparton interactions. The   
Pomeron fluxes and PDFs are mainly extracted from HERA data,
while the MPI picture (and several other relevant physics 
components) makes use of a broader spectrum of Tevatron and 
LHC data. This combination allows us, in principle, 
to predict all physical quantities of hard diffractive events,
from rapidity gap sizes to charged multiplicity distributions, 
but most importantly the fraction of diffractive events for 
any hard process. 

Reality is not quite as simple, however. In this article we have 
studied the different assumptions that go into a detailed 
framework, and explored the inherent uncertainties. One part 
concerns the assumed Pomeron flux and PDFs, where particularly 
the latter is dominated by one source only, namely the H1 analyses,
making it difficult to assess to full range of uncertainty. 
Another part concerns the MPI framework, which enters twice.
When used the first time, to determine the diffractive MPI survival, 
it involves parameters already tuned to nondiffractive data,  
so narrowly constrained in principle. There could still be leeway,
e.g.\ if we were to use other parton showers that give less/more 
activity at small $\pT$ scales, the average number of MPIs would 
have to rise/drop to compensate. Thus our studies focus on the sensitivity
of some key parameters of the framework. When the MPIs are used 
the second time, inside the diffractive subsystem itself, the 
level of uncertainty is considerably higher. A key example is 
the impact-parameter picture of the $\Pom\p$ subcollision,
notably how impact parameters are related between the $\p\p$ and 
$\Pom\p$ steps of an event. 

Our studies puts the finger on our still limited understanding 
of diffraction, also when restricted to the Pomeron framework, 
which is only one model class for diffraction. Further, we provide 
computer  code that can be used to compare with data for hard 
diffractive processes at the LHC. It thus can be used as a 
``straw man'' model, where differences between predictions and data 
can help pave the way for a deeper understanding and more accurate 
models. Specifically, with a generator it is possible both to 
emulate the experimental diffractive trigger and to compare
the resulting event properties, both of which are considerably more
complicated for analytical models.

Comparisons with data have shown qualitative agreements in many 
respects, but maybe less so than one could have hoped for. For the
Tevatron we face the problem of trying to understand 15 years old
analyses, with uncertain results. The main message probably is that 
the overall Tevatron suppression factor of $\sim 10 - 20$, relative 
to HERA-based extrapolations, agrees well with what our model gives 
from the MPI selection step. For the future it will therefore be more 
interesting to compare with LHC studies, in particular those available
in Rivet. 

It is well known that the existing \textsc{Pythia} model for soft
diffraction is not fully describing the existing LHC data; at 
places the difference can be up to a factor of two. Similarly we 
have seen less-than-perfect agreement for the hard diffractive 
processes studied in this article. There is therefore room for 
improvements in both areas, and also for work to bring the two 
approaches in closer contact. As one simple example, the soft model
currently does not involve a MPI survival step, and therefore the
Pomeron flux does not have to be normalized in the same way in the
two cases. The intention is to study such issues closer, and to
provide an improved description of diffractive cross sections,
both integrated and differential ones.   

\acknowledgments

We thank Marek Tasevsky for detailed
descriptions of the ATLAS analysis and providing the data used in
figure~\ref{Fig:ATLAScomp}. 
Work supported in part by the Swedish Research Council, contract number
621-2013-4287, and in part by the MCnetITN FP7 Marie Curie Initial 
Training Network, contract PITN-GA-2012-315877. 
This project has also received funding from the European Research
Council (ERC) under the European Union's Horizon 2020 research
and innovation programme (grant agreement No 668679).

\end{document}